\documentclass[11pt,a4paper]{article}
\pdfoutput=1
\usepackage{jheppub}
\usepackage{amsthm,amsbsy,amsfonts,mathrsfs,enumerate,float,wrapfig,amsmath}
\usepackage{array,comment,braket}
\usepackage{ytableau,tikz}
\usepackage[utf8]{inputenc}
\usepackage{subfigure}

\allowdisplaybreaks[4]

\newcommand{\be}{\begin{equation}}
\newcommand{\ee}{\end{equation}}
\newcommand{\ba}{\begin{eqnarray}}
\newcommand{\ea}{\end{eqnarray}}
\newcommand{\nn}{\nonumber}

\newcommand{\cS}{\mathcal{S}}

\newcommand{\cN}{\mathcal{N}}
\newcommand{\lt}{\left}
\newcommand{\rt}{\right}

\title{
More on topological vertex formalism for 5-brane webs with O5-plane}
\author[a]{Hirotaka Hayashi,}
\author[b,c]{Rui-Dong Zhu}

\affiliation[a]{Department of Physics, School of Science, Tokai University,\\ 4-1-1 Kitakaname, Hiratsuka-shi, Kanagawa 259-1292, Japan}
\affiliation[b]{Institute for Advanced Study \& School of Physical Science and Technology,\\ Soochow University, Suzhou 215006, China
}
\affiliation[c]{School of Theoretical Physics, Dublin Institute for Advanced Studies\\ 10 Burlington Road, Dublin, Ireland}

\emailAdd{h.hayashi@tokai.ac.jp}
\emailAdd{rdzhu@suda.edu.cn}

\abstract{
We propose a concrete form of a vertex function, which we call O-vertex, for the intersection between an O5-plane and a 5-brane in the topological vertex formalism, as an extension of the work of \cite{Kim-Yagi}. Using the O-vertex it is possible to compute the Nekrasov partition functions of 5d theories realized on any 5-brane web diagrams with O5-planes. We apply our proposal to 5-brane webs with an O5-plane and compute the partition functions of pure SO($N$) gauge theories and the pure $G_2$ gauge theory. The obtained results agree with the results known in the literature. We also compute the partition function of the pure SU(3) gauge theory with the Chern-Simons level $9$. At the end we rewrite the O-vertex in a form of a vertex operator.
}

\begin{document}
\preprint{DIAS-STP-20-10}
\maketitle

\section{Introduction}

The topological vertex formalism \cite{Iqbal:2002we, AKMV,Awata:2005fa,IKV} has long been known as an alternative method to compute the partition function of five-dimensional (5d) $\cN=1$ supersymmetric field theories on the $\Omega$-background $\mathbb{C}^2_{q,t}\times S^1$. Unlike the original way using the localization technique \cite{LMNS,Moore:1997dj,NekrasovInstanton}, the topological vertex formalism computes the partition function in a Feynman-diagram-like manner. That is to say, given a diagram corresponding to a local Calabi-Yau threefold realizing a 5d theory by an M-theory compactification, we assign a function called the topological vertex to each vertex in the diagram, and sum over all possible internal states labeled by Young diagrams and weighted by K\"ahler parameters (which plays the role of a propagator). Through a chain of string dualities \cite{Leung-Vafa}, one can see that these toric diagrams are mapped to $(p,q)$ 5-brane web diagrams in type IIB string theory. From this viewpoint 5d $\cN=1$ supersymmetric field theories are realized on the 5-brane webs \cite{Aharony:1997ju, AHK}. 

Although the topological vertex formalism is an elegant and systematic way to compute related physical quantities, the original formalism only applies to (dual) toric diagrams. In terms of the corresponding 5d gauge theories, the application had been limited mostly to linear quiver theories with SU-type gauge groups. Since then, the topological vertex formalism has been extended in various ways to compute the partititon functions of gauge theories with different gauge groups. One is the application to web diagrams which contain configurations where 5-branes jump over other 5-branes \cite{Hayashi:2013qwa,Hayashi:2014wfa, Hayashi:2015xla}. Then it is possible to compute the partition functions of Sp($N$) gauge theories \cite{Hayashi:2016jak, Sp-antisym} from the 5-brane webs constructed in \cite{Bergman:2015dpa}. Another extension is the application to web diagrams with O5-planes \cite{Kim-Yagi}. By combining the two methods the partition functions of $G_2$ gauge theories have been computed \cite{G-type}. Furthermore, web diagrams each of which are made by gluing three or four (dual) toric diagrams have been constructed in \cite{Ohmori-Hayashi} and the method developed there computes the partition functions of SO$(2N)$ gauge theories and also the pure $E_6, E_7, E_8$ gauge theories. Recently the topolgical vertex formalims has been also extended in other directions \cite{D-type,Awata:2017lqa,Zhu-elliptic,Foda-Zhu,Zenkevich:2018fzl,Bourgine:2019phm,Kimura:2019gon,Zenkevich:2019ayk,Kimura:2020lmc, Kim:2020npz}.

In this paper we consider a further extension of the formalism in \cite{Kim-Yagi} for 5-brane webs with O5-planes. To apply the topological vertex to 5-brane webs with an O5-plane ref.~\cite{Kim-Yagi} utilizes ``generalized flop transitions'' in \cite{Hayashi:2017btw}. After the transition and taking a different fundamental region, the brane configuration around an O5-plane becomes locally a strip diagram and hence we can apply the topological vertex. However the generalized flop transition is applicable to configurations that yield an Sp-type gauge group and the formalism has not been applied to for example the brane webs for pure SO($N$) gauge theories. In this paper we propose a vertex function labeled by a Young diagram which applies to an intersection point between an O5$^-$-plane, a $(2, 1)$ 5-brane and an O5$^+$-plane, where the Young diagram is assigned to the $(2, 1)$ 5-brane, by making use of the formalism in \cite{Kim-Yagi}. We call the vertex function O-vertex. With the O-vertex, it is not necessary to contain configurations of an Sp-type gauge group for applying the topological vertex and it becomes possible to apply the topological vertex to any 5-brane web diagrams with O5-planes. We will exemplify the O-vertex by computing the partition functions of various pure SO($N$) gauge theories and also the pure $G_2$ gauge theory. We also compute the partition function of the pure SU($3$) gauge theory with the Chern-Simons (CS) level $9$, which was proposed in \cite{Jefferson:2017ahm} and was also constructed from a geometry \cite{Jefferson:2018irk} and from a 5-brane web \cite{Hayashi:2018lyv} approach. Furthermore we consider integrating the O-vertex into computations using vertex operators \cite{Okounkov:2003sp,IKV}. We propose an expression of a vertex operator associated to the O-vertex.

This paper is organized as follows: in section \ref{s:O-vert}, we propose the O-vertex by utilizing the formalism in \cite{Kim-Yagi}. In section \ref{s:example}, we validate our proposal by computing the partition functions of the pure SO($N$) $(N=4, 5, 6, 7, 8)$ gauge theories and also the pure $G_2$ gauge theory from 5-brane webs with an O5-plane. We compare the results with the known expressions and then find perfect agreement. We also compute the partition function of the pure SU($3$) gauge theory with the CS level $9$. Finally in section \ref{s:O-vert-op}, we reconsider the O-vertex from vertex operator computations and we propose a vertex operator associated to the O-vertex. Section \ref{s:concl} summarizes our results and discuss possible future directions. In appendix \ref{a:O-vert} we list explicit expressions of the O-vertex and describe its properties.  Appendix \ref{a:schur} summarizes some formulae of the Nekrasov partition functions and the Schur functions, which we use in various computations.

\section{O-vertex}\label{s:O-vert}

In this section, we propose 
a vertex function, O-vertex, which can be applied to the intersection point of a 5-brane with an O5-plane for computing the partition functions of 5d theories realized on 5-brane webs with an O5-plane. 
We first give a brief review on the topological vertex formalism for 5-brane webs with an O5-plane proposed in \cite{Kim-Yagi}, 
and then extend the formalism by introducing the O-vertex in section \ref{s:prop-O-vert}.  
In section \ref{s:Higgsing-tild-O}, we further explain how to use the O-vertex for 5-brane webs with an $\widetilde{\text{O5}}$-plane by exploiting the Higgsing procedure discussed in \cite{Zafrir:2015ftn}.

\subsection{Topological vertex formalism with an O5-plane}\label{s:r-O-top}

The topological vertex formalism is a convenient method to compute the topological string partition functions associated to toric Calabi-Yau threefolds. 
Given a toric diagram, or equivalently a $(p, q)$ 5-brane web diagram in type IIB string theory, we assign the topological vertex to each vertex appearing in the diagram. The topological vertex is labeled by Young diagrams which are assigned to all the internal lines. Then we take the product of all the topological vertices weighted by K\"ahler parameters and framing factors for the internal lines and sum over all the Young diagram. 
The explicit form of the (unrefined) topological vertex labeled by Young diagrams $\lambda_1, \lambda_2, \lambda_3$ in clockwise direction  
is given by \cite{AKMV}, 
\be
C_{\lambda_1\lambda_2\lambda_3}=q^{\frac{\kappa(\lambda_2)}{2}+\frac{\kappa(\lambda_3)}{2}}s_{\lambda_3}(q^{-\rho})\sum_{\nu}s_{\lambda_1^t/\nu}(q^{-\rho-\lambda_3})s_{\lambda_2/\nu}(q^{-\rho-\lambda^t_3}),\label{topvertex}
\ee
where $\kappa(\lambda):=2\sum_{(i,j)\in\lambda}(j-i)$, and 
\be
q^{-\rho-\lambda}=\{q^{i-\frac{1}{2}-\lambda_i}\}_{i=1}^\infty=\{q^{\frac{1}{2}-\lambda_1},q^{\frac{3}{2}-\lambda_2},q^{\frac{5}{2}-\lambda_3},\dots\}.
\ee
$s_{\mu/\nu}$ is the skew Schur function. Basic properties of the skew Schur function as well as the Schur function are summarized in appendix \ref{a:Schur}. The framing factor is also needed to be assigned to each internal line depending on the local geometry. Let us first define 
\be
f_\lambda:=(-1)^{|\lambda|}q^{\frac{\kappa(\lambda)}{2}}. \label{framing}
\ee
Suppose the local geometry around an internal line with a Young diagram $\lambda$ assigned to it is given by 
\begin{align}
\begin{tikzpicture}
\draw[->] (0,0)--(0.5,0);
\draw (1,0)--(0.5,0);
\draw[->] (0,0)--(-1,0.5);
\draw (0,0)--(0.5,-0.5);
\draw (1,0)--(1,0.5);
\draw[->] (1,0)--(1.5,-0.5);
\node at (0.5,0) [above] {$\lambda$};
\node at (-1,0.5) [above] {$(a,b)$};
\node at (1.5,-0.5) [below] {$(c,d)$};
\end{tikzpicture}
\end{align}
then the framing factor associated to the internal line in the above diagram is
\ba
f_\lambda^{-(ad-bc)}.
\ea

The information above is enough to compute the topological string partition functions for toric Calabi-Yau threefolds. In fact one can apply the topological vertex to certain non-toric diagrams. Higgsing a 5d theory realized on a 5-brane web dual to a toric diagram may yield a non-toric diagram where a line in the diagram jumps over other lines \cite{Benini:2009gi}. A typical configuration in such a case is given by 
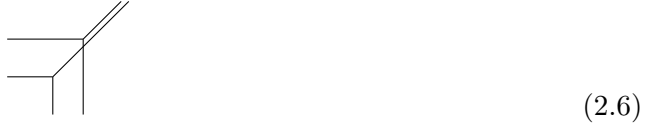
\begin{align}\label{nontoric0}
\begin{tikzpicture}
\draw (0,0)--(0,-1);
\draw (0,0)--(-1,0);
\draw (0,0)--(1/2,1/2);
\draw (-1/2+0.1,-1/2)--(-1,-1/2);
\draw (-1/2+0.1,-1/2)--(-1/2+0.1,-1);
\draw (-1/2+0.1,-1/2)--(1/2+0.1,1/2);
\end{tikzpicture}
\end{align}
where the lines with the slope $1$ are external lines. For applying the topological vertex to the diagram \eqref{nontoric0} we deform the diagram in the following way \cite{Hayashi:2013qwa, Hayashi:2015xla}
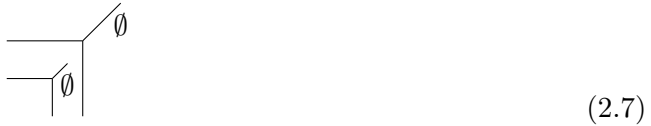
\begin{align}\label{nontoric1}
\begin{tikzpicture}
\draw (0,0)--(0,-1);
\draw (0,0)--(-1,0);
\draw (0,0)--(1/2,1/2);
\node at (1/2,1/2) [below] {$\emptyset$};
\draw (-1/2+0.1,-1/2)--(-1,-1/2);
\draw (-1/2+0.1,-1/2)--(-1/2+0.1,-1);
\draw (-1/2+0.1,-1/2)--(-1/2+0.1+0.2,-1/2+0.2);
\node at (-1/2+0.1+0.2,-1/2+0.2) [below] {$\emptyset$};
\end{tikzpicture}
\end{align}
Then we apply the topological vertex to the diagram \eqref{nontoric1} with the trivial Young diagrams assigned to both the lines with the slope $1$. K\"ahler parameters we assign in a deformed diagram need to respect the configuration before the deformation. For example, the legnth between the vertical lines in \eqref{nontoric1} should be equal to the length between the horizontal lines in the diagram. We will make use of this technique in section \ref{s:example}.

In \cite{Kim-Yagi} the topological vertex formalism has been further extended to 5-brane webs with O5-planes. We review the formalism by using an example which we will utilize later. 
We start from the following diagram, 
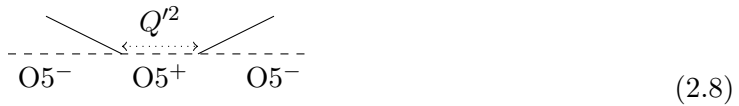
\begin{align}
\begin{tikzpicture}
\draw [dashed] (-2,0)--(2,0);
\draw (0.5,0)--(1.5,0.5);
\draw (-0.5,0)--(-1.5,0.5);
\draw [dotted,<->] (-0.5,0.1)--(0.5,0.1);
\node at (0,0) [below] {O5$^+$};
\node at (-1.5,0) [below] {O5$^-$};
\node at (1.5,0) [below] {O5$^-$};
\node at (0,0.1) [above] {$Q'^2$};
\end{tikzpicture}\label{fig:start-O-vert0}
\end{align}
with the K\"ahler paramter $Q'$. This diagram may be thought of as the one for the pure ``$Sp(0)$" gauge theory, which is trivial. Instead of directly applying the topological vertex to the diagram of \eqref{fig:start-O-vert0}, we may change the K\"ahler parameter $Q$ and deform the diagram into \cite{Hayashi:2017btw}
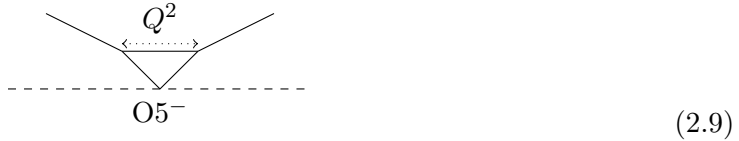
\begin{align}
\begin{tikzpicture}
\draw [dashed] (-2,0)--(2,0);
\draw (0,0)--(0.5,0.5);
\draw (0,0)--(-0.5,0.5);
\draw (-0.5,0.5)--(0.5,0.5);
\draw (0.5,0.5)--(1.5,1);
\draw (-0.5,0.5)--(-1.5,1);
\node at (0,0) [below] {O5$^-$};
\draw [dotted,<->] (-0.5,0.6)--(0.5,0.6);
\node at (0,0.6) [above] {$Q^2$};
\end{tikzpicture}
\label{fig:start-O-vert}
\end{align}
where $Q' = Q^{-1}$. Then we use the mirror image for the half of the diagram of \eqref{fig:start-O-vert}, which yields
\begin{align}
\begin{tikzpicture}
\draw [dashed] (-2,0)--(2,0);
\draw (0,0)--(0.5,0.5);
\draw (0,0)--(-0.5,-0.5);
\draw (0,0.5)--(0.5,0.5);
\draw (0,-0.5)--(-0.5,-0.5);
\draw (0.5,0.5)--(1.5,1);
\draw (-0.5,-0.5)--(-1.5,-1);
\node at (-1,0) [below] {O5$^-$};
\draw[<-] (0.1,0.5)--(0.2,0.5);
\node at (0,0.5) [left] {$\lambda$};
\node at (0,-0.5) [right] {$\lambda$};
\draw[->] (-0.2,-0.5)--(-0.1,-0.5);
\end{tikzpicture}
\label{fig:prep-Sp0s}
\end{align}
Namely we take a different fundamental region compared to \eqref{fig:start-O-vert}. Since this is a strip diagram it is possible to apply the topological vertex to \eqref{fig:prep-Sp0s}. 
The only unusual thing compared to an ordinary strip diagram is that we need to sum over all possible Young diagram configurations $\lambda$ assigned to the horizontal legs.

A subtlety is that the Young diagram $\lambda$ assigned to the horizontal leg of the mirror image is transposed compared to the original diagram and we need to be careful of the framing factor for the glued horizontal legs. 
Assuming that we can define a framing factor labeled only by the Young diagram for the gluing legs, we can determine the framing factor by requiring that the application of the topological vertex to the following local diagrams, which should represent the equivalent diagrams due to the orientifold, gives the same partition functions,
\begin{align}
\begin{tikzpicture}
\node at (0.5+1.5,1) [above] {$\emptyset$};
\node at (-1.5,1) [above] {$\emptyset$};
\node at (-0.3,0.2) [left] {$\emptyset$};
\node at (0.6+0.2,0.2) [right] {$\emptyset$};
\draw [dashed] (-2,0)--(2,0);
\draw (0.5+0.2,0.2)--(0.5+0.5,0.5);
\draw (-0.2,0.2)--(-0.5,0.5);
\draw (-0.5,0.5)--(0.5+0.5,0.5);
\draw[<-] (0.25-0.1,0.5)--(0.25+0.1,0.5);
\node at (0.25,0.5) [below] {$\lambda$};
\draw (0.5+0.5,0.5)--(0.5+1.5,1);
\draw (-0.5,0.5)--(-1.5,1);
\node at (0.25,0) [below] {O5$^-$};
\draw [dotted,<->] (-0.5,0.7)--(0.5+0.5,0.7);
\node at (0.25,0.7) [above] {$Q^2$};
\node at (6+1.5,1) [above] {$\emptyset$};
\node at (6.1+0.2,0.2) [right] {$\emptyset$};
\node at (6-0.3,-0.2) [left] {$\emptyset$};
\node at (6-1.5,-1) [below] {$\emptyset$};
\draw [dashed] (6-2,0)--(6+2,0);
\draw (6+0.2,0.2)--(6+0.5,0.5);
\draw (6-0.2,-0.2)--(6-0.5,-0.5);
\draw (6+0,0.5)--(6+0.5,0.5);
\draw (6+0,-0.5)--(6-0.5,-0.5);
\draw (6+0.5,0.5)--(6+1.5,1);
\draw (6-0.5,-0.5)--(6-1.5,-1);
\node at (6-2,0) [below] {O5$^-$};
\draw[<-] (6+0.1,0.5)--(6+0.2,0.5);
\node at (6+0,0.5) [left] {$\lambda$};
\node at (6+0,-0.5) [right] {$\lambda$};
\draw[->] (6-0.2,-0.5)--(6-0.1,-0.5);
\end{tikzpicture}
\label{fig:framingrule}
\end{align}
where we assign trivial Young diagrams to all the external legs in \eqref{fig:framingrule}. 
The diagram on the left-hand side of \eqref{fig:framingrule} is an ordinary 5-brane web and the application of the topological vertex to the diagram yields
\be
(-Q^2)^{|\lambda|}f_{\lambda}^3C_{\emptyset\emptyset\lambda}C_{\emptyset\emptyset\lambda^t} = (-Q^2)^{|\lambda|}f_\lambda^3s_\lambda(q^{-\rho})s_{\lambda^t}(q^{-\rho}). \label{horizontal.original}
\ee 
On the other hand, 
the contribution from the diagram on the right-hand side of \eqref{fig:framingrule} is given by
\be
(-Q^2)^{|\lambda|}g_\lambda C_{\emptyset\emptyset\lambda}C_{\emptyset\emptyset\lambda}=(-Q^2)^{|\lambda|}g_\lambda f_{\lambda}^2 s_\lambda(q^{-\rho})s_{\lambda}(q^{-\rho}),\label{horizontal.mirror}
\ee
with some factor $g_\lambda$, which is assigned to the glued legs with the Young diagram $\lambda$. Since 
they are the contributions from the equivalent diagrams \eqref{horizontal.original} should be equal to \eqref{horizontal.mirror}. In fact by using the identity
\be
s_{\lambda^t}(q^{-\rho})=(-1)^{|\lambda|}f_\lambda s_\lambda(q^{-\rho}),
\ee
it is possible to rewrite \eqref{horizontal.original} as
\be
(-Q^2)^{|\lambda|}(-1)^{|\lambda|}f_\lambda^4s_\lambda(q^{-\rho})s_{\lambda}(q^{-\rho}).
\ee
Therefore the factor $g_{\lambda}$ in \eqref{horizontal.mirror} turns out to be $g_{\lambda} = (-1)^{|\lambda|}f_{\lambda}^2$ and this is the framing factor assigned to the horizontal legs when we use the diagram of \eqref{fig:prep-Sp0s}.
 
It is now possible to compute the partition function for the diagram \eqref{fig:prep-Sp0s} which is equivalent to \eqref{fig:start-O-vert}. The partition function is given by
\be
Z(Q)=\sum_{\mu, \lambda}\left(-Q^2\right)^{|\mu|}\left(-Q^2\right)^{|\lambda|}(-1)^{|\lambda|}f_\lambda^{2}C_{\emptyset\mu\lambda}C_{\emptyset\mu^t\lambda}.
\ee
Inserting the explicit expressions \eqref{topvertex} and \eqref{framing} gives
\be
Z(Q) =P.E.\lt(-\frac{q}{(1-q)^2}Q^2\rt)\sum_{\lambda}\left(Q^2\right)^{|\lambda|}f^3_\lambda\frac{N_{\lambda\lambda^t}(Q^2,q)}{N_{\lambda\lambda}(1,q)},\label{ZQ}
\ee
where the Nekrasov factor in the unrefined limit is given by \eqref{Nekfactor} 
and $P.E.$ represents the Plethystic exponential defined by 
\ba
P.E.\lt(f(x_1,x_2,\dots,x_n)\rt):=\exp\lt(\sum_{k=1}^\infty \frac{1}{k}f(x_1^k,x_2^k,\dots,x_n^k)\rt).
\ea
For obtaining \eqref{ZQ} we used the Cauchy identity \eqref{Schur-twist-id-spec} to sum over the Young diagram $\mu$ and 
\be
N_{\lambda\lambda}^{-1}(1,q)=(-1)^{|\lambda|}s_\lambda(q^{-\rho})s_{\lambda^t}(q^{-\rho}).
\ee

In this article, we compute such partition functions as series expansions by K\"ahler parameters using the {\it mathematica} package developed in \cite{Ohmori-Hayashi}. 
Since the diagram \eqref{fig:prep-Sp0s} or \eqref{fig:start-O-vert} gives the pure ``Sp($0$)" gauge theory, the partition function \eqref{ZQ} should be trivial, namely
\ba
Z(Q)=1.
\ea
Indeed, one can check this statement with {\it mathematica} to find 
\ba
Z(Q)=1+ o(Q^{11}). \label{ZQexp}
\ea

\subsection{Proposal for O-vertex} \label{s:prop-O-vert}

We are interested in extending this formalism to compute the partition function for a diagram which involves the following configuration,
\begin{align}
\begin{tikzpicture}
\draw[->] (0.3,0.15)--(0.7,0.35);
\draw [dashed] (-2,0)--(2,0);
\draw (0,0)--(1,0.5);
\node at (1,0) [below] {O5$^-$};
\node at (-1,0) [below] {O5$^+$};
\node at (1,0.5) [above] {$\nu$};
\end{tikzpicture}\label{fig:O-vert0}
\end{align}
This diagram \eqref{fig:O-vert0} is obtained by sending $Q' \to 0$ with $P'$ fixed for the diagram, 
\begin{align}
\begin{tikzpicture}
\draw[->] (0.5+0.3,0.15)--(0.5+0.7,0.35);
\draw [dashed] (-2,0)--(2,0);
\draw (0.5,0)--(1.5,0.5);
\draw (-0.5,0)--(-1.5,0.5);
\draw [dotted,<->] (-0.5,0.1)--(0.5,0.1);
\node at (1.5,0.5) [above] {$\nu$};
\node at (0,0) [below] {O5$^+$};
\node at (-1.5,0) [below] {O5$^-$};
\node at (1.5,0) [below] {O5$^-$};
\node at (0,0.1) [above] {$Q'^2$};
\draw [dotted,<->] (1.5+0.05,0+0.05)--(1.5+0.05,0.5-0.05);
\node at (1.6,0.3) [right] {$P'$};
\end{tikzpicture}\label{fig:O-vert1}
\end{align}
where we also introduced the parameter $P$. 
As explained in section \ref{s:r-O-top}, the partition function of the diagram \eqref{fig:O-vert1} can be computed by deforming the diagram into the following one, 
\begin{align}
\begin{tikzpicture}
\draw[->] (0.5+0.3,0.5+0.15)--(0.5+0.7,0.5+0.35);
\draw [dashed] (-2,0)--(2,0);
\draw (0,0)--(0.5,0.5);
\draw (0,0)--(-0.5,0.5);
\draw (-0.5,0.5)--(0.5,0.5);
\draw (0.5,0.5)--(1.5,1);
\draw (-0.5,0.5)--(-1.5,1);
\node at (1.5,1) [above] {$\nu$};
\node at (0,0) [below] {O5$^-$};
\draw [dotted,<->] (-0.5,0.6)--(0.5,0.6);
\node at (0,0.6) [above] {$Q^2$};
\draw [dotted,<->] (1.5+0.05,0.5-0.05)--(1.5+0.05,1-0.05);
\node at (1.5,0.6) [right] {$P$};
\end{tikzpicture}
\label{fig:O-vert2}
\end{align}
where $Q' = Q^{-1}$ and $P'=PQ$ and taking the fundamental region as in \eqref{fig:prep-Sp0s}.  
The partition function for the diagram \eqref{fig:O-vert2} is then  
\begin{align}
Z_\nu(P,Q)=\sum_{\mu,\lambda}(-Q^2)^{|\mu|}Q^{2|\lambda|}f_\lambda^{2}(-P)^{|\nu|}C_{\nu\mu\lambda}C_{\emptyset\mu^t\lambda}. \label{OvertexV}
\end{align}
Therefore the function associated to the diagram \eqref{fig:O-vert0} is obtained by applying the limit $Q \to \infty$ with $PQ$ fixed in \eqref{OvertexV}. Then it seems that \eqref{OvertexV} diverges from the contribution coming from Young diagrams satisfying $2|\mu| + 2|\lambda| > |\nu|$. However we observe that there is non-trivial cancellation for higher orders of $Q$. When $\nu = \emptyset$, then \eqref{OvertexV} reduces to \eqref{ZQ} and it is just $1$ as we checked in \eqref{ZQexp}.  
Even when $\nu$ is non-trivial, interestingly, the result as an expansion over $Q^2$ stops at a finite order. For example, for $\nu=(5,1)$, we have
\ba
&&Z_{(5,1)}(P,Q) =\frac{q^{13}P^6}{(1-q)^6(1+q)^3(1+q^2)(1-q+q^2)(1+q+ q^2)^2}\nn\\
&&-\frac{q^9Q^2P^6}{(1-q)^5(1+q)^2(1+q^2)}+\frac{q^6Q^4P^6}{(1-q)^4 (1+q)^2}
 +\frac{q^3 Q^6P^6}{(1-q)^3(1+q)(1+q+q^2)}+ o(Q^{11}).\nn\\
\ea
For various choices of $\nu$ we have checked the function \eqref{OvertexV} terminates at the order $Q^{|\nu|}$. Hence we conjecture that the expansion by $Q$ of \eqref{OvertexV} terminates at the order $Q^{|\nu|}$ for any $\nu$. Then it is possible to take the limit $Q \to \infty$ with $PQ$ fixed and \eqref{OvertexV} becomes
\begin{align}
\lim_{\substack{Q\to \infty\\PQ\text{ fixed}}}Z_\nu(P,Q) = \sum_{\substack{\mu,\lambda\\2|\mu|+2|\lambda| = |\nu|}}(-Q^2)^{|\mu|}Q^{2|\lambda|}f_\lambda^{2}(-P)^{|\nu|}C_{\nu\mu\lambda}C_{\emptyset\mu^t\lambda}.
\end{align}
Therefore we define a vertex function $V_{\nu}$ by
\be
V_{\nu}:= \sum_{\substack{\mu,\lambda\\2|\mu|+2|\lambda| = |\nu|}}(-1)^{|\mu|}f_\lambda^{2}C_{\nu\mu\lambda}C_{\emptyset\mu^t\lambda}, \label{Vnu1}
\ee
which is associated to the diagram 
\begin{align}
\begin{tikzpicture}
\draw[->] (0.3,0.15)--(0.7,0.35);
\draw [dashed] (-2,0)--(2,0);
\draw (0,0)--(1,0.5);
\node at (1,0.5) [above] {$\nu$};
\node at (1,0) [below] {O5$^-$};
\node at (-1,0) [below] {O5$^+$};
\node at (0,0.5) {$V_\nu$};
\end{tikzpicture}
\label{fig:Vnu}
\end{align}
We will call $V_{\nu}$ O-vertex as it is associated to a line intersecting with the O5-plane.

After a little computation, we observe that $V_\nu$ takes the form of 
\ba\label{VnuPnu}
V_\nu=\frac{P_\nu(q)}{(q;q)_{|\nu|/2}},\label{Vnu}
\ea
where we defined 
\be
(q; q)_n = \prod_{k=0}^{n-1}(1-q^{k+1}).
\ee
$P_\nu$ in \eqref{VnuPnu} is a polynomial of $q$ of degree at most $m(|\nu|)=\frac{n(n+1)}{2}$ for $n=\frac{|\nu|}{2}$ and can only be non-zero when $|\nu|$ is even. Some of the explicit expressions of $P_\nu$ are listed below, 
\ba
&&P_{(2)}=-q,\quad P_{(1,1)}=1,\label{P-1}\\
&&P_{(4)}=q^3,\quad P_{(3,1)}=-q,\quad P_{(2,2)}=1+q^3,\quad P_{(2,1,1)}=-q^2,\quad P_{(1,1,1,1)}=1,\label{P-2}
\ea
and more can be found in appendix \ref{a:O-vert}. 
We will also give a candidate for the refined version of this vertex in section \ref{s:refined}.

In the same way, we can compute 
another O-vertex which is associated to the diagram given by 
\begin{align}
\begin{tikzpicture}[xscale=-1]
\draw[->] (0.3,0.15)--(0.7,0.35);
\draw [dashed] (-2,0)--(2,0);
\draw (0,0)--(1,0.5);
\node at (1,0.5) [above] {$\nu$};
\node at (1,0) [below] {O5$^-$};
\node at (-1,0) [below] {O5$^+$};
\node at (0,0.5) {$W_\nu$};
\end{tikzpicture}
\label{fig:Wnu}
\end{align}
For computing this type of O-vertex we start from the diagram in \eqref{fig:O-vert2} and take the fundamental region as
\begin{align}
\begin{tikzpicture}
\draw[->] (-0.5-0.3,0.5+0.15)--(-0.5-0.7,0.5+0.35);
\draw [dashed] (-2,0)--(2,0);
\draw (0,0)--(-0.5,0.5);
\draw (0,0)--(0.5,-0.5);
\draw (0,0.5)--(-0.5,0.5);
\draw (0,-0.5)--(0.5,-0.5);
\draw (-0.5,0.5)--(-1.5,1);
\draw (0.5,-0.5)--(1.5,-1);
\node at (-1.5,1) [above] {$\nu$};
\node at (1,0) [below] {O5$^-$};
\draw[<-] (-0.1,0.5)--(-0.2,0.5);
\node at (0.2,0.5) [right] {$\lambda$};
\node at (-0.2,-0.5) [left] {$\lambda$};
\draw[->] (0.2,-0.5)--(0.1,-0.5);
\end{tikzpicture}
\label{fig:prep-Sp0sv2}
\end{align}
The associated partition function is given by
\ba
\widetilde{Z}_\nu(P,Q)=\sum_{\mu,\lambda}(-Q^2)^{|\mu|}Q^{2|\lambda|}f_\lambda^{-4}\left(-P\right)^{|\nu|}C_{\mu\nu\lambda}C_{\mu^t\emptyset\lambda}.\label{OvertexW}
\ea
Again we observe that \eqref{OvertexW} terminates at the finite order $Q^{|\nu|}$ for various choices of $\nu$. Hence we conjecture that the expansion of \eqref{OvertexW} by $Q$ terminates at the order $Q^{|\nu|}$. Then we can take the limit $Q\to \infty$ with $PQ$ fixed to obtain another O-vertex $W_{\nu}$ assoicated the diagram of \eqref{fig:Wnu}. Namely we define
\ba
W_\nu :=\sum_{\substack{\mu,\lambda\\2|\mu| + 2|\lambda| = |\nu|}}(-1)^{|\mu|}f_\lambda^{-4}C_{\mu\nu\lambda}C_{\mu^t\emptyset\lambda}.\label{Wnu}
\ea

We also observe that 
\ba
W_\nu=\frac{\tilde{P}_\nu(q)}{(q;q)_{|\nu|/2}},\label{Wnu0}
\ea
while $\tilde{P}_\nu$ is not necessarily a polynomial here, but is still zero for $|\nu|$ odd. Some examples of $\tilde{P}_\nu$ are listed below, 
\ba
&&\tilde{P}_{(2)}=q,\quad \tilde{P}_{(1,1)}=-1,\\
&&\tilde{P}_{(4)}=q^6,\quad \tilde{P}_{(3,1)}=-q^4,\quad \tilde{P}_{2,2}=1+q^3,\quad \tilde{P}_{(2,1,1)}=-q^{-1},\quad \tilde{P}_{(1,1,1,1)}=q^{-3},\nn\\
\ea
and more details can be found in appendix \ref{a:O-vert}.

In fact \eqref{Wnu} is related to \eqref{Vnu1} in a simple way. To see that note the topological vertex satisfies, 
\begin{align}
C_{\mu\nu\lambda} = (-1)^{|\mu| + |\nu| + |\lambda|}f_{\mu}f_{\nu}f_{\lambda}C_{\nu^t\mu^t\lambda^t}. \label{Crelation}
\end{align}  
Then using \eqref{Crelation} to the two topological vertices in the sum of \eqref{Wnu} yields
\begin{align}
W_{\nu} &= \sum_{\substack{\mu,\lambda\\2|\mu| + 2|\lambda| = |\nu|}}(-1)^{|\mu|}f_\lambda^{-4}(-1)^{|\mu| + |\nu| + |\lambda|}f_{\mu}f_{\nu}f_{\lambda}C_{\nu^t\mu^t\lambda^t}(-1)^{|\mu| + |\lambda|}f_{\mu^t}f_{\lambda}C_{\emptyset\mu\lambda^t}\cr
&=\sum_{\substack{\mu,\lambda\\2|\mu| + 2|\lambda| = |\nu|}}(-1)^{|\mu|}f_\lambda^{-2}(-1)^{|\nu|}f_{\nu}C_{\nu^t\mu^t\lambda^t}C_{\emptyset\mu\lambda^t}\cr
&=q^{\frac{\kappa(\nu)}{2}}\sum_{\substack{\mu,\lambda\\2|\mu| + 2|\lambda| = |\nu|}}(-1)^{|\mu|}f_\lambda^{2}C_{\nu^t\mu\lambda}C_{\emptyset\mu^t\lambda}.\label{Wnudeform}
\end{align}
Comparing \eqref{Wnudeform} with \eqref{Vnu1}, we have 
\ba
W_\nu=q^{\frac{\kappa(\nu)}{2}}V_{\nu^t}.\label{rel-O-dual-2}
\ea

Identifying the explicit forms of the O-vertices with \eqref{Vnu} and \eqref{Wnu}, we expect that it is possible to compute the partition functions of any 5d gauge theories that are realized on 5-brane webs with an O5-plane. We will confirm this proposal by computing the partition functions of pure SO($2N$) gauge theories and compare them with the results obtained by the localization method in section \ref{s:so-2N}.

\subsection{Higgsing and $\widetilde{\text{O}}$-plane}
\label{s:Higgsing-tild-O}

\begin{figure}
\begin{center}
\subfigure[]{\label{f:so8-D7a}
\begin{tikzpicture}[scale=0.95]
\draw [dashed] (-1,0)--(5,0);
\draw (0,0)--(1,0.5);
\draw (1,0.5)--(3,0.5);
\draw (3,0.5)--(4,0);
\draw (1,0.5)--(1.5,1);
\draw (1.5,1)--(2.5,1);
\draw (2.5,1)--(3,0.5);
\draw (1.5,1)--(1.5,1.5);
\draw (1.5,1.5)--(2.5,1.5);
\draw (2.5,1)--(2.5,1.5);
\draw (1.5,1.5)--(1,2);
\draw (1,2)--(0.4,2.3);
\draw (0.4,2.3)--(0,2.7);
\draw (0.4,2.3)--(-0.2,2.3);
\draw (1,2)--(3,2);
\draw (2.5,1.5)--(3,2);
\draw (3,2)--(4,2.5);
\draw[fill=purple,draw=purple] (-0.2,2.3) circle (0.1);
\draw[purple,ultra thick,dashed] (-0.2,2.3)--(-1.2,2.3);
\node at (-0.4,2.3) [above] {D7};
\node at (-1,0) [below] {O5$^+$};
\node at (2,0) [below] {O5$^-$};
\node at (5,0) [below] {O5$^+$};
\end{tikzpicture}}
\hskip 1cm
\subfigure[]{\label{f:so8-D7b}
\begin{tikzpicture}[scale=0.95]
\draw [dashed] (-1,0)--(5,0);
\draw (0,0)--(1,0.5);
\draw (-0.2,0.5)--(3.5,0.5);
\draw (3.5,0.5)--(4.5,0);
\draw (1,0.5)--(2,1);
\draw (2,1)--(3,1);
\draw (3,1)--(3.5,0.5);
\draw (2,1)--(2.5,1.5);
\draw (3,1.5)--(2.5,1.5);
\draw (3,1)--(3,1.5);
\draw (2.5,1.5)--(2.5,2);
\draw (2.5,2)--(3.5,2);
\draw (3,1.5)--(3.5,2);
\draw (3.5,2)--(4.5,2.5);
\draw (2.5,2)--(2,2.5);
\draw[fill=purple,draw=purple] (-0.2,0.5) circle (0.1);
\draw[purple,ultra thick,dashed] (-0.2,0.5)--(-1,0.5);
\node at (-0.2,0.5) [above] {D7};
\node at (-1,0) [below] {O5$^+$};
\node at (2,0) [below] {O5$^-$};
\node at (5,0) [below] {O5$^+$};
\draw[<->,dotted] (4.6,0)--(4.6,0.5);
\node at (4.6,0.25) [right] {$T$};
\end{tikzpicture}}
\end{center}
\caption{(a): The brane web for the SO($8$) gauge theory with one flavor in the vector representation. The red dotted line represents the branch cut for the D7-brane. (b): The brane web after setting the height of the flavor D5-brane equal to the height of the bottom color D5-brane. 
}
\label{f:so8-D7}
\end{figure}
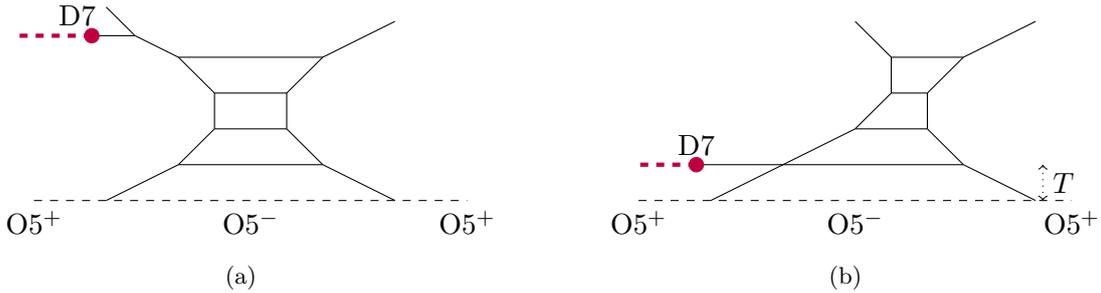

\begin{figure}
\begin{center}
\begin{tikzpicture}
\draw [green,dashed] (-1,0.55)--(5,0.55);
\draw [dashed] (-1,0.5)--(5,0.5);
\draw (1,0.5)--(3.5,0.5);
\draw (1,0.5)--(2,1);
\draw (2,1)--(3,1);
\draw (3,1)--(3.5,0.5);
\draw (2,1)--(2.5,1.5);
\draw (3,1.5)--(2.5,1.5);
\draw (3,1)--(3,1.5);
\draw (2.5,1.5)--(2.5,2);
\draw (2.5,2)--(3.5,2);
\draw (3,1.5)--(3.5,2);
\draw (3.5,2)--(4.5,2.5);
\draw (2.5,2)--(2,2.5);
\node at (0,0.5) [below] {$\widetilde{\text{O5}}^+$};
\node at (2.25,0.5) [below] {$\widetilde{\text{O5}}^-$};
\node at (4,0.5) [below] {$\widetilde{\text{O5}}^+$};
\end{tikzpicture}
\end{center}
\caption{The brane web for the pure SO($7$) gauge theory. The green dotted line represents the branch cut for a half D7-brane. }
\label{f:so7-D7}
\end{figure}
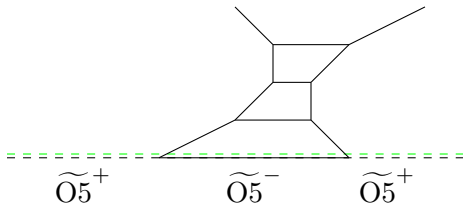
So far we have obtained the O-vertices \eqref{Vnu} and \eqref{Wnu} which are applicable to 5-brane webs with an O5-plane. In fact it turns out that it is possible to apply the O-vertices to 5-brane webs with an $\widetilde{\text{O5}}$-plane. To see that we first review the relation between an O5-plane and an $\widetilde{\text{O5}}$-plane \cite{Zafrir:2015ftn}. We start from the 5-brane web diagram for the SO($8$) gauge theory with a hypermultiplet in the vector representation, which is depicted in Figure \ref{f:so8-D7a}. The presence of the flavor is shown by the flavor D5-brane which ends on a D7-brane. This theory has a Higgs branch on which the low energy theory becomes the pure SO($7$) gauge theory. The Higgs branch opens up when we tune the mass parameter for the flavor as well as a Coulomb branch modulus. By applying the same tuning to the 5-brane web in Figure \ref{f:so8-D7a}, it is possible to see the Higgs branch in terms of the 5-brane web. We first set the height of the flavor D5-brane equal to the height of the bottom color D5-brane as in Figure \ref{f:so8-D7b}. Then we take $T \to 1$ and the D7-brane is on top of the O5-plane. When a D7-brane is on top of an O5-plane it can be split into two half D7-branes \cite{Bertoldi:2002nn}. A piece of D5-brane between the two half D7-branes can move in the direction which the D7-branes extend, which represents the Higgs branch. After removing the piece of the D5-brane, we pull the two half D7-branes in the opposite directions into the infinity. This operation leaves a configuratioin with a branch cut of one of the half D7-branes on top of the whole O5-plane and also a half D5-brane 
on top of the O5$^-$-plane, which is depicted in Figure \ref{f:so7-D7}. Then the resulting theory should be interpreted as the pure SO($7$) gauge theory. Since the SO($7$) gauge group can be realized on an $\widetilde{\text{O5}}$-plane, we may interpret the orientifold in Figure \ref{f:so7-D7} as an $\widetilde{\text{O5}}$-plane. Namely an $\widetilde{\text{O5}}$-plane is effectively realized as an O5-plane with a branch cut of a half D7-brane on it (and a half D5-brane for an $\widetilde{\text{O5}}^-$-plane).

\begin{figure}
\begin{center}
\subfigure[]{\label{f:vert-so7a}
\begin{tikzpicture}[scale=1.1]
\draw [dashed] (-1,0)--(5,0);
\draw (0.5,0)--(1.5,0.5);
\draw (1.5,0.5)--(2.75,0.5);
\draw (1.5,0.5)--(2,1);
\draw (2,1)--(2.75,1);
\draw (2,1)--(2,1.5);
\draw (2,1.5)--(3.25,1.5);
\draw (2,1.5)--(1.5,2);
\draw (2.75,1)--(3.25,1.5);
\draw (3.25,1.5)--(4.25,2);
\draw (3,0.25)--(3.25,0);
\draw (3.25,0)--(0,0);
\draw (3,0.25)--(2.75,0.5);
\draw (2.75,0.5)--(2.75,1);
\draw[fill=purple,draw=purple] (0,0) circle (0.1);
\end{tikzpicture}}
\hskip 1cm
\subfigure[]{\label{f:vert-so7b}
\begin{tikzpicture}[scale=1.1]
\draw [dashed] (-1,0)--(5,0);
\draw (0.5,0)--(1.5,0.5);
\draw (1.5,0.5)--(2.75,0.5);
\draw (1.5,0.5)--(2,1);
\draw (2,1)--(2.75,1);
\draw (2,1)--(2,1.5);
\draw (2,1.5)--(3.25,1.5);
\draw (2,1.5)--(1.5,2);
\draw (2.75,1)--(3.25,1.5);
\draw (3.25,1.5)--(4.25,2);
\draw (3,0.25)--(3.5,0);
\draw (3,0.25)--(2.25,0.25);
\draw (3,0.25)--(2.75,0.5);
\draw (2.75,0.5)--(2.75,1);
\draw[<->,dotted] (3.55,0)--(3.55,0.25);
\node at (3.55,0.25) [right] {$T \to 1$};
\node at (2.25,0.25) [left] {$\emptyset$};
\end{tikzpicture}}
\end{center}
\caption{(a): The brane web by sending $T \to 1$ from the diagram in Figure \ref{f:so8-D7b}. (b): The diagram to be used for the computations, where we shall set the leg attached to the D7-brane to be empty. }
\label{f:vert-so7}
\end{figure}
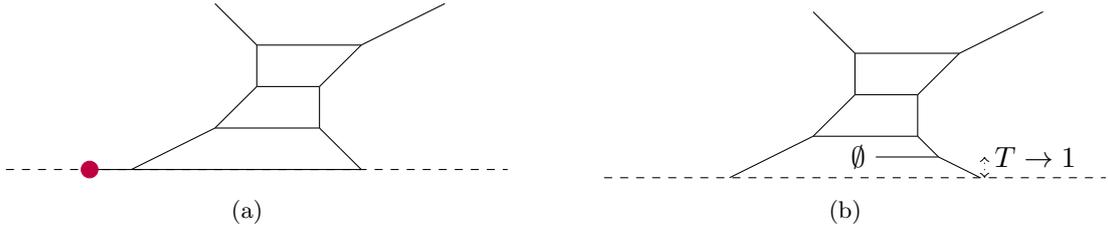

Since it is not clear how to apply the topological vertex for a configuration with the branch cut, we find it convenient to use the diagram before moving the half D7-branes into infinity. Moving the D7-branes into the left or the right direction in the diagram does not change the theory and we can equally use the diagram before the moving depicted in Figure \ref{f:vert-so7a}. For practically applying the topological vertex to the diagram in Figure \ref{f:vert-so7a}, we can use the diagram in Figure \ref{f:vert-so7b} instead of the one in Figure \ref{f:vert-so7a}, Then the partition function for the diagram in Figure \ref{f:vert-so7b} can be computed using the topological vertex and the O-vertex. The Young diagram assigned to the bottom D5-brane is empty as reviewed in section \ref{s:r-O-top} for the application of the topological vertex to certain non-toric diagrams. The K\"ahler parameter $T$ is reintroduced and it is set to $1$ at the end of the computation. Although it seems that we need to sum over infinitely many Young diagrams assigned to the line with the K\"ahler parameter $T$, it turns out that a finite sum gives the correct answer, which we will see in section \ref{s:so-2N+1}.

We can use this technique to compute the partition functions of SO($2N+1$) gauge theories and also $G_2$ gauge theories utilizing the construction in \cite{G-type}. Furthermore it is also possible to compute the pure SU($3$) gauge theory with the CS level $\kappa_{cs}=9$ by applying the brane web proposed in \cite{Hayashi:2018lyv}, which may be obtained by a twisted compactification of 6d pure SU($3$) gauge theory with a tensor multiplet \cite{Razamat:2018gro}.

\section{Examples}\label{s:example}

In this section, we give several examples of calculations that involve the usage of the O-vertex defined in section \ref{s:O-vert}. More concretely, we first check the validity of our formalism in section \ref{s:so-2N}, \ref{s:so-2N+1} and \ref{s:G2} by comparing the partition functions of the SO($2N$), SO($2N+1$) and $G_2$ gauge theories obtained in the topological vertex formalism with the O-vertex with the known one-instanton and two-instanton partition functions in the literature. In section \ref{s:su3-9}, we will further apply the O-vertex to the calculation of the partition function of the 5d pure SU($3$) gauge theory with the CS level $\kappa_{cs} = 9$, by using the brane diagram proposed for it in \cite{Hayashi:2018lyv}. 

\subsection{$\text{SO}(2N)$ gauge theories}\label{s:so-2N}

We start from the computation of the partition function of the pure SO($2N$) theory. The brane web diagram is obtained by putting a stack of $N$ D5-branes on the top of an O5$^-$-plane. Hence we can directly apply the O-vertex proposed in section \ref{s:prop-O-vert}.

\subsubsection{Pure SO(4) gauge theory}
\label{s:so4}

The first example is the simplest case $N=2$, i.e. SO(4) gauge theory. The 5-brane web diagram for the pure SO(4) theory is given by 
\begin{align}
\begin{tikzpicture}
\draw [dashed] (-1,0)--(5,0);
\draw (0,0)--(1,0.5);
\draw (4,0)--(3,0.5);
\draw (1,0.5)--(3,0.5);
\draw (1,0.5)--(1.5,1);
\draw (3,0.5)--(2.5,1);
\draw (1.5,1)--(2.5,1);
\draw (1.5,1)--(1.5,1.5);
\draw (2.5,1)--(2.5,1.5);
\draw[<->,dotted] (0,0.05)--(0,0.45);
\node at (0,0.25) [left] {$P$};
\draw[<->,dotted] (0,0.55)--(0,0.95);
\node at (0,0.75) [left] {$Q'$};
\draw[<->,dotted] (1.6,1.1)--(2.4,1.1);
\node at (2,1.1) [above] {$Q$};
\node at (2,0) [below] {O5$^-$};
\node at (-0.5,0) [below] {O5$^+$};
\node at (4.7,0) [below] {O5$^+$};
\end{tikzpicture}
\label{o-vert-so4}
\end{align}
The partition function of the pure SO(4) gauge theory realized on the web in \eqref{o-vert-so4} can be computed by applying the topological vertex formalism with the O-vertices given in \eqref{Vnu1} and \eqref{Wnu} to the diagram in \eqref{o-vert-so4}. Since the brane web has parallel external legs, the partition function computed by the topological vertex formalism contains an extra factor related to contributions from strings between the parallel external legs \cite{Bergman:2013ala, Bao:2013pwa, Hayashi:2013qwa, Bergman:2013aca}. The extra factor in this case is given by 
\begin{align}\label{so4.extra}
Z^{SO(4)}_{\text{extra}} = P.E.\left(\frac{q}{(1-q)^2}Q\right).
\end{align}
Hence the partition function of the pure SO(4) gauge theory is given by
\begin{align}\label{so4Nek}
\hat{Z}^{SO(4)}_{\text{top}} = Z^{SO(4)}_{\text{top}}/Z^{SO(4)}_{\text{extra}},
\end{align}
where $Z^{SO(4)}_{\text{top}}$ is the partition function which is directly obtained by applying the topological vertex and the O-vertex to the web diagram in \eqref{o-vert-so4} and the explicit form is given by 
\begin{align}\label{so4top}
Z^{SO(4)}_{\text{top}}&=\sum_{\mu,\nu,\alpha,\beta,\lambda,\sigma}(-Q)^{|\lambda|+|\sigma|}(-Q')^{|\alpha|+|\beta|+2|\sigma|}(-P)^{|\mu|+|\nu|}f_\lambda f_\alpha^{-1}f_\beta f_\sigma^{-3}\cr
&\hspace{2cm}\times C_{\alpha^t\emptyset\lambda}C_{\emptyset\beta^t\lambda^t}C_{\nu^t\alpha\sigma^t}C_{\beta\mu^t\sigma}V_\nu W_\mu. 
\end{align}

The K\"ahler parameters $Q, Q', P$ are related to gauge theory parameters of the pure SO(4) gauge theory. The height of the color D5-branes in \eqref{o-vert-so4} is related to the Coulomb branch moduli. The simple roots of the $\mathfrak{so}(4)$ Lie algebra are represented by $e_1- e_2$ and $e_1 + e_2$ where $e_1, e_2$ are the orthonormal basis of $\mathbb{R}^2$. A string between the top color D5-brane and the bottom color D5-brane yields a W-boson in the simple root $e_1 - e_2$. On the other hand, a string which extends from the top color brane and reflects on the O5-plane, ending on the bottom color brane corresponds to a W-boson in the simple root $e_1 + e_2$. Hence we associate the Coulomb branch moduli $A_1 = e^{-a_1}, A_2 = e^{-a_2}$ with the K\"ahler parameters by
\begin{align}
A_1A_2^{-1} = Q', \qquad A_1A_2 = Q'P^2.
\end{align}
The K\"ahler parameter $Q$ is a mass parameter of the SO(4) gauge theory and there is only one mass parameter in the theory, which is the instanton fugacity $\mathfrak{q}$. Namely we have
\be
Q = \mathfrak{q}. \label{so4.instfugacity}
\ee

Let us see if the partition function \eqref{so4Nek} reproduces the known result. The perturbative part of the partition function is obtained by taking the limit $Q \to 0$. Then the sum of the Young diagrams $\lambda$ and $\sigma$ becomes trivial and the partition function \eqref{so4top} becomes
\begin{align}
Z^{SO(4)}_{\text{top}} \to Z^{SO(4)}_{\text{left}}Z^{SO(4)}_{\text{right}} \qquad (Q \to 0),
\end{align}
where 
\begin{align}
Z^{SO(4)}_{\text{left}}&=\sum_{\alpha,\nu}(-Q')^{|\alpha|}(-P)^{|\nu|}f_\alpha^{-1}C_{\nu^t\alpha\emptyset}C_{\alpha^t\emptyset\emptyset}V_\nu,\label{so4-left}\\
Z^{SO(4)}_{\text{right}}&=\sum_{\mu,\beta}(-Q')^{|\beta|}(-P)^{|\mu|}f_\beta C_{\emptyset \beta^t\emptyset}C_{\beta\mu^t\emptyset}W_\mu.\label{so4-right}
\end{align}
$Z^{SO(4)}_{\text{left}}$ is the partition function associted with the diagram 
\begin{align}
\begin{tikzpicture}
\draw [dashed] (-1,0)--(5,0);
\draw (0,0)--(1,0.5);
\draw (1,0.5)--(2,0.5);
\draw (1,0.5)--(1.5,1);
\draw (1.5,1)--(2,1);
\draw (1.5,1)--(1.5,1.5);
\end{tikzpicture}
\label{o-vert-half}
\end{align}
and $Z^{SO(4)}_{\text{right}}$ is associated with 
\begin{align}
\begin{tikzpicture}[xscale=-1]
\draw [dashed] (-1,0)--(5,0);
\draw (0,0)--(1,0.5);
\draw (1,0.5)--(2,0.5);
\draw (1,0.5)--(1.5,1);
\draw (1.5,1)--(2,1);
\draw (1.5,1)--(1.5,1.5);
\end{tikzpicture}
\label{o-vert-half-2}
\end{align}
Using the O-vertex \eqref{Vnu1} for \eqref{so4-left} we checked that 
\be\label{so4left1}
Z^{SO(4)}_{\text{left}} = P.E.\lt(\frac{q}{(1-q)^2}(Q'+P^2Q')\rt) + o(P^6,Q^{\prime 6}),
\ee
where $o(X_1^{k_1}, X_2^{k_2}, \cdots)$ means that the computation was done until the order $X_1^{l_1}X_2^{l_2}\cdots$ with $l_1 \leq k_1, l_2 \leq k_2, \cdots$. 
Similarly applying the O-vertex \eqref{Wnu} to \eqref{so4-right} yields 
\begin{align}
Z^{SO(4)}_{\text{right}} = P.E.\lt(\frac{q}{(1-q)^2}(Q'+P^2Q')\rt) +o(P^6,Q^{\prime 6}).
\end{align}
Hence we obtained 
\begin{align}
Z^{SO(4)}_{\text{left}}Z^{SO(4)}_{\text{right}} = P.E.\lt(\frac{2q}{(1-q)^2}(A_1A_2^{-1}+A_1A_2)\rt) + o\lt(P^6, Q'^6\rt). \label{so4top.pert}
\end{align}
Since the topological vertex reproduces only the root part of the perturbative partition function we compare \eqref{so4top.pert} with \eqref{Zroot}. The root contribution to the perturbative partition funciton of the pure SO(4) gauge theory from \eqref{Zroot} is given by
\be
Z^{SO(4)}_{\text{pert}}=P.E.\lt(\frac{2q}{(1-q)^2}(A_1A_2^{-1}+A_1A_2)\rt). \label{so4nek.pert}
\ee
Hence we see the agreement between \eqref{so4top.pert} and \eqref{so4nek.pert} until the orders we computed. Namely the O-vertices \eqref{Vnu1} and \eqref{Wnu} reproduced the correct result until the orders we computed.

Let us move on to the comparison of the instanton part $\hat{Z}^{SO(4)}_{\text{top}}$ in \eqref{so4Nek}, namely,
\be\label{topso4inst}
\hat{Z}_{\text{top, inst}}^{SO(4)} = \hat{Z}_{\text{top}}^{SO(4)}/Z_{\text{pert}}^{SO(4)}.
\ee
From the identification in \eqref{so4.instfugacity}, the one-instanton partition function can be extracted from \eqref{so4Nek} as the coefficient of $Q^1$ in \eqref{topso4inst} and it is given by 
\begin{align}
\hat{Z}_{\text{top, 1-inst}}^{SO(4)}=\frac{2q}{(1-q)^2}Q'+\frac{4q}{(1-q)^2}Q^{\prime 2}+\frac{6q}{(1-q)^2}Q^{\prime 3}+\frac{8q}{(1-q)^2}Q^{\prime 4}+\frac{10q}{(1-q)^2}Q^{\prime 5}\nn\\
+\frac{12q}{(1-q)^2}Q^{\prime 6}
+\frac{2q}{(1-q)^2}P^2Q'+\frac{4q}{(1-q)^2}P^4Q^{\prime 2}+\frac{6q}{(1-q)^2}P^6Q^{\prime 3}+o(Q^{\prime 6},P^6).\nn\\
\end{align}
One can guess that the above result may be extrapolated as 
\begin{align}\label{so4.1inst0}
\hat{Z}_{\text{top, 1-inst}}^{SO(4)}=\frac{2q}{(1-q)^2}\lt(\frac{Q'}{(1-Q')^2}+\frac{P^2Q'}{(1-P^2Q')^2}\rt)+o(Q^{\prime 6},P^6).
\end{align}
or 
\begin{align}\label{so4.1inst}
\hat{Z}_{\text{top, 1-inst}}^{SO(4)}=\frac{2q}{(1-q)^2}\lt(\frac{A_1A_2^{-1}}{(1-A_1A_2^{-1})^2}+\frac{A_1A_2}{(1-A_1A_2)^2}\rt)+o(Q^{\prime 6},P^6).
\end{align}
in terms of the Coulomb branch moduli of the SO(4) gauge theory.

We can compare \eqref{so4.1inst} with the result computed by the localization method. The result of the instanton partition function for the pure SO($2N+\delta$) gauge theory for $\delta = 0, 1$ is given in \eqref{LMNS-integral}.
Applying $N=2$ to \eqref{LMNS-integral} with the unrefined $\Omega$-deformation parameter $\epsilon:=\epsilon_1 = -\epsilon_2$ yields
\begin{align}
Z_{\text{loc, 1-inst}}^{SO(4)}&=\frac{[2i\delta]}{2[\epsilon]^2} \oint\frac{{\rm d}\phi}{2\pi i}\frac{[2\phi]^4}{[\phi+a_1-i\delta][\phi-a_1-i\delta][\phi+a_1+i\delta][\phi-a_1+i\delta]}\nn\\
&\hspace{2cm}\times \frac{1}{[\phi+a_2-i\delta][\phi-a_2-i\delta][\phi+a_2+i\delta][\phi-a_2+i\delta]}\cr
&\rightarrow \frac{1}{[\epsilon]^2}\frac{[2a_1]^2+[2a_2]^2}{[a_1+a_2]^2[a_1-a_2]^2}\cr
&=\frac{q}{(1-q)^2}\frac{A_1^2+A_2^2+A_1^4A_2^2+A_1^2A_2^4-4A_1^2A_2^2}{(A_1-A_2)^2(1-A_1A_2)^2}.\label{so4nek.1inst}
\end{align}
where we took $\delta\rightarrow 0$ in the second to the last line of the evaluation. 

For the comparison using \eqref{so4nek.1inst} we also need to be careful of the extra factor in the instanton partition function. In general, the localization result $Z_{\text{loc, inst}}$ of the supersymmetric index of an ADHM quantum mechanics is related to the instanton partition function $\hat{Z}_\text{loc, inst}$ for a ultraviolet (UV) complete 5d theory by
\be
\hat{Z}_{\text{loc, inst}} = Z_{\text{loc, inst}}/Z_{\text{loc, extra}},
\ee 
where the extra factor $Z_{\text{loc, extra}}$ is the Coulomb branch moduli independent part of $Z_{\text{loc, inst}}$. Namely one needs to subtract a Coulomb branch moduli independent term from \eqref{so4nek.1inst} if there is such a term in the 1-instanton partition function.  
Indeed the expansion of \eqref{so4nek.1inst} by $A_1A_2^{-1}, A_2,$\footnote{A Weyl chamber of SO(4) is given by $a_1 \geq a_2 \geq 0$. Hence we choose the expansion parameters $A_1A_2^{-1}$ and $A_2$.} gives\footnote{$O(A_1A_2^{-1}, A_2)$ means that an expression expanded by $A_1A_2^{-1}, A_2$ starts from $A_1A_2^{-1}$ or $A_2$.}
\be
Z_{\text{loc, 1-inst}}^{SO(4)} = \frac{q}{(1-q)^2} + O(A_1A_2^{-1}, A_2),
\ee
and hence we can identify the extra term by
\be
Z_{\text{loc, extra}}^{SO(4)} = \frac{q}{(1-q)^2}. \label{so4nek.extra}
\ee
Note that \eqref{so4nek.extra} is the 1-instanton part of $Z_{\text{loc, extra}}$. Therefore we expect that the full extra factor for the partition function for the pure SO(4) gauge theory computed by \eqref{LMNS-integral} is given by \eqref{so4.extra}. Subtracting \eqref{so4nek.extra} from \eqref{so4nek.1inst} one finds
\begin{align}\label{so4nek.1inst2}
Z_{\text{loc, 1-inst}}^{SO(4)} - Z_{\text{loc, extra}}^{SO(4)}  = \frac{2q}{(1-q)^2}\lt(\frac{A_1A_2^{-1}}{(1-A_1A_2^{-1})^2}+\frac{A_1A_2}{(1-A_1A_2)^2}\rt),
\end{align}
which agrees with the 1-instanton partition function \eqref{so4.1inst} computed by the topological vertex formalism. One can also see that \eqref{so4nek.1inst2} or \eqref{so4.1inst} is given by the sum of the one-instanton partition function of the pure SU($2$) gauge theory with discrete theta angle zero, 
\be
\hat{Z}_{\text{1-inst}}^{SO(4)} = Z_{\text{1-inst}}^{SU(2)}\left(A_1A_2^{-1}\right) +  Z_{\text{1-inst}}^{SU(2)}\left(A_1A_2\right),
\ee
where
\be
Z_{\text{1-inst}}^{SU(2)}\left(A\right) = \frac{2q}{(1-q)^2}\frac{A}{(1-A)^2},
\ee
which follows from \eqref{sun} with $N=2$. This is consistent with the isomorphism of the Lie algebra $\mathfrak{so}(4) \simeq \mathfrak{su}(2) \oplus \mathfrak{su}(2)$.

Unitl the order of $P^4Q^{\prime 4}Q^2$ (namely $P^aQ'^bQ^c$ with $a \leq 4, b \leq 4, c\leq 2$), we managed to check that 
\ba
\hat{Z}^{SO(4)}_{\text{top, inst}}=Z^{SU(2)}_{\text{inst}}(\mathfrak{q},Q')Z^{SU(2)}_{\text{inst}}(\mathfrak{q},Q'P^2), \label{so4.inst}
\ea
where $Z^{SU(2)}_{\text{inst}}(\mathfrak{q}, A)$ is the instanton partition function of the pure SU(2) gauge theory with the zero discrete theta angle. The closed-form expressions of the SU($N$) instanton partition functions can be found in appendix \ref{s:Nek}. We also checked that the two-instanton partition function computed from the integral \eqref{LMNS-integral} with the extra factor removed matches with the 2-instanton part of \eqref{so4.inst} until the order $P^4Q'^4$. 

\subsubsection{Pure SO(6) and SO(8) Theories}

Let us then consider the partition functions of the pure SO(6) and SO(8) gauge theories. The method of the computation is essentally parallel to that in the case of the pure SO(4) gauge theory discussed in section \ref{s:so4}. 

\paragraph{Pure SO(6) gauge theory.} We start from the pure SO(6) gauge theory. The pure SO(6) gauge theory can be realized on the following brane diagram, 
\begin{align}
\begin{tikzpicture}
\draw [dashed] (-1,0)--(5,0);
\draw (0,0)--(1,0.5);
\draw (1,0.5)--(3,0.5);
\draw (1,0.5)--(1.5,1);
\draw (1.5,1)--(2.5,1);
\draw (1.5,1)--(1.5,1.5);
\draw (1.5,1.5)--(2.5,1.5);
\draw (1.5,1.5)--(1,2);
\draw (2.5,1)--(2.5,1.5);
\draw (2.5,1.5)--(3,2);
\draw (2.5,1)--(3,0.5);
\draw (3,0.5)--(4,0);
\draw[<->,dotted] (0,0.05)--(0,0.45);
\node at (0,0.25) [left] {$P$};
\draw[<->,dotted] (0,0.55)--(0,0.95);
\node at (0,0.75) [left] {$Q'$};
\draw[<->,dotted] (0,1.05)--(0,1.45);
\node at (0,1.25) [left] {$R$};
\draw[<->,dotted] (1.6,1.6)--(2.4,1.6);
\node at (2,1.6) [above] {$Q$};
\end{tikzpicture}
\label{o-vert-so6}
\end{align}

The relation between the K\"ahler parameters $Q, Q', P, R$ and the gauge theory parameters may be also read off from the diagram in \eqref{o-vert-so6}. The Coulomb branch moduli $A_1 = e^{-a_1}, A_2 = e^{-a_2}, A_3 = e^{-a_3}$ are related by 
\begin{align}
A_1 = PQ'R, \qquad A_2 = PQ', \qquad A_3 =  P.\label{so6.a}
\end{align}
On the other hand, the instanton fugacity $\mathfrak{q}$ is proportional to $Q$. More precisely we consider extrapolating the external $(1, 1)$ 5-brane and the $(1, -1)$ 5-brane until the O5-plane and measure the length between the 5-branes on the O5-plane. Namely, we have
\be
\mathfrak{q} = QA_1^{-2} = QQ'^{-2}P^{-2}R^{-2}.\label{so6.q}
\ee

The application of the topological vertex and the O-vertex to the diagram in \eqref{o-vert-so6} gives
\begin{align}
Z^{SO(6)}_{\text{top}}&=\sum_{\mu,\nu,\alpha,\beta,\lambda,\sigma,\gamma,\delta,\tau}(-Q)^{|\lambda|+|\sigma|+|\tau|}(-Q')^{|\alpha|+|\beta|+2|\sigma|}(-P)^{|\mu|+|\nu|}(-R)^{|\gamma|+|\delta|}\cr
&\hspace{2cm}f_\lambda f_\alpha^{-1}f_\beta f_\sigma^{-3}f_\gamma^{-1}f_\delta f_\tau^{-1} C_{\alpha^t\gamma\lambda}C_{\delta\beta^t\lambda^t}C_{\nu^t\alpha\sigma^t}C_{\beta\mu^t\sigma}C_{\gamma^t\emptyset\tau}C_{\emptyset\delta^t\tau^t}V_\nu W_\mu.\cr\label{so6.top}
\end{align}
The 5-brane web diagram \eqref{o-vert-so6} does not have parallel external legs and hence there is no extra factor in this case. The perturbative part is obtained by taking the limit $Q \to 0$. Then only $\lambda=\sigma=\tau=\emptyset$ contributes in the sum in \eqref{so6.top} and \eqref{so6.top} becomes
\begin{align}
Z^{SO(6)}_{\text{pert}}&=Z_{\text{left}}^{SO(6)}Z_{\text{right}}^{SO(6)},\label{so6.pert}
\end{align}
where 
\begin{align}
Z_{\text{left}}^{SO(6)}&=\sum_{\nu,\alpha,\gamma}(-Q')^{|\alpha|}(-P)^{|\nu|}(-R)^{|\gamma|}f_\alpha^{-1}f_\gamma^{-1} C_{\alpha^t\gamma\emptyset}C_{\nu^t\alpha\emptyset}C_{\gamma^t\emptyset\emptyset}V_\nu,\label{so6left}\\
Z_{\text{right}}^{SO(6)}&=\sum_{\mu,\beta,\delta}(-Q')^{|\beta|}(-P)^{|\mu|}(-R)^{|\delta|}f_\beta f_\delta C_{\delta\beta^t\emptyset}C_{\beta\mu^t\emptyset}C_{\emptyset\delta^t\emptyset} W_\mu.\label{so6right}
\end{align}
The explicit evaluation of the summation \eqref{so6left} yields
\begin{align}
Z_{\text{left}}^{SO(6)}=&P.E.\lt(\frac{q}{(1-q)^2}\lt(Q'+R+RQ'+P^2Q'+RQ'P^2+RQ^{\prime2}P^2\rt)\rt) + o(P^4,Q^{\prime 3},R^3)\label{so6left1}\\
=&P.E.\lt(\frac{q}{(1-q)^2}\lt(A_2A_3^{-1}+A_1A_2^{-1}+A_1A_3^{-1}+A_2A_3+A_1A_3+A_1A_3\rt)\rt) \cr
&\hspace{8cm}+ o\lt(P^4, Q'^3, R^3\rt).\label{so6left2}
\end{align}
\eqref{so6left2} is exactly the square root of the contribution of the perturbative part of the pure SO(6) gauge theory obtained from \eqref{Zroot}. We remark that one can show that 
\ba\label{so6lr}
Z_{\text{left}}^{SO(6)}=Z_{\text{right}}^{SO(6)},
\ea
since 
the property \eqref{rel-O-dual-2}, i.e. 
\ba
W_\mu=q^{\frac{\kappa(\mu)}{2}}V_{\mu^t},
\ea
is satisfied by the O-vertex.

The instanton part is obtained from \eqref{so6.top} by removing the perturabtive part \eqref{so6.pert}. Namely the instanton partition function of the pure SO(6) gauge theory is  given by 
\begin{align}\label{so6.inst}
Z^{SO(6)}_{\text{top, inst}} = Z^{SO(6)}_{\text{top}}/Z^{SO(6)}_{\text{pert}}.
\end{align}

We can compare \eqref{so6.inst} with the integral formula \eqref{LMNS-integral} for the pure SO(6) gauge theory. In fact since $\mathfrak{so}(6) \simeq \mathfrak{su}(4)$, we can also use the instanton partition function of the pure SU(4) gauge theory with the zero CS level. The one instanton partition function for the pure SO(6) gauge theory from \eqref{LMNS-integral} becomes 
\begin{align}
Z^{SO(6)}_{\text{loc, 1-inst}}&=\frac{[2i\delta]}{2[\epsilon]^2} \oint\frac{{\rm d}\phi}{2\pi i}\frac{[2\phi]^4}{\prod_{i=1}^3 [\phi +a_i\pm i\delta][\phi-a_i\pm i\delta]}\cr
&\rightarrow \frac{\lt[2a_1\rt]^2}{[\epsilon]^2\lt[a_1+a_2\rt]^2\lt[a_1-a_2\rt]^2\lt[a_1+a_3\rt]^2\lt[a_1-a_3\rt]^2}\cr
&\hspace{0.5cm}+ \frac{\lt[2a_2\rt]^2}{[\epsilon]^2\lt[a_1+a_2\rt]^2\lt[a_1-a_2\rt]^2\lt[a_2+a_3\rt]^2\lt[a_2-a_3\rt]^2}\cr
&\hspace{0.5cm}+\frac{\lt[2a_3\rt]^2}{[\epsilon]^2\lt[a_2+a_3\rt]^2\lt[a_2-a_3\rt]^2\lt[a_1+a_3\rt]^2\lt[a_1-a_3\rt]^2}. \label{so6.1inst}
\end{align}
From \eqref{sun} with $N=4$, the one instanton partition function of the pure SU(4) gauge theory with the zero CS level can be found as
\begin{align}
Z^{SU(4)}_{\text{1-inst}}&=\frac{1}{[\epsilon]^2}\frac{1}{[b_1-b_2]^2[b_1-b_3]^2[b_1-b_4]^2}+\frac{1}{[\epsilon]^2}\frac{1}{[b_1-b_2]^2[b_2-b_3]^2[b_2-b_4]^2}\cr
&\hspace{0.5cm}+\frac{1}{[\epsilon]^2}\frac{1}{[b_1-b_3]^2[b_2-b_3]^2[b_3-b_4]^2}+\frac{1}{[\epsilon]^2}\frac{1}{[b_1-b_4]^2[b_2-b_4]^2[b_3-b_4]^2},\label{su4.1inst}
\end{align}
where we denote the Coulomb branch moduli for SU(4) by $b_i$ with $\sum_{i=1}^4b_i = 0$. Since the fundamental representation of SU(4) is mapped to the spinor representation of SO(6), the relation between the Coulomb branch moduli is 
\begin{align}\label{su4so6}
&b_1 = \frac{1}{2}(a_1 + a_2 + a_3), \quad b_2 = \frac{1}{2}(a_1 - a_2 - a_3), \quad b_3 = \frac{1}{2}(-a_1 + a_2 - a_3), \cr
& b_4 = \frac{1}{2}(-a_1 - a_2 +a_3), \quad
\end{align}
With this map it is possible to check \eqref{so6.1inst} agrees with \eqref{su4.1inst}. We can then check if \eqref{so6.1inst} or \eqref{su4.1inst} agrees with the 1-instanton part of \eqref{so6.inst}. Indeed we found the agreement between them until the order $P^4Q'^4R^7$. 

We can extend the comparison to the two-instanton partition function. We checked that the two-instanton part from \eqref{so6.inst} agrees with the two-instanton part of \eqref{LMNS-integral} for the pure SO(6) gauge theory until the order of $P^4Q'^2R^2$. %

\paragraph{Pure SO(8) gauge theory.} Now we turn to the computation of the simplest non-trivial example of the SO($2N$)-type, the pure SO(8) gauge theory. The brane diagram for the pure SO(8) gauge theory is given by 
\begin{align}
\begin{tikzpicture}
\draw [dashed] (-1,0)--(5,0);
\draw (0,0)--(1,0.5);
\draw (1,0.5)--(3,0.5);
\draw (3,0.5)--(4,0);
\draw (1,0.5)--(1.5,1);
\draw (1.5,1)--(2.5,1);
\draw (2.5,1)--(3,0.5);
\draw (1.5,1)--(1.5,1.5);
\draw (1.5,1.5)--(2.5,1.5);
\draw (2.5,1)--(2.5,1.5);
\draw (1.5,1.5)--(1,2);
\draw (1,2)--(0,2.5);
\draw (1,2)--(3,2);
\draw (2.5,1.5)--(3,2);
\draw (3,2)--(4,2.5);
\draw[<->,dotted] (0,0.05)--(0,0.45);
\node at (0,0.25) [left] {$P$};
\draw[<->,dotted] (0,0.55)--(0,0.95);
\node at (0,0.75) [left] {$Q'$};
\draw[<->,dotted] (0,1.05)--(0,1.45);
\node at (0,1.25) [left] {$R$};
\draw[<->,dotted] (1.55,1.6)--(2.45,1.6);
\node at (2,1.6) [above] {$Q$};
\draw[<->,dotted] (0,1.55)--(0,1.95);
\node at (0,1.75) [left] {$T$};
\end{tikzpicture}
\label{o-vert-so8}
\end{align}
The Coulomb branch moduli $a_1, a_2, a_3, a_4$ are the height of the color D5-brane. Hence $A_i = e^{-a_i}, \; (i=1,2,3,4)$ are parametrized by
\begin{align}
A_1 = TRQ'P, \qquad A_2 = RQ'P, \qquad A_3 = Q'P, \qquad A_4 = P.
\end{align}
Extrapolating the external $(2, 1)$ and $(2, -1)$ 5-brane webs on the O5-plane, the instaton fugacity is related to the length between the 5-branes on the orientifold. Namely it is given by
\begin{align}
\mathfrak{q} = QT^2A_1^{-4} = QT^{-2}Q'^{-4}P^{-4}R^{-4}.
\end{align}

From the web diagram in \eqref{o-vert-so8} it is possible to compute the partition function using the topological vertex and the O-vertex and the partition function becomes
\begin{align}
&Z^{SO(8)}_{\text{top}}\cr
&=\sum_{\mu,\nu,\alpha,\beta,\lambda,\sigma,\gamma,\delta,\tau,\upsilon,\iota,\pi}(-Q)^{|\lambda|+|\sigma|+|\tau|+|\upsilon|}(-Q')^{|\alpha|+|\beta|+2|\sigma|}(-P)^{|\mu|+|\nu|}(-R)^{|\gamma|+|\delta|}(-T)^{|\iota|+|\pi|+2|\upsilon|}\nn\\
&\qquad\quad f_\lambda f_\alpha^{-1}f_\beta f_\sigma^{-3}f_\gamma^{-1}f_\delta f_\tau^{-1}f^{-1}_{\iota}f_{\pi}f^{-3}_{\upsilon}C_{\alpha^t\gamma\lambda}C_{\delta\beta^t\lambda^t}C_{\nu^t\alpha\sigma^t}C_{\beta\mu^t\sigma}C_{\gamma^t\iota\tau}C_{\pi\delta^t\tau^t}C_{\iota^t\emptyset\upsilon}C_{\emptyset\pi^t\upsilon^t}V_\nu W_\mu.\label{so8.top}
\end{align}
Since there is no parallel external lines, \eqref{so8.top} does not contain an extra factor.

The perturbative part is obtained by the limit $Q \to 0$ and the partition functin splits into two parts, 
\begin{align}
Z^{SO(8)}_{\text{pert}} = Z^{SO(8)}_{\text{left}}Z^{SO(8)}_{\text{right}},\label{so8.pert}
\end{align}
where
\begin{align}
Z^{SO(8)}_{\text{left}}&=\sum_{\nu,\alpha,\gamma,\iota}(-Q')^{|\alpha|}(-P)^{|\nu|}(-R)^{|\gamma|}(-T)^{|\iota|} f_\alpha^{-1}f_\gamma^{-1}f^{-1}_{\iota}C_{\alpha^t\gamma\emptyset}C_{\nu^t\alpha\emptyset}C_{\gamma^t\iota\emptyset}C_{\iota^t\emptyset\emptyset}V_\nu,\\
Z^{SO(8)}_{\text{right}}&=\sum_{\mu,\beta,\delta,\pi}(-Q')^{|\beta|}(-P)^{|\mu|}(-R)^{|\delta|}(-T)^{|\pi|}f_\beta f_\delta f_{\pi}C_{\delta\beta^t\emptyset}C_{\beta\mu^t\emptyset}C_{\pi\delta^t\emptyset}C_{\emptyset\pi^t\emptyset}W_\mu.\label{so8.right}
\end{align}
The computing the summation of the left part yields
\begin{align}
Z^{SO(8)}_{\text{left}}&=P.E.\lt(\frac{q}{(1-q)^2}\lt(Q'+R+T+Q'R+RT+Q'RT+P^2Q'+RQ'P^2+RQ^{\prime2}P^2\right.\right.\cr
&\hspace{3.5cm}\left.+Q'RTP^2+Q^{\prime2}RTP^2+Q^{\prime2}R^2TP^2\rt)\Big) + o(P^4,Q^{\prime 3},R^3,T^3)\label{so8left1}\\
&=P.E.\lt(\frac{q}{(1-q)^2}\lt(A_3A_4^{-1}+A_2A_3^{-1}+A_1A_2^{-1}+A_2A_3^{-1}+A_1A_3^{-1}+A_1A_4^{-1}+A_3A_4\right.\right.\cr
&\hspace{3.25cm}\left.+A_2A_4+A_2A_3 +A_1A_4+A_1A_3+A_1A_2\rt)\Big) + o(P^4,Q^{\prime 3},R^3,T^3).\label{so8.left}
\end{align}
We can compare \eqref{so8.left} with the localization result \eqref{Zroot} for the pure SO(8) gauge theory and \eqref{so8.left} exactly reprduces the square root of the root contribution to the peturbative partition function of the pure SO(8) gauge theory. We remark again that 
\ba
Z^{SO(8)}_{\text{left}}=Z^{SO(8)}_{\text{right}},
\ea
by the property \eqref{rel-O-dual-2}. 

The instanton partition function of the pure SO(8) gauge theory is obtained by removing the perturbative part \eqref{so8.pert} from \eqref{so8.top},
\begin{align}
Z^{SO(8)}_{\text{top, inst}} = Z^{SO(8)}_{\text{top}}/Z^{SO(8)}_{\text{pert}}. \label{so8.inst}
\end{align}
The localization result for the instanton partition function for the pure SO(8) gauge theory is obtained from \eqref{LMNS-integral}. 
For the 1-instanton part, we find the perfect match between the expression from the integral \eqref{LMNS-integral} and the result obtained from the topological vertex formalism \eqref{so8.inst} until the order $P^2Q'^2R^2T^2$. 
We also 
computed the two-instanton partition functions in the two methods 
and the results agreee with each other until the order $P^2Q'^2R^2T^2$. 

\subsection{$\text{SO}(2N+1)$ gauge theories}\label{s:so-2N+1}

As we dicussed in section \ref{s:Higgsing-tild-O} it is also possible to use the O-vertex to compute the partition functions of SO($2N+1$) gauge theories using their 5-brane web realization with an $\widetilde{\rm O5}$-plane. We will apply the formalism to the examples of the pure SO(5) and SO(7) gauge theories.

\paragraph{Pure SO(5) gauge theory.} We start from the pure SO(5) gauge theory. The 5-brane web diagram which we use for the computation of the partition function of the pure SO(5) gauge theory is given by 
\begin{align}
\begin{tikzpicture}
\draw [dashed] (-1,0)--(5,0);
\draw (0,0)--(1,0.5);
\draw (1,0.5)--(2.75,0.5);
\draw (1,0.5)--(1.5,1);
\draw (1.5,1)--(2.75,1);
\draw (1.5,1)--(1.5,1.5);
\draw (2.75,1)--(3.25,1.5);
\draw (3,0.25)--(3.5,0);
\draw (3,0.25)--(2.5,0.25);
\draw (3,0.25)--(2.75,0.5);
\draw (2.75,0.5)--(2.75,1);
\node at (2.5,0.25) [left] {$\emptyset$};
\draw[<->,dotted] (0,0.05)--(0,0.45);
\node at (0,0.25) [left] {$P$};
\draw[<->,dotted] (0,0.55)--(0,0.95);
\node at (0,0.75) [left] {$Q'$};
\draw[<->,dotted] (3.6,0)--(3.6,0.3);
\node at (3.6,0.3) [right] {$T$};
\draw[<->,dotted] (1.6,1.1)--(2.7,1.1);
\node at (2.2,1.1) [above] {$Q$};
\end{tikzpicture}
\label{o-vert-so5}
\end{align}
in the limit $T\rightarrow 1$. Note that we first introduce $T$ in \eqref{o-vert-so5} so that we can apply the O-vertex \eqref{Wnu} in the partition function computation, and then take the limit $T \to 1$ at the end of the calculation. The relation between the gauge theory parameters and the K\"ahler parameters can be again read from the diagram. Let $a_1, a_2$ be the Coulomb branch moduli of the pure SO(5) gauge theory. They are related to the height of the color D5-branes and hence $A_i = e^{-a_i}\; (i=1, 2)$ are given by
\be
A_1 = Q'P, \qquad A_2 = P. \label{so5.CB}
\ee
For the identification of the instanton fugacity we extrapolate the $(0, 1)$ and $(1, 1)$ 5-brane and measure the length between the 5-branes intersecting with the orientifold. Namely it is given by
\begin{align}
\mathfrak{q} = QA_1^{-1} = QQ'^{-1}P^{-1}.
\end{align}

Let us compute the partition function of the pure SO(5) gauge theory from the web diagram in \eqref{o-vert-so5}. Since we take $T \to 1$ at the end of the computation it seems that we need to sum over the Young diagram assigned to the line with the K\"ahler parameter $T$. It turns out that the series in terms of $T$ terminates at a fixed order of $P$. In order to see it, we consider the following diagram, 
\begin{align}
\begin{tikzpicture}
\draw [dashed] (-1,0)--(5,0);
\draw (3,0.25)--(3.5,0);
\draw (3,0.25)--(2.5,0.25);
\draw (3,0.25)--(2.5,0.75);
\draw[<-] (2.75,0.50)--(3.0,0.25);
\node at (2.5,0.25) [left] {$\emptyset$};
\draw[<->,dotted] (2,0.05)--(2,0.75);
\node at (1.7,0.25) [left] {$P$};
\draw[<->,dotted] (3.6,0)--(3.6,0.3);
\node at (3.6,0.3) [right] {$T$};
\node at (2.5,0.75) [above] {$\nu$};
\end{tikzpicture}
\label{o-vert_tilde}
\end{align}
The partition function associated to the diagram \eqref{o-vert_tilde} is given by
\be
\tilde{Z}_{\nu}(P, T) = \sum_{\mu}(-T)^{|\mu|}(-P/T)^{|\nu|}C_{\nu\mu^t\emptyset}W_{\mu}. \label{Ztilde}
\ee
We observed that the expansion of \eqref{Ztilde} by $T$ terminates for a fixed $\nu$. More precisely for a fixed $\nu$ we observed that \eqref{Ztilde} is a polynomial of $T$ with the lowest order $T^{-|\nu|}$ and the highest order $T^{|\nu|}$. We checked this feature for $|\nu| \leq 4$ with the sum taken until $|\mu|=9$. Therefore we conjecture that the expansion of \eqref{Ztilde} by $T$ terminates at the order $|\nu|$ and it is enough to compute the summation until $|\mu| = 2|\nu|$. Namely instead of \eqref{Ztilde} we can use 
\be
\tilde{Z}_{\nu}(P, T) = \sum_{\substack{\mu\\ |\mu| \leq 2|\nu|}}(-T)^{|\mu|}(-P/T)^{|\nu|}C_{\nu\mu^t\emptyset}W_{\mu}.\label{Ztilde1}
\ee
Then the number of the Young diagrams $\mu$ over which we need to sum for a fixed $|\nu|$ becomes finite. Therefore we can obtain the exact result after taking $T \to 1$ for coefficient of $P^{|\nu|}$ when we perform the summation of $\mu$ which satisfies $|\mu| \leq 2|\nu|$. 

With this prescription it is possible to compute the partition function for the 5-brane web in \eqref{o-vert-so5} and it is given by 
\begin{align}
Z^{SO(5)}_{\text{top}}&=\lim_{T\to 1}\sum_{\substack{\mu,\nu,\lambda,\sigma,\alpha,\beta,\pi\\|\mu|\leq 2|\pi|}}(-Q)^{|\lambda|+|\sigma|}(-Q')^{|\alpha|+|\beta|}Q^{\prime |\lambda|}(-P)^{|\nu|}(-T)^{|\mu|}\left(-P/T\right)^{|\pi|}\cr
&\hspace{3.5cm}f_\lambda^2f_\alpha^{-1}f_\pi f_\beta C_{\nu^t\alpha\lambda}C_{\alpha^t\emptyset\sigma}C_{\pi\mu^t\emptyset}C_{\beta\pi^t\lambda^t}C_{\emptyset\beta^t\sigma^t}V_\nu W_\mu\label{so5.top0}\\
&=\sum_{\substack{\mu,\nu,\lambda,\sigma,\alpha,\beta,\pi\\|\mu|\leq 2|\pi|}}(-Q)^{|\lambda|+|\sigma|}(-Q')^{|\alpha|+|\beta|}Q^{\prime |\lambda|}(-P)^{|\nu|+ |\pi|}(-1)^{|\mu|}\cr
&\hspace{3.5cm}f_\lambda^2f_\alpha^{-1}f_\pi f_\beta C_{\nu^t\alpha\lambda}C_{\alpha^t\emptyset\sigma}C_{\pi\mu^t\emptyset}C_{\beta\pi^t\lambda^t}C_{\emptyset\beta^t\sigma^t}V_\nu W_\mu.\label{so5.top}
\end{align}
In this case there is no extra factor as the diagram in \eqref{o-vert-so5} does not have parallel external legs.

We first work out the perturbative part of \eqref{so5.top} or \eqref{so5.top0} which is obtained by taking the limit $Q\to 0$. Then \eqref{so5.top} splits into two factors
\begin{align}
Z^{SO(5)}_{\text{pert}} = Z^{SO(5)}_{\text{left}}Z^{SO(5)}_{\text{right}},
\end{align}
where  
\begin{align}
Z^{SO(5)}_{\text{left}} &= \sum_{\nu,\alpha}(-Q')^{|\alpha|}(-P)^{|\nu|}f_\alpha^{-1}C_{\nu^t\alpha\emptyset}C_{\alpha^t\emptyset\emptyset}V_\nu,\label{so5.left}\\
Z^{SO(5)}_{\text{right}}&=\lim_{T\to 1}\sum_{\substack{\mu, \beta,\pi\\|\mu|\leq 2|\pi|}}(-Q')^{|\beta|}(-T)^{|\mu|}\left(-P/T\right)^{|\pi|}f_\pi f_\beta C_{\pi\mu^t\emptyset}C_{\beta\pi^t\emptyset}C_{\emptyset\beta^t\emptyset}W_\mu.\label{so5.right}
\end{align}
Note that \eqref{so5.left} is equal to \eqref{so4-left} and \eqref{so5.right} without taking the limit $T\to 1$ nor the restriction on the summation is equal to \eqref{so6right} with $P, Q', R$ in \eqref{so6.pert} replaced with $T, P/T, Q'$ respectively. 
When we use the result \eqref{so4left1}, \eqref{so6left1}\footnote{Note that we have the relation \eqref{so6lr}.}, \eqref{so5.left} and \eqref{so5.right} without the limit nor the restriction become 
\begin{align}
&\sum_{\nu,\alpha}(-Q')^{|\alpha|}(-P)^{|\nu|}f_\alpha^{-1}C_{\nu^t\alpha\emptyset}C_{\alpha^t\emptyset\emptyset}V_\nu=P.E.\left(\frac{q}{(1-q)^2}(Q' + P^2Q')\right), 
\label{so5.left2}\\
&\sum_{\mu, \beta,\pi}(-Q')^{|\beta|}(-P)^{|\pi|}(-T)^{|\mu|-|\pi|}f_\pi f_\beta C_{\pi\mu^t\emptyset}C_{\beta\pi^t\emptyset}C_{\emptyset\beta^t\emptyset}W_\mu\cr
&
=P.E.\left(\frac{q}{(1-q)^2}\left(P/T + Q' + PQ'/T + PT + PQ'T + P^2Q'\right)\right).
\label{so5.right2}
\end{align}
Since \eqref{so4left1}, \eqref{so6left1} have already reproduced the square root of the root contribution to the perturbative part for the pure SO(4) gauge theory and the pure SO(6) gauge theory respectively, 
we assumed that \eqref{so4left1} and \eqref{so6left1} are exact results. From \eqref{so5.right2} one can see that the order of $T$ is always less than or equal to the order of $P$. This gives another support for the the conjecture on the restriction of the Young diagram summation \eqref{Ztilde1} since the restriction also implies that the order of $T$ is less than or equal to that of $P$. Taking the limit $T\to 1$ and substituting the Coulomb branch moduli of the SO(5) into \eqref{so5.left2} and \eqref{so5.right2}, we obtain
\begin{align}
Z^{SO(5)}_{\text{left}} &=P.E.\left(\frac{q}{(1-q)^2}(A_1A_2^{-1} + A_1A_2)\right), 
\label{so5.left3}\\
Z^{SO(5)}_{\text{right}}&=P.E.\left(\frac{q}{(1-q)^2}\left(2A_2 + A_1A_2^{-1} + 2A_1 +  A_1A_2\right)\right). \label{so5.right3}
\end{align}
Then the product of \eqref{so5.left3} and \eqref{so5.right3} should correspond to the root contribution to the perturbative partition function of the pure SO(5) gauge theory and one can see that it indeed agrees with the result computed from \eqref{Zroot}.

Now we turn to the calculation of the instanton partition function from the vertex formalism. The instanton partition function is obtained by
\be
Z^{SO(5)}_{\text{top, inst}} = Z^{SO(5)}_{\text{top}}/Z^{SO(5)}_{\text{pert}}. \label{so5.inst}
\ee
We checked that the one-instanton part of \eqref{so5.inst} agrees with the known SO(5) one-instanton and two-instanton partition functions computed from \eqref{LMNS-integral} until the order $P^2Q'^2$. 

\paragraph{Pure SO(7) gauge theory.} SO(7) gauge theories are more interesting for us, as we will see later that it may be related to the construction of the pure $G_2$ gauge theory. The 5-brane web of the pure SO(7) gauge theory is given by 
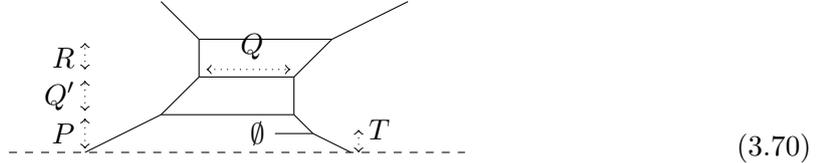
\begin{align}
\begin{tikzpicture}
\draw [dashed] (-1,0)--(5,0);
\draw (0,0)--(1,0.5);
\draw (1,0.5)--(2.75,0.5);
\draw (1,0.5)--(1.5,1);
\draw (1.5,1)--(2.75,1);
\draw (1.5,1)--(1.5,1.5);
\draw (1.5,1.5)--(3.25,1.5);
\draw (1.5,1.5)--(1,2);
\draw (2.75,1)--(3.25,1.5);
\draw (3.25,1.5)--(4.25,2);
\draw (3,0.25)--(3.5,0);
\draw (3,0.25)--(2.5,0.25);
\draw (3,0.25)--(2.75,0.5);
\draw (2.75,0.5)--(2.75,1);
\node at (2.5,0.25) [left] {$\emptyset$};
\draw[<->,dotted] (0,0.05)--(0,0.45);
\node at (0,0.25) [left] {$P$};
\draw[<->,dotted] (0,0.55)--(0,0.95);
\node at (0,0.75) [left] {$Q'$};
\draw[<->,dotted] (0,1.1)--(0,1.45);
\node at (0,1.25) [left] {$R$};
\draw[<->,dotted] (3.6,0)--(3.6,0.3);
\node at (3.6,0.3) [right] {$T$};
\draw[<->,dotted] (1.6,1.1)--(2.7,1.1);
\node at (2.2,1.1) [above] {$Q$};
\end{tikzpicture}
\label{o-vert-so7}
\end{align}
The relation between the Coulomb branch moduli $A_1 = e^{-a_1}, A_2 = e^{-a_2}, A_3 = e^{-a_3}$ and the K\"ahler parameters is
\begin{align}
A_1 = PQ'R, \qquad A_2 =PQ', \qquad A_3 = P.
\end{align}
On the other hand, the instanton fugacity is obtained in the same way as before and it is given by
\be
\mathfrak{q} = QRA_1^{-3} = QP^{-3}Q'^{-3}R^{-2}.
\ee

Applying the topological vertex and the O-vertex with the prescription \eqref{Ztilde1} to the web diagram in \eqref{o-vert-so7}, the partition function becomes 
\begin{align}
&Z^{SO(7)}_{\text{top}}\cr
&=\lim_{T \to 1}\sum_{\substack{\mu,\nu,\lambda,\sigma,\tau,\alpha,\beta,\gamma,\delta,\pi\\|\mu| \leq 2|\pi|}}(-Q)^{|\lambda|+|\sigma|+|\tau|}(-Q')^{|\alpha|+|\beta|}Q'^{|\lambda|}(-R)^{|\gamma|+|\delta|}R^{|\tau|}(-P)^{|\nu|}(-T)^{|\mu|}\left(-P/T\right)^{|\pi|}\cr
&\hspace{3.5cm}f_\lambda^2f_\tau^{-2}f_\alpha^{-1}f_\gamma^{-1}f_\pi f_\beta f_\delta C_{\nu^t\alpha\lambda}C_{\alpha^t\gamma\sigma}C_{\gamma^t\emptyset\tau}C_{\pi\mu^t\emptyset}C_{\beta\pi^t\lambda^t}C_{\delta\beta^t\sigma^t}C_{\emptyset\delta^t\tau^t}V_\nu W_\mu\cr\label{so7.top0}\\
&=\sum_{\substack{\mu,\nu,\lambda,\sigma,\tau,\alpha,\beta,\gamma,\delta,\pi\\|\mu| \leq 2|\pi|}}(-Q)^{|\lambda|+|\sigma|+|\tau|}(-Q')^{|\alpha|+|\beta|}Q'^{|\lambda|}(-R)^{|\gamma|+|\delta|}R^{|\tau|}(-P)^{|\nu|}(-1)^{|\mu|}\left(-P\right)^{|\pi|}\cr
&\hspace{3.5cm}f_\lambda^2f_\tau^{-2}f_\alpha^{-1}f_\gamma^{-1}f_\pi f_\beta f_\delta C_{\nu^t\alpha\lambda}C_{\alpha^t\gamma\sigma}C_{\gamma^t\emptyset\tau}C_{\pi\mu^t\emptyset}C_{\beta\pi^t\lambda^t}C_{\delta\beta^t\sigma^t}C_{\emptyset\delta^t\tau^t}V_\nu W_\mu.\label{so7.top}
\end{align}

The perturbative part of \eqref{so7.top} or \eqref{so7.top0} is obtained by taking the limit $Q \to 0$, 
\begin{align}
Z^{SO(7)}_{\text{pert}} &= \lim_{T \to 1}\sum_{\substack{\mu,\nu,\lambda,\sigma,\tau,\alpha,\beta,\gamma,\delta,\pi\\|\mu| \leq 2|\pi|}}(-Q')^{|\alpha|+|\beta|}(-R)^{|\gamma|+|\delta|}(-P)^{|\nu|}(-T)^{|\mu|}\left(-P/T\right)^{|\pi|}\cr
&\hspace{3.5cm}f_\alpha^{-1}f_\gamma^{-1}f_\pi f_\beta f_\delta C_{\nu^t\alpha\emptyset}C_{\alpha^t\gamma\emptyset}C_{\gamma^t\emptyset\emptyset}C_{\pi\mu^t\emptyset}C_{\beta\pi^t\emptyset}C_{\delta\beta^t\emptyset}C_{\emptyset\delta^t\emptyset}V_\nu W_\mu.\cr\label{so7.pert0}
\end{align}
Similar to the case of the pure SO(5) gauge theory, the expression \eqref{so7.pert0} before taking the limit $T \to 1$ and restricting the Young diagram in summation is the product of \eqref{so6left} and \eqref{so8.right}. Hence instead of computing \eqref{so7.pert0} directly we here simply reuse the result \eqref{so6left1} and \eqref{so8left1}, and each factor now becomes
\begin{align}
&Z^{SO(6)}_{\text{left}}=P.E.\lt(\frac{q}{(1-q)^2}\lt(Q'+R+RQ'+P^2Q'+RQ'P^2+RQ^{\prime2}P^2\rt)\rt)
\label{so7.left0}\\
&Z^{SO(8)}_{\text{right}}\cr
&=P.E.\lt(\frac{q}{(1-q)^2}\lt(P/T+Q' + R + PQ'/T + Q'R + PQ'R/T+PT+Q'PT\right.\right.\cr
&\hspace{5cm}\left.+RQ'PT+Q'P^2+RQ'P^2+RQ^{\prime 2}P^2\rt)\Big). 
\label{so7.right0}
\end{align}
Since \eqref{so6left1} and \eqref{so8left1} reproduced the square root of the contribution to the perturbative part of SO(6) and SO(8) respectively, we assumed \eqref{so6left1} and \eqref{so8left1} are the exact results. Again the restriction of the summation for the Young diagram $\mu$ in \eqref{so7.pert0} is consistent with \eqref{so7.right0} as the order of $T$ is always less than or equal to the order of $P$. The perturbative part of the pure SO(7) gauge theory is obtained by taking the $T\to 1$ limit and we have 
\begin{align}
Z^{SO(7)}_{\text{pert}}&=P.E.\lt(\frac{2q}{(1-q)^2}(R+Q'+RQ'+PQ' +P+ PQ'R + P^2Q' + RQ'P^2+RQ^{\prime 2}P^2\rt) \cr
&=P.E.\lt(\frac{2q}{(1-q)^2}(A_1A_2^{-1}+A_2A_3^{-1}+A_1A_3^{-1}+A_2 +A_3+ A_1 + A_2A_3 + A_1A_3 + A_1A_2\rt).\label{pert-so7}
\end{align}
We see that \eqref{pert-so7} agrees with the root contribution to the perturbative partition \eqref{Zroot} for the pure SO(7) gauge theory.
 
The instanton partition function is then given by removing the perturbative part,
\begin{align}\label{so7.inst}
Z^{SO(7)}_{\text{top, inst}}=Z^{SO(7)}_{\text{top}}/Z^{SO(7)}_{\text{pert}}.
\end{align}
We checked that the one-instanton part of \eqref{so7.inst} agrees with the known SO(7) one-instanton partition function calculated from the integral \eqref{uni-one-inst} until the order $P^2Q'^2R^2$. 
We also checked that the two-instaton partition function computed from \eqref{LMNS-integral} for the pure SO(7) gauge theory matches with the result computed from the topological vertex formalism with O-vertex \eqref{so7.inst} as an expansion until the order $P^2Q'^2R^2$. 

\subsection{Pure $G_2$ gauge theory}\label{s:G2}

5-brane webs with an $\widetilde{\text{O5}}$-plane yield not only SO$(2N+1)$ gauge theories but also gauge theories with an exceptional gauge group $G_2$ \cite{G-type}. The construction utilizes the fact that the pure $G_2$ gauge theory may be realized as a low energy theory on a Higgs branch of the SO(7) gauge theory with a hypermultiplet in the spinor representation. In fact the spinor matter can be introduced into 5-brane webs with an O5-plane or an $\widetilde{\text{O5}}$-plane \cite{Zafrir:2015ftn}. Therefore Higgsing the 5-brane web for the SO(7) gauge theory with spinor matter should yield a diagram for the pure $G_2$ gauge theory. In \cite{G-type}, two types of the 5-brane web diagrams were proposed for the pure $G_2$ gauge theory. It turns out that it is possible to apply the formalism of \cite{Kim-Yagi} to one of the diagrams and the partition function for the pure $G_2$ gauge theory has been computed using the topological vertex in \cite{G-type}. When we use the O-vertex it is possible to compute the partition function of the pure $G_2$ gauge theory from the other type of the diagrams. We will explicitly compute the partition function here and compare the result with the known expression of the partition function of the pure $G_2$ gauge theory.

Let us first recall the argument presented in \cite{G-type}. The brane diagram of the SO(7) theory with a hypermultiplet in the spinor representation is given by \cite{Zafrir:2015ftn}
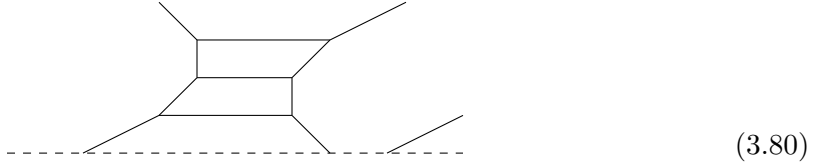
\begin{align}
\begin{tikzpicture}
\draw [dashed] (-1,0)--(5,0);
\draw (0,0)--(1,0.5);
\draw (1,0.5)--(2.75,0.5);
\draw (1,0.5)--(1.5,1);
\draw (1.5,1)--(2.75,1);
\draw (1.5,1)--(1.5,1.5);
\draw (1.5,1.5)--(3.25,1.5);
\draw (1.5,1.5)--(1,2);
\draw (2.75,1)--(3.25,1.5);
\draw (3.25,1.5)--(4.25,2);
\draw (3.25,0)--(2.75,0.5);
\draw (2.75,0.5)--(2.75,1);
\draw (4,0)--(5,0.5);
\end{tikzpicture}
\label{o-vert-so7-spinor}
\end{align}
The presence of two parallel external $(2, 1)$ 5-branes implies an SU(2) flavor symmetry associated to the spinor matter. To perform the Higgsing which breaks the SU(2) one needs to put the $(2, 1)$ 5-branes on top of each other.  For that we perform a generalized flop transition
\cite{Hayashi:2017btw} which yields 
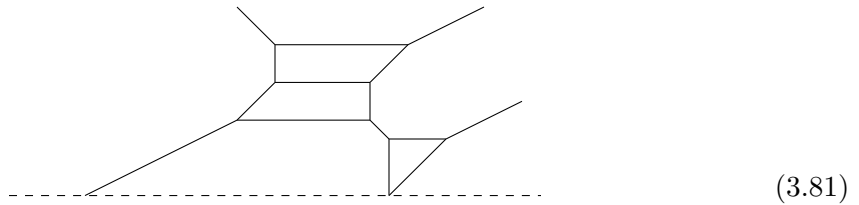
\begin{align}
\begin{tikzpicture}
\draw [dashed] (-2,-0.5)--(5,-0.5);
\draw (-1,-0.5)--(1,0.5);
\draw (1,0.5)--(2.75,0.5);
\draw (1,0.5)--(1.5,1);
\draw (1.5,1)--(2.75,1);
\draw (1.5,1)--(1.5,1.5);
\draw (1.5,1.5)--(3.25,1.5);
\draw (1.5,1.5)--(1,2);
\draw (2.75,1)--(3.25,1.5);
\draw (3.25,1.5)--(4.25,2);
\draw (3,0.25)--(3.75,0.25);
\draw (3,0.25)--(3,-0.5);
\draw (3, -0.5)--(3.75, 0.25);
\draw (3,0.25)--(2.75,0.5);
\draw (2.75,0.5)--(2.75,1);
\draw (4-0.25,0.25)--(5-0.25,0.75);
\end{tikzpicture}
\label{o-vert-so7-spinor-comp}
\end{align}
From \eqref{o-vert-so7-spinor-comp}, we perform the usual flop transition twice and arrive at 
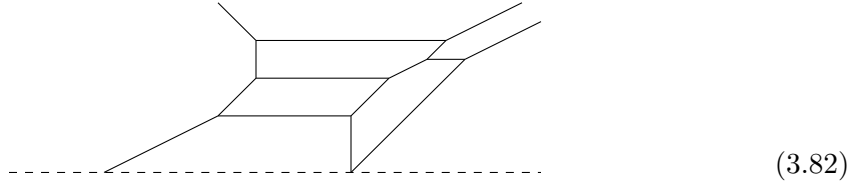
\begin{align}
\begin{tikzpicture}
\draw [dashed] (-2,-0.5)--(5,-0.5);
\draw (-0.75,-0.5)--(0.75,0.25);
\draw (0.75,0.25)--(2.5,0.25);
\draw (0.75,0.25)--(1.25,0.75);
\draw (1.25,0.75)--(3,0.75);
\draw (1.25,0.75)--(1.25,1.25);
\draw (1.25,1.25)--(3.75,1.25);
\draw (1.25,1.25)--(0.75,1.75);
\draw (3.5,1)--(3.75,1.25);
\draw (3.75,1.25)--(4.75,1.75);
\draw (3.5,1)--(4,1);
\draw (2.5,0.25)--(2.5,-0.5);
\draw (2.5,0.25)--(3,0.75);
\draw (3,0.75)--(3.5,1);
\draw (2.5,-0.5)--(4,1);
\draw (4.25-0.25,1)--(5.25-0.25,1.5);
\end{tikzpicture}
\end{align}
Then we can tune the length of 5-branes to put the two parallel external $(2, 1)$ 5-branes on top of each other. Supposing that a 7-brane ends on each external 5-brane, we can remove a piece of a $(2, 1)$ 5-brane between two $(2, 1)$ 7-branes. This corresponds to the Higgsing and sending the piece of the 5-brane far away giving rises to a diagram of the pure $G_2$ gauge theory given by 
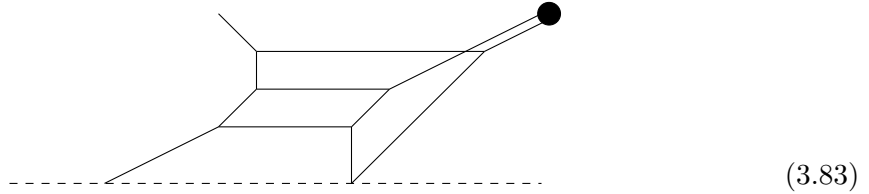
\begin{align}\label{fig:g2diag}
\begin{tikzpicture}
\draw [dashed] (-2,-0.5)--(5,-0.5);
\draw (-0.75,-0.5)--(0.75,0.25);
\draw (0.75,0.25)--(2.5,0.25);
\draw (0.75,0.25)--(1.25,0.75);
\draw (1.25,0.75)--(3,0.75);
\draw (1.25,0.75)--(1.25,1.25);
\draw (1.25,1.25)--(4.25,1.25);
\draw (1.25,1.25)--(0.75,1.75);
\draw (2.5,0.25)--(2.5,-0.5);
\draw (2.5,0.25)--(3,0.75);
\draw (3,0.75)--(5,1.75);
\draw (2.5,-0.5)--(4.25,1.25);
\draw (4.25,1.25)--(5.25,1.75);
\draw [black, fill] (5.1,1.75) circle (0.15);
\end{tikzpicture}
\end{align}
where the 7-brane on which the $(2, 1)$ 5-branes end is depicted as a black dot.

We then consider applying the topological vertex as well as the O-vertex to the diagram in \eqref{fig:g2diag}. 
At the practical level, we use the following diagram for computation. 
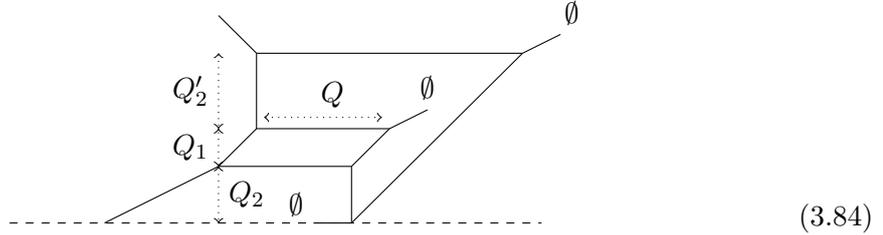
\begin{align}\label{fig:g2diag2}
\begin{tikzpicture}
\draw [dashed] (-2,-0.5)--(5,-0.5);
\draw (-0.75,-0.5)--(0.75,0.25);
\draw (0.75,0.25)--(2.5,0.25);
\draw[<->,dotted] (0.75,0.25)--(0.75,-0.5);
\node at (0.75,-0.125) [right] {$Q_2$};
\draw[<->,dotted] (0.75,0.25)--(0.75,0.75);
\node at (0.75,0.5) [left] {$Q_1$};
\draw[<->,dotted] (0.75,1.75)--(0.75,0.75);
\node at (0.75,1.25) [left] {$Q'_2$};
\draw (0.75,0.25)--(1.25,0.75);
\draw (1.25,0.75)--(3,0.75);
\draw[<->,dotted] (1.25+0.1,0.9)--(3-0.1,0.9);
\node at (2.25,0.9) [above] {$Q$};
\draw (1.25,0.75)--(1.25,1.75);
\draw (1.25,1.75)--(4.75,1.75);
\draw (1.25,1.75)--(0.75,2.25);
\draw (2.5,0.25)--(2.5,-0.5);
\draw (2,-0.5)--(2.5,-0.5);
\draw (2.5,0.25)--(3,0.75);
\draw (3,0.75)--(3.5,1);
\node at (2,-0.25) [left] {$\emptyset$};
\draw (2.5,-0.5)--(4.75,1.75);
\draw (5.25,2)--(4.75,1.75);
\node at (3.5,1) [above] {$\emptyset$};
\node at (5.4,2) [above] {$\emptyset$};
\end{tikzpicture}
\end{align}
The constraint that the two $(2,1)$ 5-branes 
with empty Young diagrams assigned coincide at the infinity forces the K\"ahler parameters to satisfy $Q'_2=Q_2$. The Coulomb branch moduli of the pure $G_2$ gauge theory can be parametrized by 
\be\label{g2CB}
Q_1 = A_1A_2^{-1}, \qquad Q_2 = A_2.
\ee
On the other hand the instanton fugacity can be obtained by extrapolating the external 5-branes to the orientifold and then measuring the length between the intersection points. The instanton fugacity $\mathfrak{q}$ is given by
\be\label{g2instf}
\mathfrak{q} = QQ_2^2(A_1A_2)^{-3} = QQ_1^{-3}Q_2^{-4}.
\ee

With the parameterization \eqref{g2CB} and \eqref{g2instf} we can compute the partition function of the pure $G_2$ gauge theory from the diagram in \eqref{fig:g2diag2}. For the part where the $(2, 1)$ 5-brane intersects with the orientifold we can use the standard O-vertex. For the right part where 5-branes intersect with the orientifold it is possible to apply the original formalism in \cite{Kim-Yagi}. Then the full partition function obtained by applying the topological vertex as well as the O-vertex to the diagram in \eqref{fig:g2diag2} is given by
\begin{equation}
\begin{split}
Z^{G_2}_{\text{top}}=\sum_{\mu,\nu,\alpha,\beta,\lambda,\sigma,\gamma,\delta,\tau}(-Q)^{|\lambda|+|\sigma|}(QQ_2^2)^{|\tau|}(-Q_1)^{|\alpha|+|\delta|}(-Q_2)^{|\nu|+|\gamma|+|\beta|}(-Q_1Q_2^2)^{|\mu|} \cr
\times f^{-1}_\alpha  f^{-1}_\gamma f_\mu f_\beta f_\sigma f_\lambda^{-1} f_\delta f^{-3}_\tau C_{\nu^t\alpha\sigma}C_{\alpha^t\gamma\lambda}C_{\gamma^t\emptyset\tau}C_{\beta\mu^t\emptyset}C_{\mu\emptyset\tau}C_{\delta\beta^t\sigma^t}C_{\emptyset\delta^t\lambda^t}V_\nu ,
\label{G2-vert-exp}
\end{split}
\end{equation}
The perturbative part can be obtained by taking the limit $Q \to 0$ since $Q$ is proportional to the instanton fugacity. In this limit, the web diagram \eqref{fig:g2diag2} splits into the left part and the right part. The left part is identical to the left part of the SO(6) diagram or that of the SO(7) diagram and the right part is the same as the half of the web diagram for a pure SU(4) gauge theory. 
Hence we can make use of \eqref{so6left} with the change $R \to Q_2$, $Q' \to Q_1$, $P \to Q_2$ and the partition function from the left part is given by
\be
Z_{\text{left}}^{G_2} = P.E.\left(\frac{q}{(1-q)^2}\left(Q_1 + Q_2 + Q_1Q_2 + Q_1Q_2^2 + Q_1Q_2^3 + Q_1^2Q_3^3\right)\right) + o(Q_1^3, Q_2^3). \label{g2left}
\ee
For the right part, it is possible to sum over the Young diagrams. In terms of the SU(4), the simple roots of SU(4) correspond to the K\"ahler parameter $Q_1, Q_2, Q_1Q_2^2$. Then the partition function from the right part yields
\begin{align}
Z_{\text{right}}^{G_2} =P.E.\left(\frac{q}{(1-q)^2}\left(Q_1 + Q_2 + Q_1Q_2^2 + Q_1Q_2 + Q_1Q_2^3 + Q_1^2Q_2^3\right)\right). \label{g2right}
\end{align}
The total perturbative part becomes the product of \eqref{g2left} and \eqref{g2right} and 
\begin{align}
Z_{\text{pert}}^{G_2}&=\lim_{Q \to 0}Z^{G_2}_{\text{top}}\cr
&=P.E.\lt(\frac{2q}{(1-q)^2}\lt(Q_1+Q_2+Q_1Q_2+Q_1Q_2^2+Q_1Q_2^3+Q_1^{2}Q_2^3\rt)\rt) + o(Q_1^3, Q_2^3)\label{G2-pert0}\\
&=P.E.\lt(\frac{2q}{(1-q)^2}\lt(A_1A_2^{-1}+A_2+A_1+A_1A_2+A_1A_2^2+A_2^2A_1\rt)\rt) + o(Q_1^3, Q_2^3).\cr \label{G2-pert}
\end{align}
In fact \eqref{G2-pert} is precisely the root contribution of the perturbative part of the partition function of the pure $G_2$ gauge theory. The simple roots of the Lie algbera of $G_2$ may be written as $\alpha_1 = [2, -3], \alpha_2 = [-1, 2]$. Then the positive roots are given by 
\be
\alpha_1 + \alpha_2 = [1, -1], \quad \alpha_1 + 2\alpha_2 = [0, 1], \quad \alpha_1 + 3\alpha_2 = [-1, 3], \quad 2\alpha_1 + 3\alpha_2 = [1, 0],
\ee
in addition to $\alpha_1, \alpha_2$. 
Associating the K\"ahler parameters $Q_1$ and $Q_2$ to the simple roots $\alpha_1$ and $\alpha_2$ respectively, we can see that \eqref{G2-pert0} or \eqref{G2-pert} reproduces the root contribution to the perturbative part of the partition function.

Let us move on to the instanton part. The instanton part is given by removing the perturbative contribution from the full partition function 
\be
Z_{\text{top, inst}}^{G_2} =Z_{\text{top}}^{G_2}/Z_{\text{pert}}^{G_2}. \label{g2inst}
\ee
It is possible to compare \eqref{g2inst} with the result in the literature. For the one-instanton part, the universal formula \eqref{uni-one-inst} gives 
\be
Z^{G_2}_{\text{1-inst}}=\frac{2q}{(1-q)^2}\frac{Q_1^3Q_2^4(1+Q_2+Q_1Q_2+3Q_1Q_2^2+Q_1Q_2^3+Q_1^2Q_2^3+Q_1^2Q_2^4)}{(1-Q_1)^2(1-Q_1Q_2^3)^2(1-Q_1^2Q_2^3)^2},\label{G_2-1ins}
\ee
where we can see that the denominator is completely determined by the set of positive long roots, $\{\alpha_1,\alpha_1+3\alpha_2,2\alpha_1+3\alpha_2\}$. The two-instanton part may be computed by using the blow up formula in \cite{Keller:2012da}. Then we found a perfect match between \eqref{g2inst} and the two-instanton part obtained from the blow up formula as well as the one-instanton part \eqref{G_2-1ins} until the order of $Q_1^5Q_2^5$.

In the diagram in \eqref{o-vert-so7-spinor}, a $(2, 1)$ 5-brane was introduced on the right side. It is also possible to realize spinor matter by having a $(1, 1)$ 5-brane on the left side of the diagram. When the diagram is reflected along a vertical axis, the following diagram 
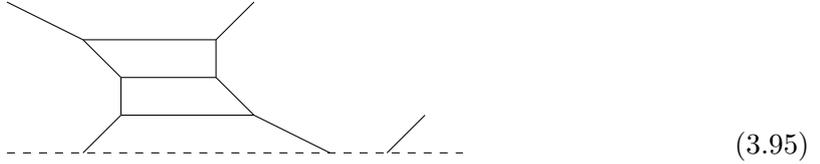
\begin{align}
\begin{tikzpicture}
\draw [dashed] (-1,0)--(5,0);
\draw (0,0)--(0.5,0.5);
\draw (0.5,0.5)--(0.5,1);
\draw (0.5,1)--(0,1.5);
\draw (0,1.5)--(-1,2);
\draw (3.25,0)--(2.25,0.5);
\draw (0.5, 0.5)--(2.25,0.5);
\draw (2.25,0.5)--(1.75,1);
\draw (1.75,1)--(0.5,1);
\draw (1.75,1)--(1.75,1.5);
\draw (1.75,1.5)--(0,1.5);
\draw (1.75,1.5)--(2.25,2);
\draw (4,0)--(4.5,0.5);
\end{tikzpicture}
\label{o-vert-so7-spinor1}
\end{align}
also realizes the SO(7) gauge theory with a hypermultiplet in the spinor representation. One can perform the Higgsing to the diagram, which gives a diagram of the pure $G_2$ gauge theory \cite{Hayashi:2018lyv}
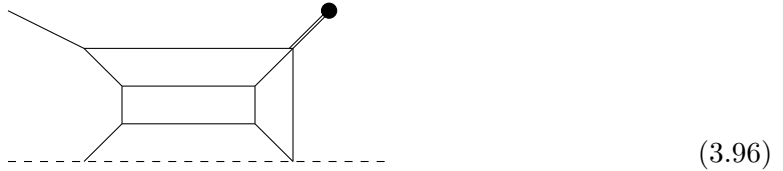
\begin{align}
\begin{tikzpicture}
\draw [dashed] (-1,0)--(4,0);
\draw (0,0)--(0.5,0.5);
\draw (0.5,0.5)--(0.5,1);
\draw (0.5,1)--(0,1.5);
\draw (0,1.5)--(-1,2);
\draw (2.75,0)--(2.25,0.5);
\draw (0.5, 0.5)--(2.25,0.5);
\draw (2.25,0.5)--(2.25,1);
\draw (2.25,1)--(0.5,1);
\draw (2.25,1)--(3.25,2);
\draw (2.75,0)--(2.75,1.5);
\draw (0,1.5)--(2.75,1.5);
\draw (2.75-0.05,1.5)--(3.25-0.05,2);
\draw [black, fill] (3.25-0.025,2) circle (0.1);
\end{tikzpicture}
\label{fig:g2diag3}
\end{align}

It is also possible to apply the topological vertex and the O-vertex to the diagram in \eqref{fig:g2diag3}. For that we use the following diagram,
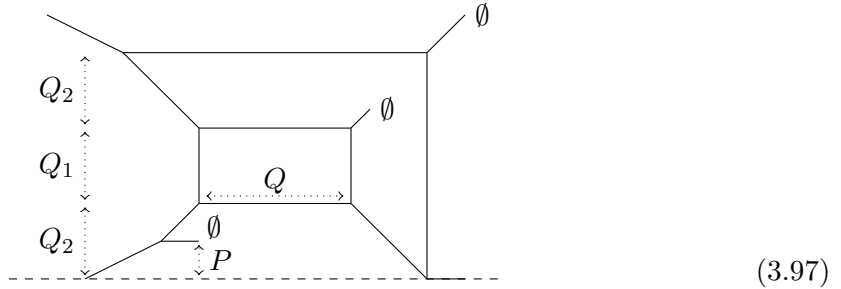
\begin{align}
\begin{tikzpicture}
\draw[dashed] (-1,0)--(5.5,0);
\draw (0,0)--(1,0.5);
\draw (1,0.5)--(1.5,0.5);
\draw (1,0.5)--(1.5,1);
\draw (1.5,1)--(1.5,2);
\draw (1.5,2)--(0.5,3);
\draw (0.5,3)--(-0.5,3.5);
\draw (1.5,1)--(3.5,1);
\draw[<->,dotted] (1.5+0.1,1+0.1)--(3.5-0.1,1+0.1);
\node at (2.5,1) [above] {$Q$};
\draw (1.5,2)--(3.5,2);
\draw (3.5,1)--(3.5,2);
\draw (3.5,2)--(3.75,2.25);
\draw (3.5,1)--(4,0.5);
\draw (4.5,0)--(5,0);
\draw (4,0.5)--(4.5,0);
\draw (4.5,0)--(4.5,3);
\draw (0.5,3)--(4.5,3);
\draw (4.5,3)--(5,3.5);
\node at (3.75,2.25) [right] {$\emptyset$};
\node at (1.7,0.4) [above] {$\emptyset$};
\draw[<->,dotted] (0,0.05)--(0,0.95);
\node at (0,0.5) [left] {$Q_2$};
\draw[<->,dotted] (0,1.05)--(0,1.95);
\node at (0,1.5) [left] {$Q_1$};
\draw[<->,dotted] (0,2.05)--(0,2.95);
\node at (0,2.5) [left] {$Q_2$};
\draw[<->,dotted] (1.5,0+0.05)--(1.5,0.5-0.05);
\node at (1.5,0.25) [right] {$P$};
\node at (5,3.5) [right] {$\emptyset$};
\end{tikzpicture}
\label{dia:G_2-alt}
\end{align}
We need to take the limit $P \to 1$ at the end of the computation. The parameterization of the Coulomb branch moduli is the same as \eqref{g2CB}. The instanton fugacity is also given by \eqref{g2instf}. Before taking the limit $P \to 1$, the computational result by the topological vertex applied to the diagram \eqref{dia:G_2-alt} gives rise to 
\begin{equation}
\begin{split}
\tilde{Z}'^{G_2}_{\text{top}}=\sum_{\nu,\lambda,\sigma,\tau,\alpha,\beta,\gamma,\delta,\mu,\eta}(-Q)^{|\lambda|+|\sigma|}(-P)^{|\nu|}(-Q_2/P)^{|\alpha|}(-Q_1)^{|\beta|+|\delta|}(-Q_2)^{|\gamma|+|\mu|}f_\alpha^{-1}f_\beta^{-1}f_\gamma^{-1}\cr
\quad f_\delta f_\sigma f_\lambda^{-1}f^{-3}_\tau
(-Q_1Q_2^2)^{|\eta|}(QQ_2^2)^{|\tau|}V_\nu C_{\nu^t\alpha\emptyset}C_{\alpha^t\beta\sigma}C_{\beta^t\gamma\lambda}C_{\gamma^t\emptyset\tau} C_{\emptyset\delta^t\lambda^t}C_{\delta\mu^t\sigma^t}C_{\eta^t\mu\emptyset}C_{\eta\emptyset\tau},
\end{split}
\end{equation}
Since we take $P \to 1$ at the end, it seems that we need to sum over $\nu$ to obtain a result at each order of $Q^aQ_1^bQ_2^c$. However, as observed in section \ref{s:so-2N+1}, it is enough to sum over $\nu$ until the order $|\nu| = 2|\alpha|$ for a fixed $|\alpha|$. Namely the partition function can be computed by 
\begin{equation}\label{g2top2}
\begin{split}
Z'^{G_2}_{\text{top}}&=\lim_{P\to 1}\tilde{Z}_{\text{top}}^{G_2}\cr
&=\sum_{\substack{\nu,\lambda,\sigma,\tau,\alpha,\beta,\gamma,\delta,\mu,\eta\\|\nu| \leq 2|\alpha|}}(-Q)^{|\lambda|+|\sigma|}(-1)^{|\nu|}(-Q_2)^{|\alpha|}(-Q_1)^{|\beta|+|\delta|}(-Q_2)^{|\gamma|+|\mu|}f_\alpha^{-1}f_\beta^{-1}f_\gamma^{-1}\cr
&\qquad f_\delta f_\sigma f_\lambda^{-1}f^{-3}_\tau
(-Q_1Q_2^2)^{|\eta|}(QQ_2^2)^{|\tau|}V_\nu C_{\nu^t\alpha\emptyset}C_{\alpha^t\beta\sigma}C_{\beta^t\gamma\lambda}C_{\gamma^t\emptyset\tau} C_{\emptyset\delta^t\lambda^t}C_{\delta\mu^t\sigma^t}C_{\eta^t\mu\emptyset}C_{\eta\emptyset\tau}.
\end{split}
\end{equation}

The perturbative part is obtained by applying the limit $Q \to 0$ to \eqref{g2top2}. The diagram splits into the left part and the right part. The left half of the diagram before the limit $P \to 1$ is the left part of the pure SO(8) diagram whereas the right part diagram is the half of the diagram of an SU(3) gauge theory with a flavor. The partition function from the left part of the pure SO(8) diagram is given in \eqref{so8left1} and it becomes in this case 
\begin{equation}\label{ZprimeG2left}
\begin{split}
\tilde{Z}'^{G_2}_{\text{left}} &= P.E.\left(\frac{q}{(1-q)^2}\left(Q_2/P + Q_1 + Q_2 + Q_1Q_2/P + Q_1Q_2 + Q_1Q_2^2/P + Q_2P + Q_1Q_2P\right.\right.\cr
&\hspace{7cm}\left.+Q_1Q_2^2+ Q_1Q_2^2P + Q_1Q_2^3P + Q_1^2Q_2^3P\right)\Big), 
\end{split}
\end{equation}
where we identified $Q' = Q_2/P, R = Q_1, T= Q_2$ and we assumed that \eqref{so8left1} gives the exact result. 
Then $P \to 1$ limit gives 
\begin{equation}
\begin{split}
Z'^{G_2}_{\text{left}}&= P.E.\left(\frac{q}{(1-q)^2}\left(Q_1 + 3Q_2 + 3Q_1Q_2 + 3 Q_1Q_2^2 + Q_1Q_2^3 + Q_1^2Q_2^3\right)\right) .\label{g2left2}
\end{split}
\end{equation}
On other hand the square root of the perturbative part of the partition function of an SU(3) gauge theory with a flavor is given by
\begin{equation}
Z^{SU(3)}_{\text{pert}} = P.E.\left(\frac{q}{(1-q)^2}\left(A'_1A_2'^{-1} - A'_2M'^{-1} - M'A_3'^{-1} - A'_1M'^{-1} + A'_2A_3'^{-1} + A'_1A_3'^{-1}\right)\right),
\end{equation}
where $A'_1, A'_2, A'_3\; (A'_1A'_2A'_3=1)$ are K$\ddot{\text{a}}$hler parameters for the Coulomb branch moduli of the SU(3) and $M'$ is the mass parameter for the flavor. Identifying $A'_1A_2'^{-1} = Q_1, A'_2M^{-1} = Q_2, MA_3'^{-1} = Q_1Q_2^2$ yields
\begin{equation}
Z^{SU(3)}_{\text{pert}} = P.E.\left(\frac{q}{(1-q)^2}\left(Q_1 - Q_2 -Q_1Q_2^2 -Q_1Q_2 +Q_1Q_2^3 + Q_1^2Q_2^3\right)\right).\label{g2right2}
\end{equation}
Combining \eqref{g2left2} with \eqref{g2right2}, we obtain
\begin{equation}
\begin{split}
Z'^{G_2}_{\text{pert}}= P.E.\lt(\frac{2q}{(1-q)^2}\lt(Q_1+Q_2+Q_1Q_2+Q_1Q_2^2+Q_1Q_2^3+Q_1^2Q_2^3\rt)\rt),\label{g2pert2}
\end{split}
\end{equation}
which reproduces the perturbative part of the pure $G_2$ partition function \eqref{G2-pert0}.

The instanton part of the pure $G_2$ partition function is given by removing the perturbative part \eqref{g2pert2} from \eqref{g2top2},
\begin{align}
Z'^{G_2}_{\text{top, inst}} = Z'^{G_2}_{\text{top}}/Z'^{G_2}_{\text{pert}}. \label{g2inst-a}
\end{align}
We checked that both the one-instanton partition function and the two-instanton partition function extracted out from the above expression agree with the result \eqref{G_2-1ins} and the blow up result until the order $Q_1^2Q_2^2$.

\subsection{Pure SU(3) gauge theory with the Chern-Simons level $9$}\label{s:su3-9}
Lastly we compute the partition function of the 5d pure SU(3) gauge theory with the Chern-Simons level $9$ found in \cite{Jefferson:2017ahm, Jefferson:2018irk}. This theory may be obtained by a circle compactification with a twist of the 6d pure SU(3) gauge theory with a tensor multiplet \cite{Razamat:2018gro}. A 5-brane web diagram for the theory has been also found in \cite{Hayashi:2018lyv} and it is given by 
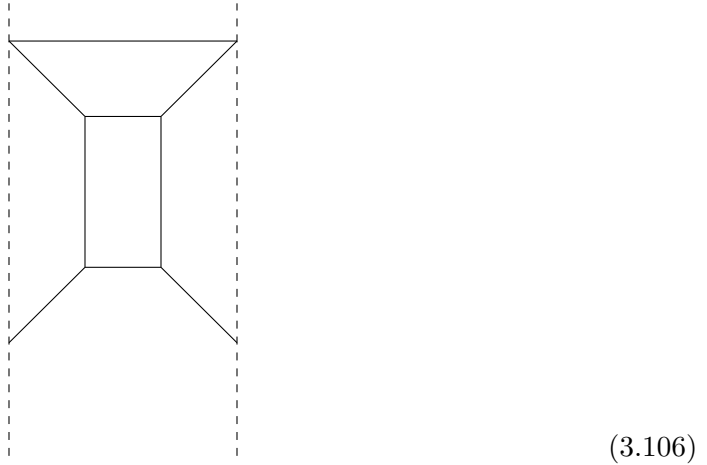
\begin{align}
\begin{tikzpicture}[rotate=90, transform shape]
\draw[dashed] (-1,0)--(5,0);
\draw[dashed] (-1,3)--(5,3);
\draw (0.5,0)--(1.5,1);
\draw (1.5,1)--(1.5,2);
\draw (1.5,2)--(0.5,3);
\draw (1.5,1)--(3.5,1);
\draw (1.5,2)--(3.5,2);
\draw (3.5,1)--(3.5,2);
\draw (3.5,2)--(4.5,3);
\draw (3.5,1)--(4.5,0);
\draw (4.5,0)--(4.5,3);
\end{tikzpicture}
\label{dia:SU(3)9_0}
\end{align}
where we have an $\widetilde{\text{ON}}$-plane on the left side and another $\widetilde{\text{ON}}$-plane on the right side. The three color D5-branes imply the presence of the SU(3) gauge group. The S-duality of type IIB string theory exchanges a D5-brane with an NS5-brane and it amounts to rotating a brane web diagram by 90 degrees. Hence the S-dual version of the diagram \eqref{dia:SU(3)9_0} becomes 
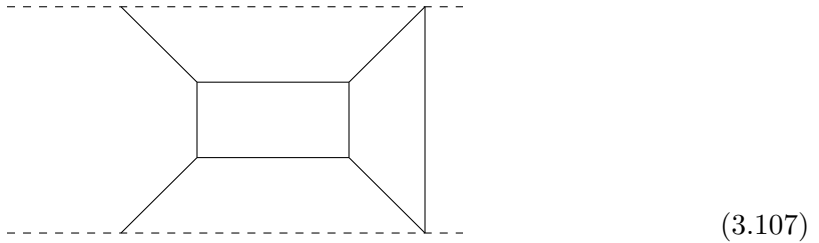
\begin{align}
\begin{tikzpicture}
\draw[dashed] (-1,0)--(5,0);
\draw[dashed] (-1,3)--(5,3);
\draw (0.5,0)--(1.5,1);
\draw (1.5,1)--(1.5,2);
\draw (1.5,2)--(0.5,3);
\draw (1.5,1)--(3.5,1);
\draw (1.5,2)--(3.5,2);
\draw (3.5,1)--(3.5,2);
\draw (3.5,2)--(4.5,3);
\draw (3.5,1)--(4.5,0);
\draw (4.5,0)--(4.5,3);
\end{tikzpicture}
\label{dia:SU(3)9_1}
\end{align}
In \eqref{dia:SU(3)9_1} the dotted lines represent $\widetilde{\text{O5}}$-planes which is the S-dual of the $\widetilde{\text{ON}}$-planes in \eqref{dia:SU(3)9_0}. Here the vertical direction becomes periodic and it implies that this theory is a marginal theory whose UV completion is related to a 6d theory. 

Using the O-vertex, it is also possible to compute the partition function of the pure SU(3) gauge theory with the CS level $9$ from the diagram \eqref{dia:SU(3)9_1}. For that, first note that the upper half or the lower half of the diagram \eqref{dia:SU(3)9_1} is the same as the lower half of the diagram \eqref{fig:g2diag3} for the pure $G_2$ gauge theory. Hence we can compute the partition function in the same way as we have done in the latter part of section \ref{s:G2}. Namely for applying the topological vertex and O-vertex we use the following diagram,
\begin{align}
\begin{tikzpicture}
\draw[dashed] (-1,0)--(6,0);
\draw[dashed] (-1,3)--(6,3);
\draw (0,0)--(1,0.5);
\draw (1,0.5)--(1.5,0.5);
\draw (1,0.5)--(1.5,1);
\draw (1.5,1)--(1.5,2);
\draw (1.5,2)--(1,2.5);
\draw (1,2.5)--(0,3);
\draw (1,2.5)--(1.5,2.5);
\draw (1.5,1)--(3.5,1);
\draw (1.5,2)--(3.5,2);
\draw (3.5,1)--(3.5,2);
\draw (3.5,2)--(4.5,3);
\draw (3.5,1)--(4.5,0);
\draw (4.5,0)--(5,0);
\draw (4.5,3)--(5,3);
\draw (4.5,0)--(4.5,3);
\node at (1.5,0.5) [right] {$\emptyset$};
\node at (1.5,2.5) [right] {$\emptyset$};
\draw[<->,dotted] (0,0.05)--(0,0.95);
\node at (0,0.5) [left] {$Q_2$};
\draw[<->,dotted] (0,1.05)--(0,1.95);
\node at (0,1.5) [left] {$Q$};
\draw[<->,dotted] (0,2.05)--(0,2.95);
\node at (0,2.5) [left] {$Q_2$};
\draw[<->,dotted] (1.5+0.1,1.1)--(3.5-0.1,1.1);
\node at (2.5,1.1) [above] {$Q_1$};
\draw[<->,dotted] (2,0.05)--(2,0.5);
\node at (2,0.3) [right] {$P$};
\draw[<->,dotted] (2,3-0.05)--(2,2.5);
\node at (2,2.7) [right] {$P$};
\end{tikzpicture}
\label{dia:SU(3)-9}
\end{align}
where we need again to take the $P\rightarrow 1$ limit at the end. $Q_1$ and $Q_2$ are related to the Coulomb branch moduli $A_1, A_2, A_3\; (A_1A_2A_3 = 1)$ of SU(3) by 
\begin{equation}\label{su3CB}
Q_2 = A_1A_2^{-1}, \qquad Q_1 = A_2A_3^{-1},
\end{equation}
and the length between the $\widetilde{\text{O5}}$-planes is the instanton fugacity of the SU(3) gauge theory and it is 
\be\label{su3instf}
\mathfrak{q} = QQ_2^2.
\ee

In this case, the diagram also involves the vertex of 
\begin{align}
\begin{tikzpicture}[yscale=-1]
\draw [dashed] (-2,0)--(2,0);
\draw (0,0)--(1,0.5);
\draw[->] (0,0)--(0.75,0.375);
\node at (1,0.5) [below] {$\nu$};
\node at (1,0) [above] {$O5^-$};
\end{tikzpicture}
\label{o-vert-upsidedown}
\end{align}
which has not appeared before. We can also determine the function associated to the vertex \eqref{o-vert-upsidedown} by redoing the argument in section \ref{s:prop-O-vert}. 
For that we consider
\begin{align}\label{fig:overtex3}
\begin{tikzpicture}[yscale=-1]
\draw [dashed] (-2,0)--(2,0);
\draw (0,0)--(0.5,0.5);
\draw (0,0)--(-0.5,0.5);
\draw (-0.5,0.5)--(0.5,0.5);
\draw (0.5,0.5)--(1.5,1);
\draw[->] (0.5,0.5)--(1.25, 0.875);
\draw (-0.5,0.5)--(-1.5,1);
\node at (1.5,1) [below] {$\nu$};
\draw [dotted,<->] (-0.5,0.6)--(0.5,0.6);
\node at (0,0.6) [below] {$Q^2$};
\draw [dotted,<->] (1.5+0.05,0.5-0.05)--(1.5+0.05,1-0.05);
\node at (1.5,0.6) [right] {$P$};
\end{tikzpicture}
\end{align}
Taking the limit $Q \to \infty$ with $PQ$ fixed for the diagram \eqref{fig:overtex3} reproduces the diagram \eqref{o-vert-upsidedown}. The partition function for the diagram \eqref{fig:overtex3} can be computed by the method in \cite{Kim-Yagi}, which was reviewed in section \ref{s:r-O-top}, and it gives
\begin{align}\label{overtex3}
\tilde{Z}_\nu(P,Q)=\sum_{\mu,\lambda}(-Q^2)^{|\mu|}Q^{2|\lambda|}f_\lambda^{-4}(-P)^{|\nu|}C_{\mu\nu\lambda}C_{\mu^t\emptyset\lambda}.
\end{align}
One can see that \eqref{overtex3} is exactly the same as \eqref{OvertexW} and the limit $Q\to \infty$ with $PQ$ fixed yields the O-vertex $W_{\nu}$. Therefore we can use $W_{\nu}$ for \eqref{o-vert-upsidedown}.  
Then by applying the topological vertex and the O-vertex, we find  that the partition function of the diagram \eqref{dia:SU(3)-9} 
is given by
\begin{equation}\label{su3top}
\begin{split}
Z^{SU(3)_9}_{\text{top}}&=\lim_{P\to 1}\sum_{\substack{\lambda,\sigma,\nu,\mu,\alpha,\beta,\gamma,\delta,\eta,\xi,\zeta\\|\nu|\leq 2|\alpha|, |\mu|\leq 2|\gamma|}}(-Q_1)^{|\lambda|+|\sigma|}(-P)^{|\nu|+|\mu|}(-Q_2/P)^{|\alpha|+|\gamma|}(-Q)^{|\beta|+|\delta|}f_\sigma f_\lambda^{-1}f_\alpha^{-1}f_\beta^{-1}f_{\gamma}^{-1}\cr
&\quad f_\delta  f_\zeta^{-1}(-Q_2)^{|\eta|+|\xi|}(-QQ_2^2)^{|\zeta|}V_\nu W_{\mu^t}C_{\nu^t\alpha\emptyset}C_{\alpha^t\beta\sigma}C_{\beta^t\gamma\lambda}C_{\gamma^t\mu\emptyset}C_{\eta\delta^t\lambda^t}C_{\delta\xi^t\sigma^t}C_{\eta^t\zeta\emptyset}C_{\zeta^t\xi\emptyset},
\end{split}
\end{equation}
where the summation of $\nu$ and $\mu$ is restricted since it gives the exact result for a fixed order of $|\alpha|$ and $|\gamma|$ as argued in section \ref{s:so-2N+1}.

One can extract out the perturbative part in the sense of SU(3) by setting $Q\rightarrow 0$ for \eqref{su3top}, which forces $\beta$, $\delta$ and $\zeta$ in the above summation to be $\emptyset$. Then the diagram \eqref{dia:SU(3)9_1} splits into the upper part and the lower part, and the two diagrams are related to each other by the reflection with respect to a horizontal axis. Hence the partition function from the upper part will be the same as that from the lower part, and the perturbative part of the partition function is given by  
\begin{align}
Z^{SU(3)_9}_{\text{pert}}=\lt(\sum_{\substack{\nu,\alpha,\sigma,\xi\\|\nu| \leq 2|\alpha|}}(-Q_1)^{|\sigma|}(-1)^{|\nu|}(-Q_2)^{|\alpha|}(-Q_2)^{|\xi|}f_\sigma f_\alpha^{-1} V_\nu C_{\nu^t\alpha\emptyset}C_{\alpha^t\emptyset\sigma}C_{\emptyset\xi^t\sigma^t}C_{\emptyset\xi\emptyset}\rt)^2.\nn\\
\end{align}
We checked until the order of $Q_1^2Q_2^2$ that the above expression agrees with 
\begin{align}
Z^{SU(3)_9}_{\text{pert}}&=P.E.\lt(\frac{2q}{(1-q)^2}(Q_1+Q_2+Q_1Q_2)\rt)\cr
&=P.E.\lt(\frac{2q}{(1-q)^2}(A_1A_2^{-1}+A_2A_3^{-1}+A_1A_3^{-1})\rt).
\end{align}
This is the standard perturbative part of the partition function of a pure SU(3) gauge theory. 

Let us then evaluate the full partition function, 
There are several sub-diagrams in this theory that can be relatively easily computed. The first one is the following half diagram, 
\begin{align}
\begin{tikzpicture}
\draw[dashed] (-1,0)--(2.5,0);
\draw[dashed] (-1,3)--(2.5,3);
\draw (0,0)--(1,0.5);
\draw (1,0.5)--(1.5,0.5);
\draw (1,0.5)--(1.5,1);
\draw (1.5,1)--(1.5,2);
\draw (1.5,2)--(1,2.5);
\draw (1,2.5)--(0,3);
\draw (1,2.5)--(1.5,2.5);
\draw (1.5,1)--(2.5,1);
\draw (1.5,2)--(2.5,2);
\node at (1.5,0.5) [right] {$\emptyset$};
\node at (1.5,2.5) [right] {$\emptyset$};
\draw[<->,dotted] (0,0.05)--(0,0.95);
\node at (0,0.5) [left] {$Q_2$};
\draw[<->,dotted] (0,1.05)--(0,1.95);
\node at (0,1.5) [left] {$Q$};
\draw[<->,dotted] (0,2.05)--(0,2.95);
\node at (0,2.5) [left] {$Q_2$};
\end{tikzpicture}
\label{dia:SU(3)-9-left}
\end{align}
From the periodicity in the vertical direction, one would expect the exact partition function of the above diagram to be given by an infinite product of plethystic exponentials. For example we checked with our explicit computation that the partition function from the diagram \eqref{dia:SU(3)-9-left} gives 
\begin{align}
&Z_{\text{left}}^{SU(3)_9}(Q_2,Q)\cr
&=
\sum_{\substack{\nu,\mu,\alpha,\beta,\gamma\\|\nu| \leq 2|\alpha|, |\mu| \leq 2|\gamma|}}(-1)^{|\nu|+|\mu|}(-Q_2)^{|\alpha|+|\gamma|}(-Q)^{|\beta|} f_\alpha^{-1}f_\beta^{-1}f_{\gamma}^{-1}V_\nu W_{\mu^t}C_{\nu^t\alpha\emptyset}C_{\alpha^t\beta\emptyset}C_{\beta^t\gamma\emptyset}C_{\gamma^t\mu\emptyset}\cr
&=P.E.\lt(\frac{q}{(1-q)^2}(Q+4Q_2+4QQ_2+6QQ_2^2)\rt)+o(Q^2,Q_2^2).\label{su3left}
\end{align}
We note that the above expression can alternatively be reproduced from the $P\rightarrow 1$ limit of the positive root system of affine $D_4$ Lie algebra. In general the positive roots of an affine Lie algebra $\hat{\mathfrak{g}}$ associated to a Lie algebra $\mathfrak{g}$ is given by
\begin{align}
\hat{\Delta}_+ = \left\{\alpha + n\delta |n > 0, \alpha \in \Delta \right\} \cup \left\{\alpha |\alpha \in \Delta_+\right\},
\end{align}
where $\Delta$ is the set of the roots of $\mathfrak{g}$, $\Delta_+$ is the set of the positive roots of $\mathfrak{g}$ and $\delta$ is the null (or imaginary) root. In this case, the K$\ddot{\text{a}}$hler parameters for the simple roots of SO(8) are $Q_2P^{-1}, Q, Q_2P^{-1}, Q_2P$ and the K$\ddot{\text{a}}$hler parameter for the affine node is $Q_2P$.  From this we can see that the K$\ddot{\text{a}}$hler parameter for the null root becomes $Q^2Q_2^4$. Then, until the order $Q^2Q_2^2$, the positive roots with $n=0$ gives 
\be
P.E.\left(\frac{q}{(1-q)^2}(Q + 3Q_2 + 3QQ_2 + 3QQ_2^2)\right),
\ee
while the positives roots with $n=1$ gives
\be
P.E.\left(\frac{q}{(1-q)^2}(Q_2 + QQ_2 + 3QQ_2^2)\right),
\ee
after taking the $P \to 1$ limit, and their product indeed reproduces \eqref{su3left}. 

To have more insight let us consider the following diagram, 
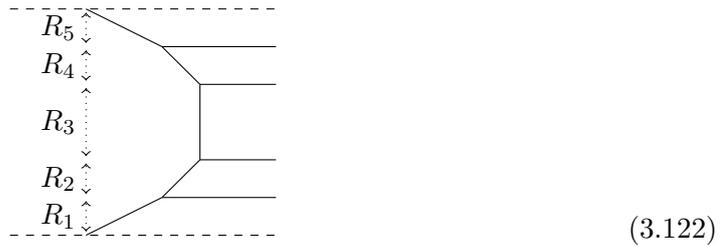
\begin{align}
\begin{tikzpicture}
\draw[dashed] (-1,0)--(2.5,0);
\draw[dashed] (-1,3)--(2.5,3);
\draw (0,0)--(1,0.5);
\draw (1,0.5)--(2.5,0.5);
\draw (1,0.5)--(1.5,1);
\draw (1.5,1)--(1.5,2);
\draw (1.5,2)--(1,2.5);
\draw (1,2.5)--(0,3);
\draw (1,2.5)--(2.5,2.5);
\draw (1.5,1)--(2.5,1);
\draw (1.5,2)--(2.5,2);
\draw[<->,dotted] (0,0.05)--(0,0.45);
\draw[<->,dotted] (0,0.55)--(0,0.95);
\node at (0,0.25) [left] {$R_1$};
\node at (0,0.75) [left] {$R_2$};
\draw[<->,dotted] (0,1.05)--(0,1.95);
\node at (0,1.5) [left] {$R_3$};
\draw[<->,dotted] (0,2.05)--(0,2.45);
\draw[<->,dotted] (0,2.55)--(0,2.95);
\node at (0,2.25) [left] {$R_4$};
\node at (0,2.75) [left] {$R_5$};
\end{tikzpicture}
\label{dia:affined4}
\end{align}
where there is no restriction on the K$\ddot{\text{a}}$hler parameters $R_i\;(i=1, \cdots, 5)$. 
The application of the topological vertex and the O-vertex to the diagram \eqref{dia:affined4} gives
\begin{align}
&\sum_{\nu,\mu,\alpha,\beta,\gamma}(-R_1)^{|\nu|}(-R_2)^{|\alpha|}(-R_3)^{|\beta|}(-R_4)^{|\gamma|}(-R_5)^{|\nu|} f_\alpha^{-1}f_\beta^{-1}f_{\gamma}^{-1}V_\nu W_{\mu^t}C_{\nu^t\alpha\emptyset}C_{\alpha^t\beta\emptyset}C_{\beta^t\gamma\emptyset}C_{\gamma^t\mu\emptyset}\cr
&=P.E.\left(\frac{q}{(1-q)^2}\left(R_2 + R_1^2 R_2 + R_3 + R_2 R_3 + R_1^2 R_2 R_3 + R_1^2 R_2^2 R_3 + R_4 + R_3 R_4\right.\right.\cr
&\hspace{3cm} + R_2 R_3 R_4 + R_1^2 R_2 R_3R_4 + R_1^2 R_2^2 R_3 R_4 + R_1^2 R_2^2 R_3^2 R_4 + R_4 R_5^2 + R_3 R_4 R_5^2\cr
&\hspace{3cm}+ R_2 R_3 R_4 R_5^2 + R_1^2 R_2 R_3 R_4 R_5^2 + R_1^2 R_2^2 R_3 R_4 R_5^2 + R_1^2 R_2^2 R_3^2 R_4 R_5^2\cr
&\hspace{3cm}+ R_3 R_4^2 R_5^2 +R_2 R_3 R_4^2 R_5^2 + R_1^2 R_2 R_3 R_4^2 R_5^2 + R_1^2 R_2^2 R_3 R_4^2 R_5^2 + R_2 R_3^2 R_4^2 R_5^2\cr
&\hspace{3cm}\left.+ R_1^2 R_2 R_3^2 R_4^2 R_5^2 + 6R_1^2R_2^2R_3^2R_4^2R_5^2\right) + R_1^2R_2^2R_3^2R_4^2R_5^2\Big)\cr
&\hspace{11cm} + o(R_1^2,R_2^2,R_3^2,R_4^2,R_5^2).\label{affined4full}
\end{align}
When we associate $R_4, R_3, R_2, R_2R_1^2$ for the simple roots of SO(8) and $R_4R_5^2$ for the affine node of the Dynkin diagram of the affine $D_4$ Lie algebra, we can see that \eqref{affined4full} reproduces the contribution of the positive roots of the affine $D_4$ Lie algebra for $n=0, 1$ except for the terms proportional to $\prod_{i=1}^5R_i^2$ which is the K$\ddot{\text{a}}$hler parameter corresponding to the imaginary root. We expect that the property that the contribution is associated to the positive roots of the affine $D_4$ Lie algebra holds for higher orders except for terms associated to $\left(\prod_{i=1}^5R_i^2\right)^k$ or $\left(Q^2Q_2^4\right)^k$ for $k=1, 2, \cdots$ when we take the limit to the diagram \eqref{dia:SU(3)-9-left}. 

The presence of the affine $D_4$ structure is in fact expected from the two orientifolds.  When the two orientifolds in \eqref{dia:SU(3)-9-left} are decoupled, then the diagram gives a half of the diagram for a pure SU(4) gauge theory. When we introduce one orientifold, then the diagram is a half of the diagram of the pure SO(8) gauge theory and the change of the Dynkin diagram from SU(4) to SO(8) is given by
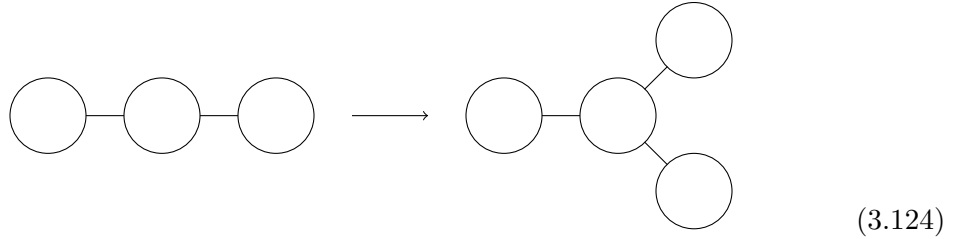
\begin{align}
\begin{tikzpicture}
\draw (0,0) circle [radius=0.5];
\draw (0.5,0)--(1,0);
\draw (1.5,0) circle [radius=0.5];
\draw (2,0)--(2.5,0);
\draw (3,0) circle [radius=0.5];
\draw[->] (4,0)--(5,0);
\draw (6,0) circle [radius=0.5];
\draw (6.5,0)--(7,0);
\draw (7.5,0) circle [radius=0.5];
\draw (8.5,1) circle [radius=0.5];
\draw (7.5+0.35, 0.35)--(8.5-0.35,1-0.35);
\draw (8.5,-1) circle [radius=0.5];
\draw (7.5+0.35,-0.35)--(8.5-0.35,-1+0.35);
\end{tikzpicture}
\end{align}
Hence having another orientifold further changes the Dynkin diagram \cite{Hanany:2001iy},
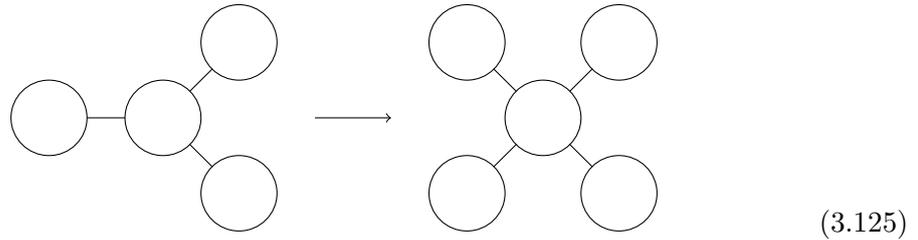
\begin{align}
\begin{tikzpicture}
\draw (6,0) circle [radius=0.5];
\draw (6.5,0)--(7,0);
\draw (7.5,0) circle [radius=0.5];
\draw (8.5,1) circle [radius=0.5];
\draw (7.5+0.35, 0.35)--(8.5-0.35,1-0.35);
\draw (8.5,-1) circle [radius=0.5];
\draw (7.5+0.35,-0.35)--(8.5-0.35,-1+0.35);
\draw[->] (9.5,0)--(10.5,0);
\draw (11.5,1) circle [radius=0.5];
\draw (11.5+0.35,1-0.35)--(12.5-0.35,0.35);
\draw (11.5,-1) circle [radius=0.5];
\draw (11.5+0.35,-1+0.35)--(12.5-0.35,-0.35);
\draw (12.5,0) circle [radius=0.5];
\draw (13.5,1) circle [radius=0.5];
\draw (13.5 -0.35,1-0.35)--(12.5+0.35,0.35);
\draw (13.5,-1) circle [radius=0.5];
\draw (13.5-0.35,-1+0.35)--(12.5+0.35,-0.35);
\end{tikzpicture}
\end{align}
which is nothing but the Dynkin diagram of the affine $D_4$ Lie algebra. 

As for the second sub-web let us consider the right half of the diagram \eqref{dia:SU(3)-9}, which is given by
\begin{align}
\begin{tikzpicture}
\draw[dashed] (2,0)--(5.5,0);
\draw[dashed] (2,3)--(5.5,3);
\draw (2.5,1)--(3.5,1);
\draw (2.5,2)--(3.5,2);
\draw (3.5,1)--(3.5,2);
\draw (3.5,2)--(4.5,3);
\draw (4.5,3)--(5,3);
\draw (3.5,1)--(4.5,0);
\draw (4.5,0)--(5,0);
\draw (4.5,0)--(4.5,3);
\draw[<->,dotted] (5,0.05)--(5,0.95);
\node at (5,0.5) [right] {$Q_2$};
\draw[<->,dotted] (5,2.05)--(5,2.95);
\node at (5,2.5) [right] {$Q_2$};
\draw[<->,dotted] (5,1.05)--(5,1.95);
\node at (5,1.5) [right] {$Q$};
\end{tikzpicture}
\label{dia:SU(3)-9-right}
\end{align}
Taking into account the mirror images, this diagram is like an infinitely long strip diagram. 
The partition function for the diagram \eqref{dia:SU(3)-9-right} becomes
\begin{align}\label{su3right}
Z_{\text{right}}^{SU(3)_9}(Q_2,Q)&=\sum_{\delta,\eta,\xi,\zeta}(-Q)^{|\delta|}(-Q_2)^{|\eta|+|\xi|}(-QQ_2^2)^{|\zeta|}f_\delta  f_\zeta^{-1}C_{\eta\delta^t\emptyset}C_{\delta\xi^t\emptyset}C_{\eta^t\zeta\emptyset}C_{\zeta^t\xi\emptyset}
\end{align}
Here the Young diagram $\zeta$ is assigned to the line which connects the diagram back to the original one through the orientifold. 
The partition function of the form \eqref{su3right} has appeared in \cite{Haghighat:2013gba, Haghighat:2013tka, Hohenegger:2013ala} and it is given by
\begin{equation}
\begin{split}
&Z_{\text{right}}^{SU(3)_9}(Q_2,Q)\cr
=&P.E.\left(\left[\frac{q}{(1-q)^2}\left(Q-2Q_2 - 2QQ_2 + 2QQ_2^2 - 2QQ_2^3 + QQ_2^4 - 2Q^2Q_2^3 + 4Q^2Q_2^4\right)\right.\right.\cr
&\hspace{10.5cm}\left.\left.+ Q^2Q_2^4\right]\frac{1}{1-Q^2Q_2^4}\right).
\end{split}
\end{equation}
The factor $\frac{1}{1-Q^2Q_2^4}$ accounts for strings wound around the periodic direction.


Lastly we compute the full partition function. Until some orders of the parameters the explicit form of the partition function is given by
\begin{align}
Z^{SU(3)_9}_{\text{top}}=&1+\frac{2q}{(1-q)^2}Q_2+\frac{q^2(3+2q+3q^2)}{(1-q)^2(1-q^2)^2}Q_2^2+\frac{2q}{(1-q)^2}Q_1\cr
&+\frac{2(q+q^3)}{(1-q)^4}Q_1Q_2
+\frac{2q^2(2+3q-2q^2+3q^3+2q^4)}{(1-q)^4(1-q^2)^2}Q_1Q_2^2\cr
&+\frac{q^2(3+2q+3q^2)}{(1-q)^2(1-q^2)^2}Q_1^2 +\frac{2q^2(2+3q-2q^2+3q^3+2q^4)}{(1-q)^4(1-q^2)^2}Q_1^2Q_2
\cr
&+\frac{q^2(3+4q+21q^2+8q^3-8q^5+21q^6+4q^7+3q^8)}{(1-q)^4(1-q^2)^4}Q_1^2Q_2^2\cr
&+\frac{2q}{(1-q)^2}Q+\frac{2(q+q^3)}{(1-q)^4}Q_2Q+\frac{2q(4-6q-q^2+14q^3-q^4-6q^5+4q^6)}{(1-q)^4(1-q^2)^2}Q_2^2Q\cr
&+\frac{4q(1-q+q^2)}{(1-q)^4}Q_1Q +\frac{2q(3-4q+6q^2-4q^3+3q^4)}{(1-q)^6}Q_1Q_2Q\cr
&+\frac{2q(3-2q+6q^3-2q^5+3q^6)}{(1-q)^4(1-q^2)^2}Q_1^2Q\cr
&+\frac{q^2(3+2q+3q^2)}{(1-q)^2(1-q^2)^2}Q^2+\frac{2q^2(2+3q-2q^2+3q^3+2q^4)}{(1-q)^4(1-q^2)^2}Q_2 Q^2\cr
&+\frac{2q(3-2q+6q^3-2q^5+3q^6)}{(1-q)^4(1-q^2)^2}Q_1Q^2+\cdots,
\end{align}
where $\cdots$ stands for higher order terms. 
Unfortunately, not much is known about the partition function of the pure SU(3) gauge theory with the CS level $9$. 
It has been argued in \cite{Razamat:2018gro} that the theory is obtained by a circle compactification with a twist of the 6d pure SU(3) gaue theory with a tensor multiplet. The elliptic genus of the 6d SU(3) strings has been also computed in \cite{Kim:2016foj,Gu:2018gmy}, but its relation with the partition function of the 5d pure SU(3) gauge theory with the CS level $9$ remains unclear at the moment. In particular, one can see from the brane diagram (\ref{dia:SU(3)-9}) that there are only three K\"ahler parameters in the 5d theory, while there are 4 parameters (including the string parameter that weights different number of strings) in the 6d case. We leave the comparison between these partition functions after a proper twisting to a future work.

We can also obtain the plethystic exponential form of \eqref{su3top} and it is given by 
\begin{equation}
\begin{split}
Z^{SU(3)_9}_{\text{top}}=&P.E.\left(\frac{2q}{(1-q)^2}\lt(Q_2+Q_1+Q_1Q_2+Q+Q_2Q+2Q_1Q+3Q_1Q_2Q\rt.\right.\cr
&\left.+4Q_2^2Q+Q_2^3Q+Q_2^4Q+3Q_1^2Q+3Q_1Q^2+16Q_1^2Q^2\right)+9Q_1^2Q^2 + \cdots \Big).\label{GV-su39}
\end{split}
\end{equation}
From the expression \eqref{GV-su39}, it is possible to extract the Gopakumar-Vafa (GV) invariants. 
The GV invariant $n^g_\beta$, which was originally introduced in the series of papers \cite{Gopakumar:1998ii,Gopakumar:1998jq,Gopakumar:1998ki,Gopakumar:1998vy}, is defined so that the partition can be expressed as 
\begin{align}
Z(g_s,\{Q\})&=\exp\lt(\sum_{g=0}^\infty \sum_{\beta \in H_2(X,\mathbb{Z})} \sum_{d=1}^\infty n^g_\beta \frac{1}{d}\lt(2\sin\frac{dg_s}{2}\rt)^{2g-2}Q^{d}_{\beta}\rt)\cr
&=P.E.\left(\sum_{g=0}^{\infty}\sum_{\beta \in H_2(X,\mathbb{Z})}n_{\beta}^g\frac{(-q)^{1-g}}{(1-q)^{2-2g}}Q_\beta\right) \label{GV}
\end{align}
where $q=:e^{ig_s}$, $X$ is the corresponding Calabi-Yau manifold and $Q_{\beta}$ is the K$\ddot{\text{a}}$hler parameter for a curve $\beta \in H_2(X, \mathbb{Z})$. These topological invariants are related to the number of BPS states in the underlining theory. Then comparing \eqref{GV} with \eqref{GV-su39}, one can obtain the GV invariants for various lines in the brane web.

The partition function \eqref{su3top} is the one we obtain when we apply the topological vertex and the O-vertex to the diagram \eqref{dia:SU(3)-9} or \eqref{dia:SU(3)9_1}. To obtain the Nekrasov partition function of the pure SU(3) gauge theory with the CS level $9$, we need to remove a factor which does not depend on the Coulomb branch moduli defined in \eqref{su3CB} \cite{Bergman:2013ala, Bao:2013pwa, Hayashi:2013qwa, Bergman:2013aca}. Therefore the Nekrasov partition function of the pure SU(3) gauge theory with the CS level $9$ is 
\begin{equation}
\begin{split}
\hat{Z}^{SU(3)_9}_{\text{top}}=&P.E.\left(\frac{2q}{(1-q)^2}\lt(A_1/A_2 + A_1^2A_2 + A_1A_2^2 + \mathfrak{q}\left(A_1/A_2 +A_1^2/A_2^2 + A_2/A_1 + A_2^2/A_1^2\right.\rt.\right.\cr
&\left.\left.+ 3A_2^3 + 2A_2^4/A_1 +3A_2^6 \right) + \mathfrak{q}^2\left(3A_2^6/A_1^3 + 16A_2^8/A_1^2\right)\right)+9\mathfrak{q}^2A_2^8/A_1^2 + \cdots \Big), \label{GV-su39v2}\cr
\end{split}
\end{equation}
where we erased $A_3$ by $A_3 = A_1^{-1}A_2^{-1}$, and the extra factor is 
\begin{align}
Z^{SU(3)_9}_{\text{extra}} = P.E.\left(\frac{8q\mathfrak{q}}{(1-q)^2} + \cdots \right).
\end{align}
We remark that 
the instaton part of the partition function is expected to be symmetric under the Weyl group of SU(3). 
Until the order we computed we can for example see a part of the symmetry from the part 
\begin{align}
\hat{Z}^{SU(3)_9}_{\text{top}} =P.E.\left(\dots+\frac{2q\mathfrak{q}}{(1-q)^2}\lt(A_1/A_2+A_1^2/A_2^2+ A_2/A_1 + A_2^2/A^2_1\rt)+\dots\right),
\end{align}
in \eqref{GV-su39v2}, which is symmetric under the exchange $A_1 \leftrightarrow A_2$\footnote{We need to compute higher order terms for seeing the paired part under the exchange $A_1 \leftrightarrow A_2$ for the other terms in the instanton part in \eqref{GV-su39v2}.}.

\section{O-vertex as a vertex operator}\label{s:O-vert-op}
The topological vertex computation may be rephrased by expectation values of some vertex operators \cite{Okounkov:2003sp,IKV}. 
In this section we propose a vertex operator corresponding to the O-vertex introduced in section \ref{s:O-vert} along the line of the vertex operator formalism. Some technical details of the vertex operator formalism are summarized in appendix \ref{a:Schur}.

\subsection{O-vertex operator}

We consider reformulating the computation using the O-vertex obtained in section \ref{s:O-vert} by using operators and their expectation values. 
This may be regarded as a first step to develop a purely analytic method to compute SO($N$) and $G_2$ partition functions in a closed form at each order of the instanton number. 
More details will be presented in a future work \cite{Hayashi-Zhu-2}. Let us try to express $V_\mu$ as an expectation value of a vertex operator $\mathbb{O}(q)$, 
\be\label{vertexO0}
V_\nu (-P)^{|\nu|}=\bra{0}\mathbb{O}(P,q)\ket{\nu}.
\ee
The ket state $\ket{\nu}$ here is the fermion basis labeled by the Frobenius coordinates of $\nu$ (see appendix \ref{a:Schur} for more details). It is straightforward to obtain the operator $\mathbb{O}(P, q)$ as a form in the expansion of the fermion basis. Since the fermion basis is orthonormal $\Braket{\nu'|\nu} = \delta_{\nu'\nu}$, 
one way to express the operator $\mathbb{O}(P, q)$ is given by
\begin{align}\label{vertexO}
\mathbb{O}(P,q) = \sum_{\mu}(-P)^{|\mu|}V_{\mu}\psi_{\mu}^{\ast},
\end{align}
with
\begin{align}
\psi_{\mu}^{\ast} = (-1)^{\beta_1 + \beta_2 + \cdots + \beta_{s} + \frac{s}{2}}\psi^{\ast}_{\alpha_1}\psi^{\ast}_{\alpha_2} \cdots \psi^{\ast}_{\alpha_s}\psi_{\beta_s} \cdots \psi_{\beta_2}\psi_{\beta_1},
\end{align}
where $\left(\alpha_1, \alpha_2, \cdots, \alpha_s|\beta_1, \beta_2, \cdots, \beta_s\right)$ is the Frobenius coordinate of a Young diagram $\mu$ (see Figure \ref{f:Frobenius}).
For example, the expression of the operator $\mathbb{O}(P,q)$ until the order $P^4$ is given by 
\begin{equation}
\begin{split}
\mathbb{O}(P,q)=&1+\frac{qP^2}{1-q}\psi^\ast_{3/2}\psi_{1/2}+\frac{P^2}{1-q}\psi^\ast_{1/2}\psi_{3/2}-\frac{q^3P^4}{(1-q)(1-q^2)}\psi^\ast_{7/2}\psi_{1/2}\cr
&-\frac{qP^4}{(1-q)(1-q^2)}\psi^\ast_{5/2}\psi_{3/2}-\frac{(1+q^3)P^4}{(1-q)(1-q^2)}\psi^\ast_{3/2}\psi^\ast_{1/2}\psi_{1/2}\psi_{3/2}\cr
&+\frac{q^2P^4}{(1-q)(1-q^2)}\psi^\ast_{3/2}\psi_{5/2}+\frac{P^4}{(1-q)(1-q^2)}\psi^\ast_{1/2}\psi_{7/2}+o(P^4),
\end{split}
\end{equation}
from the explicit form of $V_{\nu}$, which can be computed by \eqref{Vnu1}.

In fact one can infer a candidate for another form of the vertex operator $\mathbb{O}(P, q)$ which satisfies \eqref{vertexO0}. In order to see that, let us compute the perturbative part of the pure SO($2N+2$) gauge theory from the O-vertex. We focus on the contribution from the left half of the 5-brane web diagram of the pure SO($2N+2$) gauge theory as it contains the O-vertex $V_{\nu}$. The partition function from the left half of the diagram contains the square root of the perturbative part of the partition function of a pure SU($N+1$) gauge theory. From the topological vertex computation, it is possible to factor out such a part by using the identity \eqref{Schur-nor-id-spec} and the remaining contribution is given by 
\be\label{sopert0}
\begin{split}
&\sum_{\nu,\eta_1,\eta_2,\dots \eta_N}V_\nu(-P)^{|\nu|}\left(\prod_{i=1}^NQ_i^{|\eta_i|}\right)s_{\nu/\eta_1}(q^{-\rho})s_{\eta_1/\eta_2}(q^{-\rho})\dots s_{\eta_{N-1}/\eta_N}(q^{-\rho})s_{\eta_N}(q^{-\rho})\cr
&=\sum_{\nu,\eta_1,\eta_2,\dots \eta_N}V_\nu (-P)^{|\nu|}s_{\nu/\eta_1}(q^{-\rho})s_{\eta_1/\eta_2}(\tilde{Q}_1q^{-\rho})\dots s_{\eta_{N-1}/\eta_N}(\tilde{Q}_{N-1}q^{-\rho})s_{\eta_N}(\tilde{Q}_Nq^{-\rho}),
\end{split}
\ee
where we defined 
\be
\tilde{Q}_i=\prod_{j=1}^iQ_j.
\ee
It is possible to write \eqref{sopert0} in terms of the expectation value of vertex operators. First note that the skew Schur function may be written by
\be\label{SS}
s_{\lambda/\mu}(\vec{x})=\bra{\lambda}V_-(\vec{x})\ket{\mu},
\ee
where $V_-(\vec{x})$ is defined in \eqref{Vpm}. 
Then using \eqref{vertexO0} and \eqref{SS}, \eqref{sopert0} can be written as
\begin{align}
&\sum_{\nu,\eta_1,\eta_2,\dots \eta_N}V_\nu (-P)^{|\nu|}s_{\nu/\eta_1}(q^{-\rho})s_{\eta_1/\eta_2}(\tilde{Q}_1q^{-\rho})\dots s_{\eta_{N-1}/\eta_N}(\tilde{Q}_{N-1}q^{-\rho})s_{\eta_N}(\tilde{Q}_Nq^{-\rho})\cr
&=\Braket{0|\mathbb{O}(P, q)|\nu}\Braket{\nu|V_-(q^{-\rho})|\eta_1}\Braket{\eta_1|V_-(\tilde{Q}_1q^{-\rho})|\eta_2}\cdots\Braket{\eta_N|V_-\left(\tilde{Q}_Nq^{-\rho}\right)|0}\cr
&=\bra{0}\mathbb{O}(P,q)V_-(q^{-\rho})\prod_{i=1}^NV_-(\tilde{Q}_iq^{-\rho})\ket{0}\cr
&=\Braket{0|\mathbb{O}(P,q)\exp\lt(\sum_{n=1}^\infty \frac{1}{n}\frac{(1+\sum_{i=1}^N\tilde{Q}_i^{n})q^{\frac{n}{2}}}{1-q^n}J_{-n}\rt)|0},\label{sopert}
\end{align}
where we used the completeness of the Frobenius basis to sum over the Young diagrams from the second line to the third line. 

Since \eqref{sopert} gives a part of the perturbative part of the partition function of the pure SO($2N+2$) gauge theory except for the pure SU($N+1$) part, \eqref{sopert} should be equal to 
\be
P.E. \lt(\frac{P^2q}{2(1-q)^2}\lt(-(1+\sum_{i=1}^N\tilde{Q}^{2}_i)+(1+\sum_{i=1}^N\tilde{Q}_i)^2\rt)\rt).\label{pert-part}
\ee
Then comparing \eqref{sopert} with \eqref{pert-part}, we find that a candidate for another form of the vertex operator $\mathbb{O}(P, q)$ in the bosonic basis may be given by
\be
\mathbb{O}(P,q)=\exp\lt(\sum_{n=1}^\infty\left(-\frac{P^{2n}(1+q^n)}{2n(1-q^n)}J_{2n}+\frac{P^{2n}}{2n}J_nJ_n\right)\rt). \label{OV}
\ee
Indeed we can see that inserting \eqref{OV} into \eqref{sopert} yields
\begin{align}
&\Braket{0|\mathbb{O}(P,q)\exp\lt(\sum_{n=1}^\infty \frac{1}{n}\frac{(1+\sum_{i=1}^N\tilde{Q}_i^{n})q^{\frac{n}{2}}}{1-q^n}J_{-n}\rt)|0}\cr
&=\lt\langle0\lt|\exp\lt(\sum_{n=1}^\infty \frac{1}{n}\frac{(1+\sum_{i=1}^N\tilde{Q}_i^{n})q^{\frac{n}{2}}}{1-q^n}J_{-n}\rt)\exp\lt(\sum_{n=1}^\infty -\frac{P^{2n}(1+q^n)}{2n(1-q^n)}J_{2n}+\frac{P^{2n}}{2n}J_nJ_n\rt.\rt.\rt.\cr
&\lt.\lt.\lt.-\frac{P^{2n}q^n(1+\sum_{i=1}^N\tilde{Q}_i^{2n})}{2n(1-q^n)^2}+\frac{P^{2n}q^{\frac{n}{2}}(1+\sum_{i=1}^N\tilde{Q}_i^n)}{n(1-q^n)}J_n+\frac{P^{2n}q^n(1+\sum_{i=1}^N\tilde{Q}_i^n)^2}{2n(1-q^n)^2}\rt)\rt|0\rt\rangle\cr
&=\exp\lt(\sum_{n=1}^\infty -\frac{P^{2n}q^n(1+\sum_{i=1}^N\tilde{Q}_i^{2n})}{2n(1-q^n)^2}+\frac{P^{2n}q^n(1+\sum_{i=1}^N\tilde{Q}_i^n)^2}{2n(1-q^n)^2}\rt),
\end{align}
where we used a variant of the Baker-Campbell-Hausdorff formula
\be
e^Xe^Y = e^{Y}e^{X-\lt[Y,X\rt]+\frac{1}{2}\lt[Y,\lt[Y,X\rt]\rt]-\frac{1}{3!}\lt[Y,\lt[Y,\lt[Y,X\rt]\rt]\rt]+\dots},
\ee
and that $J_n\ket{0}=\bra{0}J_{-n}=0$ for $n>0$.

Similarly, we can associate a vertex operator to the O-vertex $W_{\nu}$ in \eqref{Wnu}. Note that the partition function from the right half of the 5-brane web diagram for the pure SO($2N+2$) gauge theory is given by
\ba
\sum_{\mu,\xi}(-P)^{|\mu|}\left(\prod_{i=1}^NQ_i^{|\xi_i|}\right)W_\mu q^{\frac{\kappa(\mu^t)}{2}}s_{\mu^t/\xi_1}(q^{-\rho}) s_{\xi_1/\xi_2}(q^{-\rho})\dots s_{\xi_{N-1}/\xi_N}(q^{-\rho})s_{\xi_N}(q^{-\rho}),\label{W-pert-comp}
\ea
where we again factored out the square root of the perturbative part of the partition function of a pure SU($N+1$) gauge theory. 
Since $W_{\mu}$ comes with a factor $q^{\frac{\kappa(\mu^t)}{2}}$, it will be useful to define a vertex operator $\tilde{\mathbb{O}}(P,q)$ by
\be
W_{\mu} (-P)^{|\mu|}q^{\frac{\kappa(\mu^t)}{2}}=\bra{0}\tilde{\mathbb{O}}(P,q)\ket{\mu^t}. \label{tildeO1}
\ee
From the relation \eqref{rel-O-dual-2}, the righthand side of \eqref{tildeO1} is 
\be
W_{\mu} (-P)^{|\mu|}q^{\frac{\kappa(\mu^t)}{2}} = V_{\mu^t}(-P)^{|\mu|} = \Braket{0|\mathbb{O}(P, q)|\mu^t}.
\ee
Therefore $\tilde{\mathbb{O}}(P, q)$ may be identified with $\mathbb{O}(P, q)$.

\subsection{Towards refinement of O-vertex}
\label{s:refined}
So far we have focused on the unrefined cases where the O-vertices $V_{\nu}, W_{\nu}$ and the vertex operator $\mathbb{O}(P, q)$ depends only on $q$. Let us see how the vertex operator may be modified in the refined case. We here assume that there exist vertex functions $V_{\nu}(t, q), W_{\nu}(t, q)$ associated with the intersection between an O5-plane and a 5-brane and also the partition function from a diagram away from the orientifold can be computed from the standard refined topological vertex. This refined version of the topological vertex is known to take the form \cite{IKV}
\ba
C_{\mu\nu\lambda}(t,q)=\lt(\frac{q}{t}\rt)^{\frac{||\nu||^2+||\lambda||^2}{2}}t^{\frac{\kappa(\nu)}{2}}P_{\lambda^t}(t^{-\rho},q,t)\sum_\eta \lt(\frac{q}{t}\rt)^{\frac{|\eta|+|\mu|-|\nu|}{2}}s_{\mu^t/\eta}(q^{-\lambda}t^{-\rho})s_{\nu/\eta}(t^{-\lambda^t}q^{-\rho}),\cr\label{ref-top}
\ea
where $||\lambda||^2=\sum_i\lambda_i^2$ for $\lambda=\{\lambda_i\}$ and $P_\lambda$ is the Macdonald function with the normalization such that 
\ba
P_{\lambda^t}(t^{-\rho},q,t)=t^{\frac{||\lambda||^2}{2}}\prod_{(i,j)\in\lambda}(1-t^{\lambda_j^t-i+1}q^{\lambda_i-j})^{-1}.
\ea
To use the refined topological vertex, we need to choose a preferred direction and we assign the refined topological vertex \eqref{ref-top} so that $\lambda$ is the preferred direction. The framing factor is also refined \cite{IKV, Taki07} and we assign
\be
\tilde{f}_\lambda(t,q)=(-1)^{|\lambda|}q^{-\frac{||\lambda^t||^2}{2}}t^{\frac{||\lambda||^2}{2}}\left(\frac{t}{q}\right)^{\frac{|\lambda|}{2}},
\ee
for non-preferred directions and 
\be
f_{\lambda}(t, q) = (-1)^{|\lambda|}t^{-\frac{||\lambda^t||^2}{2}}q^{\frac{||\lambda||^2}{2}},
\ee
for preferred directions.

Then, suppose that the refined version of the O-vertex $V_{\nu}(q)$ is given by $V_{\nu}(t, q)$, the partition function corresponding to \eqref{sopert0} will become\footnote{
The direct application of the refined topological vertex gives a factor $\left(\frac{t}{q}\right)^{\frac{|\nu|}{2}}$ and we defined $V_{\nu}(t, q)$ with the factor included.} 
\begin{align}
&\sum_{\nu,\eta_1,\eta_2,\dots \eta_N}V_\nu(t,q)(-P)^{|\nu|}(\prod_{i=1}^NQ_i^{|\eta_i|})s_{\nu/\eta_1}(q^{-\rho})s_{\eta_1/\eta_2}(q^{-\rho})\dots s_{\eta_{N-1}/\eta_N}(q^{-\rho})s_{\eta_N}(q^{-\rho})\cr
&=\Braket{0|\mathbb{O}(P,t, q)\exp\lt(\sum_{n=1}^\infty \frac{1}{n}\frac{(1+\sum_{i=1}^N\tilde{Q}_i^{n})q^{\frac{n}{2}}}{1-q^n}J_{-n}\rt)|0},\label{sopert0refined}
\end{align}
where we defined
\be
V_\nu(t,q) (-P)^{|\nu|}=\bra{0}\mathbb{O}(P,t,q)\ket{\nu}.
\ee
From the localization result with the two $\Omega$-deformation parameters turned on, \eqref{sopert0refined} should be equal to
\ba
P.E. \lt(\frac{P^2q}{2(1-q)(1-t)}\lt(-(1+\sum_{i=1}^N\tilde{Q}^{2}_i)+(1+\sum_{i=1}^N\tilde{Q}_i)^2\rt)\rt).\label{ref-pert-part}
\ea
Then from the comparison of \eqref{sopert0refined} with \eqref{ref-pert-part}, we obtain 
\be\label{refinedOV}
\begin{split}
\mathbb{O}(P,t,q)=&1+\frac{qP^2}{1-t}\psi^\ast_{3/2}\psi_{1/2}+\frac{P^2}{1-t}\psi^\ast_{1/2}\psi_{3/2}-\frac{q^2tP^4}{(1-t)(1-t^2)}\psi^\ast_{7/2}\psi_{1/2}\cr
&-\frac{(q-q^2+qt)P^4}{(1-t)(1-t^2)}\psi^\ast_{5/2}\psi_{3/2}-\frac{(1+q^2t)P^4}{(1-t)(1-t^2)}\psi^\ast_{3/2}\psi^\ast_{1/2}\psi_{1/2}\psi_{3/2}\cr
&+\frac{(q-t+qt)P^4}{(1-t)(1-t^2)}\psi^\ast_{3/2}\psi_{5/2}+\frac{P^4}{(1-t)(1-t^2)}\psi^\ast_{1/2}\psi_{7/2}
+o(P^5).
\end{split}
\ee
At first sight, it seems difficult to obtain each coefficient in \eqref{refinedOV}. However we can use the fact that $\tilde{Q}_i$ dependence is summarized as $(1 + \sum_{i=1}^N\tilde{Q}_i^n)$ in \eqref{sopert0refined}. The contraction of fermions in \eqref{refinedOV} in the Frobenius basis with the vertex operator given by $\exp\lt(\sum_{n=1}^\infty \frac{1}{n}\frac{(1+\sum_{i=1}^N\tilde{Q}_i^{n})q^{\frac{n}{2}}}{1-q^n}J_{-n}\rt)$ 
yields polynomials generated by $(1 + \sum_{i=1}^N\tilde{Q}_i^n)$ for each order of $P^{2k}$. 
Also each coefficient of $P^{2k}$ in \eqref{ref-pert-part} can be also written by a polynomial of $(1 + \sum_{i=1}^N\tilde{Q}_i^n)$. 
We can then compare the two expressions to determine $\mathbb{O}(P, t, q)$. Until the orders we computed we have enough equations to determine the parameters for $\mathbb{O}(P, t, q)$.

We remark that in the unrefined case, the vertex operators for 
$V_{\nu}$ can be the same as that for $W_\nu q^{\frac{\kappa(\nu^t)}{2}}$, i.e. $\tilde{\mathbb{O}}(P,q)=\mathbb{O}(P,q)$. In the refined case, there is one more step we need to take. Note that the perturbative part (\ref{Zroot}) we ought to reproduce in the topological vertex formalism has two parts. 
The perturbative part we use for obtaining the vertex operator $\mathbb{O}(P, t, q)$ from the left part is given by \eqref{ref-pert-part} and this comes from the the first term in the exp in \eqref{Zroot}. The the partition function from the right half of the diagram needs to reproduce the second term in \eqref{Zroot}, namely
\be
P.E. \lt(\frac{P^2t}{2(1-q)(1-t)}\lt(-(1+\sum_{i=1}^N\tilde{Q}^{2}_i)+(1+\sum_{i=1}^N\tilde{Q}_i)^2\rt)\rt).\label{ref-pert-part1}
\ee
On the other hand the partition function computed by applying the refined topological vertex to the right half of the diragram with the factors for the perturabtive part of the pure SU($N+1$) gauge theory removed 
gives 
\begin{equation}
\begin{split}
&\sum_{\nu,\eta_1,\eta_2,\dots \eta_N}W_\nu(t,q)(-P)^{|\nu|}(\prod_{i=1}^NQ_i^{|\eta_i|})s_{\nu^t/\eta_1}(q^{-\rho})s_{\eta_1/\eta_2}(q^{-\rho})\dots s_{\eta_{N-1}/\eta_N}(q^{-\rho})s_{\eta_N}(q^{-\rho})\cr
&=\Braket{0|\tilde{\mathbb{O}}(P,t, q)\exp\lt(\sum_{n=1}^\infty \frac{1}{n}\frac{(1+\sum_{i=1}^N\tilde{Q}_i^{n})q^{\frac{n}{2}}}{1-q^n}J_{-n}\rt)|0},\label{sopert0refined1}
\end{split}
\end{equation}
where we defined
\be
W_\nu(t,q) (-P)^{|\nu|}=\bra{0}\tilde{\mathbb{O}}(P,t,q)\ket{\nu^t}.
\ee
We also included the factor $t^{-\frac{||\nu||^2}{2}}q^{\frac{||\nu^t||^2}{2}}\left(\frac{q}{t}\right)^{-\frac{|\nu|}{2}}$ which comes from the refined topological vertex computation in the definition of $W_{\nu}(t, q)$. 
Then \eqref{sopert0refined1} has the same form as \eqref{sopert0refined} but the final result obtained from \eqref{sopert0refined1} needs to be equal to \eqref{ref-pert-part1}, rather than \eqref{ref-pert-part}. Note that \eqref{ref-pert-part1} is reproduced by changing $P$ into $P\left(\frac{t}{q}\right)^{\frac{1}{2}}$ in \eqref{ref-pert-part}. Hence a vertex operator associated to $W_{\nu}(P, t, q) $ is expected to be given by $\tilde{\mathbb{O}}(P,t,q)=\mathbb{O}\left(P\left(\frac{t}{q}\right)^{\frac{1}{2}},t, q\right)$. 

\section{Conclusion}
\label{s:concl}

In this paper, we introduced a new type of the topological vertex called O-vertex, which is associated to the intersection point between an O5$^-$-plane, $(2, 1)$ (or $(2, -1)$) 5-brane and an O5$^+$-plane, by making use of the formalism in \cite{Kim-Yagi}. The O-vertex is labeled by a Young diagram assigned to the $(2, 1)$ (or $(2, -1)$) 5-brane. With the O-vertex we computed the partition functions of the pure SO($N$) $(N=4, 5, 6, 7, 8)$ gauge theories and also of the pure $G_2$ gauge theory. The comparison between the results obtained by the O-vertex and the known results showed the perfect agreement until some orders for the perturbative part, the one-instanton part and the two-instanton part, which legitimates our proposal. We then applied the O-vertex to the 5-brane web diagram of the pure SU($3$) gauge theory with the CS level $9$ and computed its partition function. It would be interesting to perform more consistency checks for the result of the pure SU$(3)$ gauge theory with the CS level $9$. We also presented the GV-invariants of low orders. Furthermore we also provided a vertex operator formalism for the O-vertex. We proposed an explicit expression for the vertex operator associated to the O-vertex. 

There are two crucial steps in the study of the O-vertex and the instanton partition function in the future. One is to find a closed form for the O-vertex. The O-vertex $V_{\nu}$ we proposed is given by the sum of finite terms for each order of the associated K$\ddot{\text{a}}$hler parameter. Since it is given by the sum of finite terms,  we can compute the O-vertex exactly at each order. However it would be still interesting to find a closed formula for the O-vertex. We also proposed a candidate in the form of the vertex operator in section \ref{s:O-vert-op}, but we do not have a proof of this proposal at the moment. The polynomial $P_\nu$ appearing in the O-vertex has various properties as observed in appendix \ref{a:O-vert}. 
It will also be interesting to prove these observed properties. The link of the O-vertex with the dimer model and the melting crystal \cite{Okounkov:2003sp,Kenyon:2003uj,Ooguri:2008yb,Ooguri:2009ri,Aganagic:2009kf,Li:2020rij} will also be a topic to work on in the future. 

The other is to establish the refinement of this generalized topological vertex formalism. 
By achieving that, we can start to discuss the relation with the formalism constructed in the S-dual setup \cite{D-type,Kimura:2019gon}, and to derive the qq-characters \cite{NPS,BPS/CFT,Kimura-Pestun,BMZ,Kim-qq,5dBMZ,Kimura:2016dys,Kimura-r,BFHMZ,Kimura:2017auj}\footnote{The unrefined limit of the qq-characters is certainly also interesting, as one can expect things to be simplified in this limit from the computation of \cite{Haouzi:2020yxy}.} associated to BCD-type gauge theories. The fundamental qq-characters of BCD-type gauge theories was discussed in a recent work \cite{Haouzi:2020yxy}, and unlike the beautiful results obtained for A-type gauge theories, they contain infinite number of terms and there is no known closed form for these qq-characters. Since the topological vertex formalism computes the partition function in a different basis (labeled by Young diagrams), it might simplify the expression of the qq-characters beyond the A-type gauge group. The relation with the blow-up equation \cite{Nakajima:2003pg,Nakajima:2005fg,Huang:2017mis,Gu:2018gmy,Kim:2019uqw}, which requires the refinement, would be also of great interest in the current context.

\acknowledgments
We thank J.-E. Bourgine, Z. Duan, T. Kimura, S.-S. Kim, K. Lee, D. O'Connor, K. Sun, F. Yagi, H. Zhang for useful discussions. We are grateful to APCTP for hospitality during the initial stage of this work. 
The work of HH is supported in part by JSPS KAKENHI Grant Number JP18K13543. 

\appendix

\section{Explicit form of O-vertex}\label{a:O-vert}

In this appendix, 
we provide the explicit form of the O-vertices $V_{\nu}, W_{\nu}$ defined in \eqref{Vnu1} \eqref{Wnu} respectively. As discussed in section \ref{s:O-vert}, we observed that the O-vertex $V_{\nu}$ may be written as \eqref{Vnu}. We here write down the explicit form of $P_{\nu}(q)$ in \eqref{Vnu} and see an interesting pattern.  
The expression of $P_{\nu}(q)$ for $|\nu| = 2, 3$ has been written in \eqref{P-1} and \eqref{P-2} and the explicit form for $6 \leq |\nu| \leq 10$ is 
\begin{equation}
\begin{split}
&P_{(6)}=-q^6,\quad P_{(5,1)}=q^3,\quad P_{(4,2)}=-(q+q^5+q^6),\quad P_{(4,1,1)}=q^4+q^5,\quad P_{(3,3)}=1+q^4+q^5,\cr
&P_{(3,2,1)}=0,\quad P_{(3,1,1,1)}=-(q+q^2),\quad P_{(2,2,2)}=-(q+q^2+q^6),\quad P_{(2,2,1,1)}=1+q+q^5,\cr
&P_{(2,1,1,1,1)}=-q^3,\quad P_{(1,1,1,1,1,1)}=1,
\end{split}
\end{equation}
\begin{equation}
\begin{split}
&P_{(8)}=q^{10},\quad P_{(7,1)}=-q^6,\quad P_{(6,1,1)}=-(q^7+q^8+q^9),\quad P_{(6,2)}=q^3+q^8+q^9+q^{10},\cr
&P_{(5,3)}=-(q+q^6+q^7+q^8),\quad P_{(5,1,1,1)}=q^3+q^4+q^5,\quad P_{(5,2,1)}=P_{(3,2,1,1,1)}=0,\cr
&P_{(4,2,1,1)}=-(q+q^2+q^3+q^7+q^8+q^9),\quad P_{(4,3,1)}=-(q^6+q^8), P_{(4,1,1,1,1)}=q^5+q^6+q^7,\cr
&P_{(4,4)}=1+q^5+q^6+q^7+q^8+q^{10},\quad P_{(4,2,2)}=q^2+q^3+q^4+q^5+q^7+q^8+q^9+q^{10},\cr
&P_{(3,3,2)}=-(q+q^2+q^3+q^7+q^8+q^9),\quad P_{(3,2,2,1)}=-(q^2+q^4),\quad P_{(3,1,1,1,1,1)}=-(q+q^2+q^3),\cr
&P_{(3,3,1,1)}=1+q+q^2+q^3+q^5+q^6+q^7+q^8,\quad P_{(2,2,2,2)}=1+q^2+q^3+q^4+q^5+q^{10},\cr
&P_{(2,2,2,1,1)}=-(q^2+q^3+q^4+q^9)\quad  P_{(2,2,1,1,1,1)}=1+q+q^2+q^7,\cr
&P_{(2,1,1,1,1,1,1)}=-q^4,\quad P_{(1,1,1,1,1,1,1,1)}=1,
\end{split}
\end{equation}
\begin{equation}
\begin{split}
&P_{(10)}=q^{15},\quad P_{(9,1)}=-q^{10},\quad P_{(8,2)}=q^6+q^{12}+q^{13}+q^{14}+q^{15},\quad P_{(8,1,1)}=-(q^{11}+q^{12}+q^{13}+q^{14}),\cr
&P_{(7,3)}=-q^3-q^9-q^{10}-q^{11}-q^{12},\quad P_{(7,2,1)}=0,\quad P_{(7,1,1,1)}=q^6+q^7+q^8+q^9,\cr
&P_{(6,4)}=q+q^7+q^8+q^9+q^{10}+q^{11}+q^{12}+q^{13}+q^{14}+q^{15},\cr
&P_{(6,3,1)}=-(q^9+q^{10}+q^{11}+q^{12}+q^{13}),\quad P_{(6,1,1,1,1)}=q^8+q^9+2q^{10}+q^{11}+q^{12},\cr
&P_{(6,2,1,1)}=-(q^3+q^4+q^5+q^6+q^{10}+q^{11}+2q^{12}+q^{13}+q^{14}),\cr
&P_{(5,5)}=-(1+q^6+q^7+q^8+q^9+q^{10}+q^{11}+q^{12}+q^{13}+q^{14}),\quad P_{(5,4,1)}=0,\cr
&P_{(5,3,1,1)}=q+q^2+q^3+q^4+q^5+q^6+q^7+2q^8+2q^9+2q^{10}+q^{11}+q^{12},\quad P_{(5,2,1,1,1)}=0,\cr
&P_{(5,1,1,1,1,1)}=-(q^3+q^4+2q^5+q^6+q^7),\cr
&P_{(4,4,2)}=q+q^2+q^3+q^4+q^5+q^6+q^7+2q^8+2q^9+2q^{10}+2q^{11}+2q^{12}+q^{13}+q^{14}+q^{15}.
\end{split}
\end{equation}
Then we observe that the polynomial $P_{\nu}(q)$ satisfies the following properties,
\begin{itemize}
\item $P_\nu(q)$ is a polynomial of $q$ of degree at most $m(|\nu|)=\frac{n(n+1)}{2}$ for $n=\frac{|\nu|}{2}$ and is zero for $|\nu|$ odd. 

\item When we denote 
\ba
P_\nu(q)=\sum_{i=0}^{m(|\nu|)}a_iq^i,
\ea
then we have 
\ba
P_{\nu^t}(q)=(-)^{n}\sum_{i=0}^{m(|\nu|)}a_iq^{m(|\nu|)-i}.
\ea
All coefficients $a_i$ seem to have the same sign, and the overall sign is determined from the property described below. 

\item For $\nu=(|\nu|)$, $P_\nu(q)=(-1)^nq^m$. For $\nu=(1,1,\dots,1)$, $P_{\nu}(q)=1$. 

\item As for the sign of $P_\nu$, when we move a box in the $i$-th line of $\nu$ to the $(i+j)$-th line to form a new Young diagram $\nu'$, then the sign of $P_{\nu'}$ differs from that of $P_\nu$ by $(-1)^{j}$.

\item In the $q=1$ limit, we found that $P_\nu(1)$ gives the number of ways to generate the Young diagram $\nu$ from only $(2)$ and $(1,1)$. For example, $P_{(2,2)}(1)=2$, and it can indeed be generated from the addition of a $(2)$ on the top of another $(2)$, or putting $(1,1)$ on the right of another $(1,1)$. More explicitly, we have 
\ba
\ydiagram{2,2}\ =\ \ydiagram{2}+\ydiagram{2}\ , \quad {\rm or}\quad \ydiagram{2,2}\ =\ \ydiagram{1,1}+\ydiagram{1,1}\ .
\ea
We checked this recursive relation until 
$|\nu| = 8$. As a further example 
for $|\nu| = 10$, we consider $(4,4,2)$, for which $P_{(4,4,2)}(1)=20$. It can be generated from $(4,4)$ and $(4,2,2)$ by adding $(2)$ respectively at a suitable position, and also from $(3,3,2)$ by adding $(1,1)$ on the right. Indeed we can check that 
\ba
P_{(4,4,2)}(1)=P_{(4,4)}(1)+P_{(4,2,2)}(1)+P_{(3,3,2)}(1)=6+8+6=20.
\ea

\end{itemize}

On the other hand $W_{\nu}$ is written by \eqref{Wnu0}
and the explicit form of $\tilde{P}_{\nu}(q)$ for $|\nu| = 6$ in \eqref{Wnu0} 
is given by 
\begin{equation}
\begin{split}
&\tilde{P}_{(6)}=q^{15},\quad \tilde{P}_{(5,1)}=-q^{12}\quad \tilde{P}_{(4,2)}=q^{5}+q^6+q^{10},\cr
&\tilde{P}_{(4,1,1)}=-(q^{4}+q^5),\quad \tilde{P}_{(3,3)}=q^{4}+q^5+q^9,\quad \tilde{P}_{(3,2,1)}=0,\cr
&\tilde{P}_{(3,1,1,1)}=q+q^2,\quad \tilde{P}_{(2,2,2)}=q^{-3}+q+q^2,\quad \tilde{P}_{(2,2,1,1)}=-(q^{-4}+1+q),\cr
&\tilde{P}_{(2,1,1,1,1)}=q^{-6},\quad \tilde{P}_{(1,1,1,1,1,1)}=-q^{-9}.
\end{split}
\end{equation}
In general $W_{\nu}$ is related to $V_{\nu}$ by \eqref{rel-O-dual-2} as shown in \eqref{Wnudeform}.

\section{Some useful formulae}
\label{a:schur}
In this appendix, we summarize some useful formulae that we make use of in the computations in this paper.

\subsection{Nekrasov partition functions}
\label{s:Nek}

The partition functions of certain 5d $\mathcal{N}=1$ supersymmetric gauge theories may be computed by the localization method. It consists of two factors, the perturbative part and the instanton part. The perturbative part of a pure gauge theory with a gauge group $G$ is in general given by 
\begin{align}
Z^{G}_{\text{pert}} = Z_{\text{Cartan}}^GZ_{\text{root}}^G,
\end{align}
where 
\be
Z_{\text{root}}^G = P.E.\left(\left(\frac{q}{(1-q)(1-t)} + \frac{t}{(1-q)(1-t)}\right)\sum_{\alpha \in \Delta_+}e^{-\alpha\cdot a}\right),\label{Zroot}
\ee
and
\be
Z_{\text{Cartan}}^G = P.E.\left(\frac{\text{rank}(G)}{2}\left(\frac{q}{(1-q)(1-t)} + \frac{t}{(1-q)(1-t)}\right)\right).
\ee
$\Delta_+$ is the set of positive roots of the Lie algebra $\mathfrak{g}$ of $G$ and $a = (a_1, \cdots, a_{\text{rank}(G)})$ is the Coulomb branch moduli in the Cartan subalgebra. $q, t$ are related to the $\Omega$-deformation parameters $\epsilon_1, \epsilon_2$ by $q=e^{-\epsilon_1}, t = e^{\epsilon_2}$. The unrefined case corresponds to $q = t$. 

On the other hand, the partition function of the instanton part is more involved. The instanton part of the pure SU($N$) gauge theory with the zero CS level is given by 
\be\label{locsun}
Z^{SU(N)}_{\text{loc, inst}}=\sum_{k=0}^\infty \mathfrak{q}^k\frac{1}{|W(SU(k))|}\oint\lt(\prod_{i=1}^k\frac{{\rm d}\phi_i}{2\pi i}\rt)Z_{k}^{SU(N)},
\ee
with 
\ba
|W(G)|=\lt\{\begin{array}{cc}
n! & G=SU(n)\\
2^{n-1+\delta}n! & G=O(2n+\delta)\\
2^nn! & G=Sp(n)\\
\end{array},\rt.
\ea
and 
\ba
Z^{SU(N)}_k=\frac{[2\epsilon_+]^k}{[\epsilon_{1,2}]^k}\prod_{i=1}^k\prod_{j=1}^N[\phi_i-a_j\pm\epsilon_+]^{-1}\prod_{\substack{i,j=1\\i< j}}^k\frac{[\phi_{ij}]^2[\phi_{ij}\pm2\epsilon_+]}{[\phi_{ij}\pm\epsilon_1][\phi_{ij}\pm \epsilon_2]},
\ea
where $\phi_{ij}=\phi_i-\phi_j$, 
\be
\lt[x\rt]:=2\sinh\frac{x}{2}=e^{\frac{x}{2}}-e^{-\frac{x}{2}},\quad \epsilon_\pm=\frac{\epsilon_1\pm\epsilon_2}{2}.
\ee
$a_i\; (i=1, \cdots, N)$ with $\sum_{i=1}^Na_i = 0$ are the Coulomb branch moduli. $\mathfrak{q}$ is the instanton fugacity. 
In fact there is a well-known closed-form formula for the instanton partition function \eqref{locsun} of the pure SU($N$) gauge theory with the CS level zero, 
\ba\label{sun}
Z^{SU(N)}_{\text{loc, inst}}=\sum_{k=0}^\infty \mathfrak{q}^{k}\sum_{\substack{\lambda_{1,2,\dots,N}:\ {\rm partitions}\\\sum_{i=1}^N|\lambda_i|=k}} \prod_{i,j=1}^N N_{\lambda_i\lambda_j}(a_{i}-a_j;\epsilon_1,\epsilon_2)^{-1},
\ea
where the Nekrasov factor $N_{\lambda\nu}(a;\epsilon_1,\epsilon_2)$ is given by \cite{NekrasovInstanton,nakajimalectures,AW08}
\ba
N_{\lambda\nu}(a;\epsilon_1,\epsilon_2):=\prod_{(i,j)\in\lambda}[a+\epsilon_1(\lambda^t_i-j)+\epsilon_2(-\nu_j+i-1)]\prod_{(i,j)\in\nu}[a+\epsilon_1(-\nu_i^t+j-1)+\epsilon_2(\lambda_j-i)].\cr
\ea

The instanton part of the partition function of the pure SO($2N+\delta$) gauge theory for $\delta = 0, 1$ is also given by a contour integral. The Losev-Moore-Nekrasov-Shatashvili (LMNS) integrand of the integral for the $k$-instanton partition function of the pure SO($2N+\delta$) gauge theory  
is 
\cite{Marino:2004cn, Nekrasov-Shadchin,Fucito:2004gi,Shadchin:2005mx,Hollands:2010xa}
\begin{align}
&Z^{SO(2N+\delta)}_k\cr
&=(-1)^k\frac{[2\epsilon_+]^k}{[\epsilon_{1,2}]^k}\prod_{i<j}\cS(\pm \phi_i\pm\phi_j-\epsilon_+)^{-1}\prod_{i=1}^k\prod_{j=1}^N[\pm \phi_i\pm a_j-\epsilon_+]^{-1}\prod_{i=1}^k\frac{[\pm 2\phi_i][\pm 2\phi_i+2\epsilon_+]}{[\pm \phi_i-\epsilon_+]^\delta},\cr
\end{align}
where 
\be
\cS(\phi):=\frac{[\phi\pm\epsilon_-]}{[\phi\pm\epsilon_+]}.
\ee
Then the instanton partition function is found via the contour integral 
\be
Z^{SO(2N+\delta)}_{\text{loc, inst}}=\sum_{k=0}^\infty \mathfrak{q}^k\frac{1}{|W(Sp(k))|}\oint\lt(\prod_{i=1}^k\frac{{\rm d}\phi_i}{2\pi i}\rt)Z_{k}^{SO(2N+\delta)}.\label{LMNS-integral}
\ee
The contour integral \eqref{LMNS-integral} may be evaluated by the Jeffrey-Kirwan residues \cite{Benini:2013xpa, Hwang:2014uwa, Hori:2014tda}. Or alternatively one can choose the contour in the following way in the unrefined cases with $\epsilon_1 + \epsilon_2=0$. 
We first assume that ${\rm Im}(\epsilon_1)={\rm Im}(\epsilon_2)=\delta$, $\epsilon_+=i\delta\in i\mathbb{R}$ and evaluate the contour integral by picking up the poles in the upper half plane. 
Then we take the limit $\delta\rightarrow 0$ at the end. As argued in \cite{Marino:2004cn,Fucito:2004gi}, the poles may be labeled by colored Young diagrams in general. 

There is in fact a subtlety for the instanton partition function computed by the localization method. The instanton partition function may contain a factor which does not depend on Coulomb branch moduli 
In that case, one needs to remove the extra factor $Z_{\text{loc, extra}}$ from the instanton partition function $Z_{\text{loc, inst}}$ to obtain the correct instanton partition function of a UV complete 5d theory \cite{Bergman:2013ala, Bao:2013pwa, Hayashi:2013qwa, Bergman:2013aca, Hwang:2014uwa}. Namely we consider 
\be
\hat{Z}_{\text{loc, inst}} =Z_{\text{loc, inst}}/Z_{\text{loc, extra}}.
\ee
Similarly we also need to remove an extra factor from the instanton partition function computed by the topological vertex. We see that this indeed happens for the pure SO(4) gauge theory in section \ref{s:so4}.

As for the one-instanton partition function there is a universal formula for  
an arbitrary gauge group $G$ and it is given by \cite{Benvenuti:2010pq, ABCDEFG-instanton, Zafrir:2015uaa, Billo:2015pjb, Billo:2015jyt}, 
\ba
Z^G_{1\text{-inst}}=\sum_{\gamma \in\Delta_l}\frac{e^{(h^\vee -1)\gamma/2}}{(1-e^{(\epsilon_1+\epsilon_2)+\gamma})(e^{\gamma/2}-e^{-\gamma/2})\prod_{\alpha,\gamma^\vee\cdot\alpha=1}
(e^{\alpha/2}-e^{-\alpha/2})},\label{uni-one-inst}
\ea
where $\Delta_l$ is the set of long roots in the Lie algebra, $\mathfrak{g}$ of $G$, and for each of simple root $\alpha_i$, $e^{\alpha_i}$ here is identified with the Coulomb branch 
moduli $A_i$. This formula will be repeatedly used in the consistency check in this article. Furthermore, the higher-instanton partition function can be obtained using the blow up formula in \cite{Keller:2012da, Kim:2019uqw}. Again, if the partition function contains an extra factor which is independent of the Coulomb branch moduli, one needs to remove the extra factor to obtain the correct instanton partition function.

\subsection{Schur functions}\label{a:Schur}

The definition of a Schur function $s_\lambda(\{x_i\}_{i=1}^n)$ is 
\ba
s_\lambda(x_1,\dots,x_n):=\frac{{\rm det}(x_i^{\lambda_j+n-j})}{{\rm det}(x_i^{n-j})},
\ea
for a Young diagram $\lambda=\{\lambda_j\}$. It is, by definition, a symmetric polynomial and the set of all Schur functions forms a complete basis of symmetric polynomials. Therefore, the product of two Schur functions can be expressed as a linear combination of Schur functions,
\ba
s_\mu s_\nu=\sum_\lambda c^\lambda_{\mu\nu}s_\lambda.
\ea
With this fusion coefficient $c^\lambda_{\mu\nu}$, we further define the skew Schur function, 
\ba
s_{\lambda/\mu}:=\sum_\nu c^\lambda_{\mu\nu}s_\nu.
\ea

\begin{figure}[t]
\begin{center}
\begin{tikzpicture}
\draw (0,0)--(2.5,0);
\draw (0,-0.5)--(2.5,-0.5);
\draw (0,-1)--(2,-1);
\draw (0,-1.5)--(2,-1.5);
\draw (0,-2)--(1,-2);
\draw (0,-2.5)--(1,-2.5);
\draw (0,-3)--(0.5,-3);
\draw (0,0)--(0,-3);
\draw (0.5,0)--(0.5,-3);
\draw (1,0)--(1,-2.5);
\draw (1.5,0)--(1.5,-1.5);
\draw (2,0)--(2,-1.5);
\draw (2.5,0)--(2.5,-0.5);
\draw[dashed] (0,0)--(2.5,-2.5);
\draw[<->] (2.5,-0.25)--(0.25,-0.25);
\draw[<->] (2,-0.75)--(0.75,-0.75);
\draw[<->] (2,-1.25)--(1.25,-1.25);
\draw[<->] (0.25,-0.25)--(0.25,-3);
\draw[<->] (0.75,-0.75)--(0.75,-2.5);
\draw[<->] (1.25,-1.25)--(1.25,-1.5);
\node at (2.5,-0.25) [right] {$\alpha_1$};
\node at (2,-0.75) [right] {$\alpha_2$};
\node at (2,-1.25) [right] {$\alpha_3$};
\node at (0.25,-3) [below] {$\beta_1$};
\node at (0.75,-2.5) [below] {$\beta_2$};
\node at (1.25,-1.5) [below] {$\beta_3$};
\end{tikzpicture}
\end{center}
\caption{
The Frobenius coordinates can be read as the length of each row and column to the diagonal line for a Young diagram $\lambda$. 
As an example, we show the Young diagram $\lambda=(5,4,4,2,2,1)$ here. 
Then the Frobenius coordinates are given by $(\alpha_1,\alpha_2,\alpha_3|\beta_1,\beta_2,\beta_3)=\left(\frac{9}{2},\frac{5}{2},\frac{3}{2}\big|\frac{11}{2},\frac{7}{2},\frac{1}{2}\right)$.}
\label{f:Frobenius}
\end{figure}
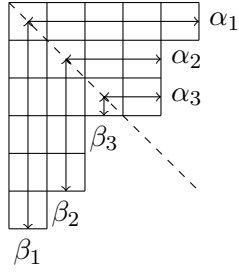

An important fact about the skew Schur function is that it can be expressed as a fermion correlation \cite{Macdonald-book,kac_1990}, 
\ba
s_{\lambda/\mu}(\vec{x})=\bra{\mu}V_+(\vec{x})\ket{\lambda}=\bra{\lambda}V_-(\vec{x})\ket{\mu},\label{skew-schur}
\ea
where 
\ba\label{Vpm}
V_\pm (\vec{x})=\exp\lt(\sum_{n=1}^\infty \frac{1}{n}\sum_i x_i^n J_{\pm n}\rt),
\ea
with
\be
J_n:=\sum_{j\in \mathbb{Z}+1/2}\psi_{-j}\psi^\ast_{j+n}.
\ee
$J_n$ and $\psi_n, \psi_n^{\ast}$ satisfy the following commutation relations, 
\begin{align}
&\{\psi_n,\psi_m\}=\{\psi_n^\ast,\psi_m^\ast\}=0,\quad \{\psi_n,\psi^\ast_m\}=\delta_{n+m,0}, \\
&\lt[J_n,\psi_k\rt]=\psi_{n+k},\quad \lt[J_n,\psi^\ast_k\rt]=-\psi^\ast_{n+k},\quad \lt[J_n,J_m\rt]=n\delta_{n+m,0}.
\end{align}
$\ket{\lambda}$ is a fermion basis with the label of the Frobenius coordinate (see Figure \ref{f:Frobenius}) 
of a Young diagram $\lambda$. When the Frobenius coordinate of a Young diagram $\lambda$ is $\lambda=(\alpha_1,\alpha_2,\dots|\beta_1,\beta_2\dots)$, then $\ket{\lambda}$ is given by
\ba
\ket{\lambda}=(-1)^{\beta_1+\beta_2+\dots+\beta_s+\frac{s}{2}}\psi^\ast_{-\beta_1}\psi^\ast_{-\beta_2}\dots\psi^\ast_{-\beta_s}\psi_{-\alpha_s}\psi_{-\alpha_{(s-1)}}\dots \psi_{-\alpha_1}\ket{0},\label{Frobenius-basis}
\ea
where $s$ is the number of boxes on the diagonal line in $\lambda$ and the vacuum state $\ket{0}$ satisfies $\psi_\alpha\ket{0}=\psi^\ast_\beta\ket{0}=0$ for any $\alpha>0$, $\beta>0$. 

There are two Cauchy identities known for the skew Schur functions. 
\ba
\sum_\lambda s_{\lambda/\mu}(x)s_{\lambda/\nu}(y)=\prod_{i,j}(1-x_iy_j)^{-1}\sum_\eta s_{\nu/\eta}(x)s_{\mu/\eta}(y),\label{Schur-nor-id}\\
\sum_\lambda s_{\lambda/\mu^t}(x)s_{\lambda^t/\nu}(y)=\prod_{i,j}(1+x_iy_j)\sum_\eta s_{\nu^t/\eta}(x)s_{\mu/\eta^t}(y)\label{Schur-twist-id}.
\ea
To see the first identify \eqref{Schur-nor-id} we first note that the lefthand side of \eqref{Schur-nor-id} can be written as
\be
\sum_\lambda \bra{\mu}V_+(\vec{x})\ket{\lambda}\bra{\lambda}V_-(\vec{y})\ket{\nu} =\bra{\mu}V_+(\vec{x})V_-(\vec{y})\ket{\nu}.\label{sidv1}
\ee
Then by using the commutation relation 
\be
V_+(\vec{x})V_-(\vec{y})=\prod_{i,j}\frac{1}{1-x_iy_j}V_-(\vec{y})V_+(\vec{x}),\label{V-V+}
\ee
\eqref{sidv1} can be further written as
\begin{align}
\bra{\mu}V_+(\vec{x})V_-(\vec{y})\ket{\nu} &= \prod_{i,j}(1-x_iy_j)^{-1}\bra{\mu}V_-(\vec{y})V_+(\vec{x})\ket{\nu}\cr
&=\prod_{i,j}(1-x_iy_j)^{-1}\sum_{\eta}\bra{\mu}V_-(\vec{y})\ket{\eta}\bra{\eta}V_+(\vec{x})\ket{\nu}, \label{sidv2}
\end{align}
and then the righthand side of \eqref{sidv2} precisely reproduces the righthand side of \eqref{Schur-nor-id}.

The second identity can be derived using the automorphism of the fermionic algebra, $\mho:\psi\leftrightarrow \psi^\ast$, under which $\lambda$ transforms to $\lambda^t$ as $\mho$ exchanges $\{\alpha_i\}$ and $\{\beta_i\}$. In particular, we have 
\ba
\ket{\lambda^t}=(-1)^{|\lambda|}\mho(\ket{\lambda}).\label{mho-identity}
\ea
The second Cauchy identity then follows directly from the fact that the transformation $\mho$ does not change the expectation value of a correlator and $\mho^2={\bf id}$. i.e. 
\begin{align}
\bra{\lambda^t}V_-(\vec{y})\ket{\nu}&=(-1)^{|\lambda|}\mho(\bra{\lambda})V_-(\vec{y})\ket{\nu}\cr
&=(-1)^{|\lambda|}\mho^2(\bra{\lambda})\mho(V_-(\vec{y}))\mho(\ket{\nu})\cr
&=(-1)^{|\lambda|-|\nu|}\bra{\lambda}\mho(V_-(\vec{y}))\ket{\nu^t}\cr
&=\bra{\lambda}V^{-1}_-(-\vec{y})\ket{\nu^t},
\end{align}
where we used $\mho(J_n)=-J_n$. Therefore 
\begin{align}
\sum_\lambda\bra{\mu^t}V_+(\vec{x})\ket{\lambda}\bra{\lambda^t}V_-(\vec{y})\ket{\nu}&=\sum_\lambda\bra{\mu^t}V_+(\vec{x})\ket{\lambda}\bra{\lambda}V_-^{-1}(-\vec{y})\ket{\nu^t}\cr
&=\bra{\mu^t}V_+(\vec{x})V_-^{-1}(-\vec{y})\ket{\nu^t}\cr
&=\prod_{i,j}(1+x_iy_j)\bra{\mu^t}V_-^{-1}(-\vec{y})V_+(\vec{x})\ket{\nu^t}\cr
&=\prod_{i,j}(1+x_iy_j)\sum_{\eta}\bra{\mu^t}V_-^{-1}(-\vec{y})\ket{\eta}\bra{\eta}V_+(\vec{x})\ket{\nu^t}\cr
&=\prod_{i,j}(1+x_iy_j)\sum_\eta \bra{\mu}V_-(\vec{y})\ket{\eta^t}\bra{\eta}V_+(\vec{x})\ket{\nu^t},\label{sidv3}
\end{align}
where we used
\be
V_+(\vec{x})V_-^{-1}(-\vec{y})=\prod_{i,j}(1+x_iy_j)V_-^{-1}(-\vec{y})V_+(\vec{x}).\label{V-V+v2}
\ee
We can see that \eqref{sidv3} yields \eqref{Schur-twist-id}.

The specification of the variables, $\{x\}=\{q^{-\rho-\sigma}\}$ and $\{y\}=\{q^{-\rho-\tau}\}$, is very useful in this article, and in this case (\ref{Schur-nor-id}) and (\ref{Schur-twist-id}) become
\ba
&&\sum_\lambda Q^{|\lambda|} s_{\lambda/\mu}(q^{-\rho-\sigma})s_{\lambda/\nu}(q^{-\rho-\tau})\nn\\
&&=P.E.\lt(\frac{q}{(1-q)^2}Q\rt)N^{-1}_{\sigma^t\tau}(Q,q)\sum_\eta Q^{|\mu|+|\nu|-|\eta|}s_{\nu/\eta}(q^{-\rho-\sigma})s_{\mu/\eta}(q^{-\rho-\tau}),\label{Schur-nor-id-spec}\\
&&\sum_\lambda (-Q)^{|\lambda|}s_{\lambda/\mu^t}(q^{-\rho-\sigma})s_{\lambda^t/\nu}(q^{-\rho-\tau})\nn\\
&&=P.E.\lt(-\frac{q}{(1-q)^2}Q\rt)N_{\sigma^t\tau}(Q,q)\sum_\eta (-Q)^{|\mu|+|\nu|-|\eta|} s_{\nu^t/\eta}(q^{-\rho-\sigma})s_{\mu/\eta^t}(q^{-\rho-\tau}),\nn\\\label{Schur-twist-id-spec}
\ea
where the unrefined Nekrasov factor is defined by 
\ba\label{Nekfactor}
N_{\lambda\nu}(Q,q):=\prod_{(i,j)\in\lambda}(1-Q q^{\lambda_i+\nu^t_j-i-j+1})\prod_{(i,j)\in\nu}(1-Qq^{-\nu_i-\lambda^t_j+i+j-1}).
\ea

From \eqref{skew-schur} and the completeness of the fermionic basis, the following relation also holds,
\ba
\sum_\eta s_{\mu/\eta}(x)s_{\eta/\nu}(y)=\sum_\eta \bra{\mu}V_-(\vec{x})\ket{\eta}\bra{\eta}V_-(\vec{y})\ket{\nu}=\bra{\mu}V_-(\{\vec{x},\vec{y}\})\ket{\nu}.
\ea

\bibliographystyle{JHEP}
\bibliography{O-vert}
\end{document}